\documentclass[twocolumn, journal]{IEEEtran}

\IEEEoverridecommandlockouts

\usepackage{cite}
\usepackage{amsmath,amssymb,amsfonts}
\usepackage{algorithmic}
\usepackage{graphicx}
\usepackage{textcomp}
\usepackage{xr}
\def\BibTeX{{\rm B\kern-.05em{\sc i\kern-.025em b}\kern-.08em
    T\kern-.1667em\lower.7ex\hbox{E}\kern-.125emX}}

\usepackage{tikz}
\usetikzlibrary{backgrounds}
\usetikzlibrary{spy}
\usetikzlibrary{calc,arrows.meta}
\usepackage{pgfplots}
\usetikzlibrary{plotmarks}
\usetikzlibrary{arrows.meta}
\usetikzlibrary{positioning,fit}
\usetikzlibrary{intersections}
\usetikzlibrary{patterns}
\usetikzlibrary{decorations.shapes}

\usepgfplotslibrary{patchplots}
\usepgfplotslibrary{colormaps}

\usepackage{cancel}

\usepackage{pdfpages} %

\usepackage{multirow}
\usepackage{makecell}

\usepackage{algorithmic}
\usepackage[ruled,vlined]{algorithm2e}
\usepackage{setspace}
\SetKwRepeat{Do}{do}{while}
\SetKwRepeat{Until}{do}{until}

\pgfdeclarelayer{back1}
\pgfdeclarelayer{lay1}
\pgfdeclarelayer{back2}
\pgfdeclarelayer{lay2}
\pgfsetlayers{back2,lay2,back1,lay1,main}

\newcommand{\exportFigures}{true}
\newcommand{\exportFiguresAsPNG}{true}

\ifthenelse{\equal{\exportFigures}{true}}
{
  \usepgfplotslibrary{external}
  \tikzexternalize[prefix=compiled_tikz_figures/,optimize command away=\includepdf]
  \ifthenelse{\equal{\exportFiguresAsPNG}{true}}
  {
    \tikzset
    {   png export/.style={
        external/system call={
        pdflatex \tikzexternalcheckshellescape -halt-on-error --extra-mem-top=10000000 -interaction=batchmode -jobname "\image" "\texsource" && pdftops -eps "\image.pdf" && convert -density 700 -transparent white "\image.pdf" "\image.png"
    }}}
  \tikzset{png export}
  }
  {}
}
{}

\usepackage{url}

\usepackage{bm}
\usepackage[caption=false,font=footnotesize]{subfig}
\usepackage[normalem]{ulem}

\usepackage{mathtools, cuted}

\usepackage{acronym}

\usepackage{booktabs}

\usepackage[latin1]{inputenc}
\usepackage{tikz}
\usetikzlibrary{shapes,arrows}
\usetikzlibrary{arrows.meta}
\usetikzlibrary{positioning}

\usepackage{ellipsis}

\usetikzlibrary{calc}
\usetikzlibrary{decorations.pathreplacing,decorations.markings,shapes.geometric}
\usetikzlibrary{decorations.pathmorphing}
\usetikzlibrary{fit}
\usetikzlibrary{pgfplots.groupplots}

\usetikzlibrary{calc,arrows.meta}

\usetikzlibrary{backgrounds} %

\definecolor{green(pigment)}{rgb}{0.0, 0.65, 0.31}
\definecolor{frenchblue}{rgb}{0.0, 0.45, 0.73} 
\definecolor{mediumcandyapplered}{rgb}{0.89, 0.02, 0.17}

\usepackage[most]{tcolorbox}
\tcbuselibrary{breakable}
\tcbset{every box/.style={enhanced,breakable}}
\tcbset{colframe=black,colback=red!10,enhanced,breakable,sharp corners}

\usepackage{enumitem} %

\usepackage{notation}

\usepackage{adjustbox}

\definecolor{alex}{RGB}{51,183,150}
\definecolor{erik}{RGB}{235,134,52}
\newcommand{\alex}[1]{{\color{alex}#1}}

\newcommand{\ticked}{$\text{\rlap{$\checkmark$}}\square$}
\newcommand{\unticked}{{$\square$}}
\newcommand{\tick}[1]{\ifthenelse{#1=1}{\ticked}{\unticked}}

\newcommand{\rmv}{\hspace*{-.3mm}}
\newcommand{\norm}[2]{\ensuremath{\lVert #1 \rVert^{#2}}}%

\newcommand{\minus}{\rmv - \rmv}

\newcommand{\s}{\hspace*{0.5pt}}

\newcommand{\pd}{p_{\text{d}}({u}_{k,n}^{(j)})}
\newcommand{\pdrv}{p_{\text{d}}(\rv{u}_{k,n}^{(j)})}

\providecommand{\norm}[1]{\lVert#1\rVert}

\newcommand{\rd}{\textcolor{red}}
\newcommand{\ist}{\hspace*{.3mm}}
\newcommand{\iist}{\hspace*{1mm}}

\newcommand{\nn}{\nonumber}

\DeclareMathOperator*{\argmin}{arg\,min}

\newcommand{\zd}{\ensuremath{{z_\mathrm{d}^{(j)}}_{\rmv\rmv\rmv\rmv\rmv\rmv\rmv  m,n}}}
\newcommand{\zu}{\ensuremath{{z_\mathrm{u}^{(j)}}_{\rmv\rmv\rmv\rmv\rmv\rmv\rmv m,n}}}
\newcommand{\zdr}{\ensuremath{{\rv{z}_\mathrm{d}^{(j)}}_{\rmv\rmv\rmv\rmv\rmv\rmv\rmv  m,n}}}
\newcommand{\zur}{\ensuremath{{\rv{z}_\mathrm{u}^{(j)}}_{\rmv\rmv\rmv\rmv\rmv\rmv\rmv m,n}}}

\newcommand{\zuZero}{\ensuremath{{z_\mathrm{u}^{(j)}}_{\rmv\rmv\rmv\rmv\rmv\rmv\rmv m,0}}}
\newcommand{\zdZero}{\ensuremath{{z_\mathrm{d}^{(j)}}_{\rmv\rmv\rmv\rmv\rmv\rmv\rmv  m,0}}}
\newcommand{\zuZeroMax}{\ensuremath{{{z}_\mathrm{u}^{(j)}}_{\rmv\rmv\rmv\rmv\rmv\rmv\rmv \text{max}, 0}}}

\newcommand{\noise}{\ensuremath{n}}

\newlength{\figureheight}
\newlength{\figurewidth}
\graphicspath{{./figures/}}

\allowdisplaybreaks

\definecolor{mycolor01}{rgb}{0.00000,0.00000,1.00000}
\definecolor{mycolor02}{rgb}{0.133,0.545,0.133}
\definecolor{mycolor03}{rgb}{0.50000,0.00000,0.50000}
\definecolor{mycolor05}{rgb}{1.00000,0.83984,0.00000}
\definecolor{mycolor04}{rgb}{0.92969,0.50781,0.92969}
\definecolor{mycolor06}{rgb}{1.00000,0.64453,0.00000}
\definecolor{mycolor07}{rgb}{0.50000,0.50000,0.50000}
\definecolor{mycolor08}{rgb}{1.00000,0.00000,0.00000}
\definecolor{mycolor09}{rgb}{0.2510 ,0.8784, 0.8157}
\definecolor{mycolor10}{rgb}{0.54297,0.00000,0.00000}
\definecolor{mycolor11}{rgb}{0.6445, 0.1641,0.1641}
\definecolor{mycolor12}{rgb}{1, 0, 1}

 \pgfdeclarelayer{back1}
 \pgfdeclarelayer{lay1}
 \pgfdeclarelayer{back2}
 \pgfdeclarelayer{lay2}
 \pgfsetlayers{back2,lay2,back1,main,lay1}

\makeatletter
  
\tikzset{
  nomorepostactions/.code={\let\tikz@postactions=\pgfutil@empty},
  decmark/.style 2 args={decoration={markings, pre length=#2,
    mark= between positions 0 and 1 step (1/6)*\pgfdecoratedpathlength with{%
        \tikzset{solid, every mark, line width=0.5pt}\tikz@options
        \pgftransformresetnontranslations
        \pgfuseplotmark{#1}%
      },  
    },
    postaction={decorate},
    /pgfplots/legend image post style={
        mark=#1, mark options={solid}, every path/.append style={nomorepostactions}
    },
  },
  posmark/.style 2 args={decoration={markings,
		mark= at position #2*\pgfdecoratedpathlength with{%
			\tikzset{solid,every mark, line width=0.5pt}\tikz@options
			\pgftransformresetnontranslations
			\pgfuseplotmark{#1}%
		},  
	},
	postaction={decorate},
	/pgfplots/legend image post style={
		mark=#1,mark options={solid},every path/.append style={nomorepostactions}
	},
  },
markbeginend/.style 2 args={decoration={markings,
		mark= between positions 0 and 1 step (1)*\pgfdecoratedpathlength with{%
			\tikzset{#2,every mark}\tikz@options
			\pgfuseplotmark{#1}%
		},  
	},
	postaction={decorate},
	/pgfplots/legend image post style={
		mark=#1,mark options={#2},every path/.append style={nomorepostactions}
	},
},
markend/.style 2 args={decoration={markings,
		mark= at position \pgfdecoratedpathlength with{%
			\tikzset{#2,every mark}\tikz@options
			\pgfuseplotmark{#1}%
		},  
	},
	postaction={decorate},
	/pgfplots/legend image post style={
		mark=#1,mark options={#2},every path/.append style={nomorepostactions}
	},
},
}

\makeatother

\pgfplotsset{
resultStyle1/.style={mark=none, line width=0.5pt, mycolor01, decmark={oplus}{0}},
resultStyle2/.style={mark=none, line width=0.5pt, mycolor02, decmark={triangle}{0}},%
resultStyle3/.style={mark=none ,line width=0.5pt, mycolor03, decmark={+}{0}},
resultStyle4/.style={mark=none, line width=0.5pt, mycolor06, decmark={star}{0}},
resultStyle5/.style={mark=none, line width=0.5pt, mycolor08, decmark={o}{0}},
resultStyle6/.style={mark=none, line width=0.5pt, mycolor05, decmark={square}{0}}, 
resultStyle7/.style={mark=none, line width=0.5pt, mycolor09, decmark={diamond}{0}}, 
resultStyle8/.style={mark=none, line width=0.5pt, mycolor11, decmark={otimes}{0}}, 
resultStyle9/.style={mark=none, line width=0.5pt, mycolor12, decmark={x}{0}}, 
resultStyleBase/.style={mark=none, line width=0.5pt,}, 
compareStyle1/.style={mark=none, line width=0.5pt, mycolor01},
compareStyle2/.style={mark=none, line width=0.5pt, mycolor02},%
compareStyle3/.style={mark=none ,line width=0.5pt, mycolor03},
compareStyle4/.style={mark=none, line width=0.5pt, mycolor06},
compareStyle5/.style={mark=none, line width=0.5pt, mycolor08},
compareStyle6/.style={mark=none, line width=0.5pt, mycolor05}, 
compareStyle7/.style={mark=none, line width=0.5pt, mycolor09}, 
compareStyle8/.style={mark=none, line width=0.5pt, mycolor11}, 
compareStyle9/.style={mark=none, line width=0.5pt, mycolor12}, 
}
  
  \pgfplotsset{
        compat=newest,
        simple style group/.style={
                label style={font=\scriptsize},
                legend style={font=\scriptsize},
                tick label style={font=\scriptsize},
                nodes near coords style={font=\scriptsize},
                title style={font=\scriptsize},
                scale only axis,
                grid style={dotted},
                mark options={solid}, %
        },
        simple style/.style={
                label style={font=\scriptsize},
                legend style={font=\scriptsize},
                tick label style={font=\scriptsize},
                nodes near coords style={font=\scriptsize},
                title style={font=\scriptsize},
                width=\figurewidth,
                height=\figureheight,
                at={(0\figurewidth,0\figureheight)},
                scale only axis,
                grid style={dotted},
                mark options={solid}, %
        },
        base style/.style={
                label style={font=\scriptsize},
                legend style={font=\scriptsize},
                tick label style={font=\scriptsize},
                nodes near coords style={font=\scriptsize},
                title style={font=\scriptsize},
                width=\figurewidth,
                height=\figureheight,
                at={(0\figurewidth,0\figureheight)},
                scale only axis,
                cycle list={
                {mark=none, line width=0.5pt, mycolor01, solid},
                {mark=none, line width=0.5pt, mycolor02, dash dot},
                {mark=none ,line width=0.5pt, mycolor03, densely dashed},
                {mark=none, line width=0.5pt, mycolor04, dash dot dot},
                {mark=x   , line width=0.5pt, mycolor05},
                {mark=.   , line width=0.7pt, mycolor06}, 
                {mark=square,only marks, mark size = 0.8pt, mycolor07,
                mark options = {line width = 0.4pt}},
                {mark=x,     only marks, mark size = 1.3pt, mycolor08,
                mark options = {line width = 0.4pt}},
                {mark=o,     only marks, mark size = 0.8pt, mycolor09,
                mark options = {line width = 0.4pt}},
                {mark=o, mycolor10},
                },
                grid style={dotted},
                xmajorgrids,
                ymajorgrids,
                mark options={solid}, %
        },
        base style group/.style={
        	label style={font=\scriptsize},
        	legend style={font=\scriptsize},
        	tick label style={font=\scriptsize},
        	nodes near coords style={font=\scriptsize},
        	title style={font=\scriptsize},
        	scale only axis,
        	grid style={dotted},
        	xmajorgrids,
        	ymajorgrids,
        	mark options={solid}, %
        },
        std graph style new/.style={
                xlabel style={yshift=1mm},
                ylabel style={yshift=-1.5mm},
                yticklabel style={xshift=1mm},
        },
        color lines style/.style={
                cycle list={
                    {mark=none, mycolor01, decmark={oplus}{0} },
                    {mark=none, mycolor02, decmark={+}{0} }, 
                    {mark=none, mycolor03, decmark={triangle}{0} }, 
                    {mark=none, mycolor04, decmark={star}{0} }, 
                    {mark=none, mycolor05, decmark={o}{0} },
                    {mark=none, mycolor06, decmark={square}{0} },
                },
        },
        meas graph style/.style={
                xlabel style={yshift=1mm},
                ylabel style={yshift=-1mm},
                xmajorgrids,
                ymajorgrids,
                mark repeat = 1,
                mark phase = 0,
                cycle list={
                    {color=black, only marks, mark=*, mark size=0.5pt, mark options={solid, black}},
                    {color=red, only marks, mark=*, mark size=0.1pt, line width=0.25pt},
                },
                ylabel={},
        }, 
        ci graph style/.style={
                xlabel style={yshift=1mm},
                ylabel style={yshift=-1.5mm},
                yticklabel style={xshift=1mm},
                mark repeat = 1,
                mark phase = 0,
                ymin=1e-3,
                ymax=100,
                ytick = {100, 50, 10, 1, 0.1, 0.01, 1e-3, 1e-4},
                yticklabels = {$0$, $50$, $90$, $99$, $99.9$, $99.99$, $99.999$, $99.9999$},
                y dir=reverse,
        },     
        bp coeff style/.style={
               scale only axis=true,
               width=0.225*.9\linewidth,
               height=0.225*.9\linewidth,
               scale only axis,
               xmin=-4.000,
               xmax=4.000,
               xlabel={$\ell${\color{white}$\aod$}},
               ticklabel style={font=\footnotesize},
               ymin=0.000, ymax=0.9,
               ylabel={$c_\ell$},
               xlabel style={font=\footnotesize},
               ylabel style={font=\footnotesize},
               major tick length=2pt%
        },
        bp graph style/.style={        
               scale only axis=true,
               width=0.35*1.1\linewidth,
               height=0.225*.9\linewidth,
               scale only axis,
               xmin=-3.14, xmax=3.14,
               xlabel={$\aod${\color{white}$\ell$}},
               ticklabel style={font=\footnotesize},
               xtick={-3.14,-1.57,0.0,1.57,3.14},
               xticklabels={$-\pi$,$-\tfrac{\pi}{2}$,$0$,$\tfrac{\pi}{2}$,$\pi$},
               ymin=0.000, ymax=3,
               ylabel={Beampattern},
               xlabel style={font=\footnotesize}, ylabel style={font=\footnotesize},
               major tick length=2pt
        },
        peb graph style/.style={        
               width=0.66\linewidth,%
               scale only axis,
               point meta min=-2.583,
               point meta max=-0.300,
               axis on top,
               xmin=0.000,
               xmax=12.000,
               xlabel={x in meter},
               y dir=reverse,
               ymin=0.000,
               ymax=8.000,
               ylabel={y in meter},
               ytick={7.0,6.0,...,0.0},
               xtick={0.0,1.0,...,12.0},
               yticklabels={$1$,$2$,$3$,$4$,$5$,$6$,$7$,$8$},
               xlabel style={font=\scriptsize,yshift=0.125cm},
               ylabel style={font=\scriptsize,yshift=-0.125cm},
               ticklabel style={font=\scriptsize},
               unit vector ratio*=1 1 1,
               yticklabel pos=left,
               major tick length=2pt,
               colormap={mymap}{[1pt] rgb(0pt)=(1,1,1); rgb(1pt)=(0.858903,0.984776,0.839302); rgb(2pt)=(0.777958,0.94143,0.649487); rgb(3pt)=(0.755504,0.864264,0.463393); rgb(4pt)=(0.777509,0.754439,0.310168); rgb(5pt)=(0.820314,0.619497,0.21003); rgb(6pt)=(0.854796,0.471879,0.170327); rgb(7pt)=(0.851327,0.326629,0.183322); rgb(8pt)=(0.784671,0.198575,0.225774); rgb(9pt)=(0.637629,0.0993149,0.259577); rgb(10pt)=(0.400067,0.0343393,0.229819); rgb(11pt)=(0,0,0)},
               colorbar style={ylabel={Position Error Bound in centimeter (logscale)}, ytick={-0.4,-0.82,...,-2.92}, yticklabels={$39.8$, $15.1$, $5.8$, $2.2$, $0.8$, $0.3$},ylabel style={yshift=0.5mm,font=\scriptsize,scale=0.8},width=2.0mm,xshift=-4.25mm,ticklabel style={font=\scriptsize},major tick length=0pt}, %
               colormap access=piecewise constant
        },
        peb ellipses/.style={color=white, line width=0.4pt, forget plot}
    }

\tikzset{naming/.style={align=center,font=\small}}
\tikzset{antenna/.style={insert path={-- coordinate (ant#1) ++(0,0.25) -- +(135:0.25) + (0,0) -- +(45:0.25)}}}
\tikzset{station/.style={naming,draw,shape=dart,shape border rotate=90, minimum width=10mm, minimum height=10mm,outer sep=0pt,inner sep=3pt}}
\tikzset{mobile/.style={naming,draw,shape=rectangle,minimum width=12mm,minimum height=6mm, outer sep=0pt,inner sep=3pt}}
\tikzset{radiation/.style={{decorate,decoration={expanding waves,angle=90,segment length=4pt}}}}

\tikzset{
  pobl/.style={
    inner sep=0pt, outer sep=0pt, fill=#1,
  },
  pobl gron/.style n args={2}{
    pobl=#1, rounded corners=#2,
  },
  pics/person/.style n args={3}{
    code={
      \node (-corff) [pobl=#1, minimum width=.25*#2, minimum height=.375*#2, rotate=#3, pic actions] {};
      \node (-pen) [minimum width=.3*#2, circle, pobl=#1, outer sep=.01*#2, anchor=south, rotate=#3, pic actions] at (-corff.north) {};
      \node (-coes dde) [pobl gron={#1}{1pt}, anchor=north west, minimum width=.12125*#2, minimum height=.25*#2, rotate=#3, pic actions] at (-corff.south west) {};
      \node [pobl=#1, anchor=north, minimum width=.12125*#2, minimum height=.15*#2, rotate=#3, pic actions] at (-coes dde.north) {};
      \node (-coes chwith) [pobl gron={#1}{1pt}, anchor=north east, minimum width=.12125*#2, minimum height=.25*#2, rotate=#3, pic actions] at (-corff.south east) {};
      \node [pobl=#1, anchor=north, minimum width=.12125*#2, minimum height=.15*#2, rotate=#3, pic actions] at (-coes chwith.north) {};
      \node (-braich dde) [pobl gron={#1}{.75pt}, minimum width=.075*#2, minimum height=.325*#2, outer sep=.0064*#2, anchor=north west, rotate=#3, pic actions] at (-corff.north east)  {};
      \node [pobl=#1, minimum width=.05*#2, minimum height=.2*#2, outer sep=.0064*#2, anchor=north west, rotate=#3, pic actions] at (-corff.north east) {};
      \node (-braich chwith) [pobl gron={#1}{.75pt}, minimum width=.075*#2, minimum height=.325*#2, outer sep=.0064*#2, anchor=north east, rotate=#3, pic actions] at (-corff.north west) {};
      \node [pobl=#1, minimum width=.0375*#2, minimum height=.2*#2, outer sep=.0064*#2, anchor=north east, rotate=#3, pic actions] at (-corff.north west) {};
      \node (-fit person) [fit={(-pen.north) (-braich dde.east) (-coes chwith.south) (-braich chwith.west)}] {};
    },
  },
  pics/SBS/.style={code={
      \begin{scope}[local bounding box=#1]
      \fill [pic actions/.try] (-1,0) -- (-1/2,3) -- (1/2, 3) -- (1,0) -- cycle;
      \fill [pic actions/.try] (-1/16,2) rectangle (1/16,4);
      \fill [pic actions/.try] (0,4) circle [radius=1/4];
      \foreach \i in {-1,1}
        \fill [shift=(90:4), xscale=\i]
          \foreach \r in {1,3/2,2}{
            (-45:\r) arc (-45:45:\r) -- (45:\r-1/10)
            arc(45:-45:\r-1/10) -- cycle
          };
       \end{scope}
  }},
}

\newcommand*\meqref[1]{\cite[Eq.~\eqref{M-#1}]{Supplement}}
\newcommand*\mref[2]{\cite[#1~\ref{M-#2}]{Supplement}}

\begin{document}

\title{\huge{Graph-based Simultaneous Localization and Bias Tracking\\[-1mm]}}	
\author{\normalsize \IEEEauthorblockN{Alexander Venus,~\IEEEmembership{\normalsize  Member,~IEEE}, Erik Leitinger,~\IEEEmembership{\normalsize Member,~IEEE}, Stefan Tertinek,\\[-0mm] Florian~Meyer~\IEEEmembership{\normalsize Member,\hspace{-.3mm} IEEE}, and Klaus Witrisal,~\IEEEmembership{\normalsize Member,~IEEE}\\[0mm]}
\vspace*{-8mm}
\thanks{
A.\ Venus and E.\ Leitinger are with the Signal Processing and Speech Communication Laboratory, Graz University of Technology, Graz, Austria, and the Christian Doppler Laboratory for Location-aware Electronic Systems (e-mail: (erik.leitinger,a.venus)@tugraz.at). 
S.\ Tertinek is with NXP Semiconductors, Gratkorn, Austria (stefan.tertinek@nxp.com). 
F.\ Meyer is with the Department of Electrical and Computer Engineering and Scripps Institution of Oceanography, University of California San Diego, San Diego, CA, USA (e-mail: fmeyer@ucsd.edu). 
This paper is an extension of work originally presented in \cite{VenusAsilomar2022}. 
The financial support by the Christian Doppler Research Association, the Austrian Federal Ministry for Digital and Economic Affairs and the National Foundation for Research, Technology and Development is gratefully acknowledged.}
}

\maketitle
\frenchspacing

\renewcommand{\baselinestretch}{0.96}\small\normalsize %

\begin{abstract}
We present a factor graph formulation and particle-based sum-product algorithm for robust localization and tracking in multipath-prone environments. 
The proposed sequential algorithm jointly estimates the mobile agent's position together with a time-varying number of \acp{mpc}. The \acp{mpc} are represented by ``delay biases" corresponding to the offset between \ac{los} component delay and the respective delays of all detectable \acp{mpc}. The delay biases of the \acp{mpc} capture the geometric features of the propagation environment \acl{wrt} the mobile agent. Therefore, they can provide position-related information contained in the \acp{mpc} without explicitly building a map of the environment. We demonstrate that the position-related information enables the algorithm to provide high-accuracy position estimates even in fully \ac{olos} situations. 
Using simulated and real measurements in different scenarios we demonstrate that the proposed algorithm significantly outperforms state-of-the-art multipath-aided tracking algorithms and 
show that the performance of our algorithm constantly attains the \ac{pcrlb}. 
Furthermore, we demonstrate the implicit capability of the proposed method to identify unreliable measurements and, thus, to mitigate lost tracks. 
\end{abstract}

\acresetall %

\IEEEpeerreviewmaketitle

\vspace{-2.5mm} 
\section{Introduction}\label{sec:introduction}

\begin{figure}[t]
	
	\centering
	\setlength{\abovecaptionskip}{0mm}
	\setlength{\belowcaptionskip}{0pt}
	
	\setlength{\figurewidth}{0.38\textwidth}
	\setlength{\figureheight}{0.1\textwidth}

	\tikzsetnextfilename{eye_catcher}
	\hspace{-3mm}\includegraphics{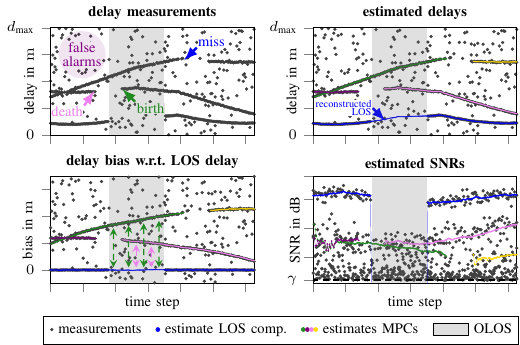}
	\caption{In \ac{olos} situations the \ac{los} path is reconstructed using the position-related information contained in the delay biases of all detected \acp{mpc}. The delay biases as well as their SNRs are jointly estimated together with the agent's state. The estimation problem is complicated by false alarms (clutter), missed detections (miss), object birth (birth), and object death (death).
	}\label{fig:eye_catcher}
	\vspace{-6mm}
\end{figure}

Localization of mobile agents using radio signals is still a challenging task in indoor or urban scenarios\cite{WitrisalSPM2016Copy, Mendrzik2019}. %
Here, the environment is characterized by strong multipath propagation (commonly referred to as ``\ac{nlos} propagation'') and frequent \ac{olos} situations, which can prevent the correct extraction of the information contained in the \ac{los} component (see Fig.~\ref{fig:eye_catcher}). 
There exist many safety and security-critical applications, such as autonomous driving \cite{Karlsson2017}, or medical services \cite{KoEMBMag2010}, %
 where robustness of the position estimate\footnote{We define robustness as the percentage of cases in which a system can achieve its given potential accuracy. I.e., a robust sequential localization algorithm can keep the agent's track in a very high percentage of cases, even in challenging environments.} is of critical importance. 
\vspace{-2.5mm}
\subsection{State-of-the-Art Methods} \label{sec:sota}
Joint sensing and communication systems will be a defining feature of future 6G communication networks \cite{Chaccour2022_6G,Wymeersch2022_LocSense1}. To enable integrated sensing, computationally feasible algorithms that provide accurate location information, even in challenging environments such as indoor or urban scenarios, are of paramount importance. 
New localization and tracking approaches within the context of 6G networks take advantage of large measurement apertures as provided by \ac{uwb} systems \cite{DardariProcIEEE2009, TaponeccoTWC2011} and the large number of antennas in mmWave systems \cite{RusekSPM2013}, allowing to resolve the received radio signal into a superposition of a finite number of specular \acp{mpc} \cite{DardariProcIEEE2009,ShenTIT2010,AdityaProc2018}. 
These approaches mitigate the effect of multipath propagation \cite{GiffordTSP2022} aiming to obtain unbiased estimates of the \ac{los} component \cite{AdityaProc2018, DardariProcIEEE2009, WymeerschIEEE2012} 
 or even take advantage of \acp{mpc} by exploiting inherent position information.  Thus, multipath propagation is turned from an impairment to an asset \cite{GentnerTWC2016,ShahmansooriTWC2018, LeitingerTWC2019}. %
Prominent examples are \ac{mpslam} methods  \cite{LeitingerTWC2019, LeitingerICC2019, GentnerTWC2016, KimTWC2020}  
that estimate \acp{mpc} and associate them to \acp{va} representing the locations of the mirror images of anchors on reflecting surfaces \cite{PedersenJTAP2018}. The locations of \acp{va} are %
estimated jointly with the position of the mobile agent.  %
In this way, \ac{mpslam} can provide high-accuracy position estimates, even in \ac{olos} situations \cite{VenusTWC2023}. %
However, \ac{mpslam} requires \acp{mpc} that can be resolved and correspond to specular reflections on flat, even surfaces in the environment with sufficient extent \cite{PedersenJTAP2018}. 
This is why the method introduced in \cite{WieVenWilLeiArxiv2023} performs \ac{mpslam} considering antenna dispersion and diffuse / non-resolvable \acp{mpc} at the cost of increased problem complexity. 
Similarly, the method introduced in \cite{Yu2020,VenusTWC2023} exploits the positional information of \acp{mpc} using a low-complexity model featuring a single bias to a stochastically modeled multipath ``cluster'' to perform robust positioning and tracking. 
Although \ac{mpslam} can be straightforwardly extended to %
 three dimensions\cite{KimWym:TVT2022}, this significantly  increases the complexity of the inference model and, thus, complicates the numerical representation. Furthermore, in scenarios where %
the number of detectable \acp{va} is low (sparse information), geometric ambiguity can lead to a multimodal state distribution and, thus, cause the algorithm to follow wrong modes \cite{KrekovicTSP2020}. 
Machine learning methods avoid model-based representations, relying on data to capture details of the actual environment. Yet, applying deep machine learning to complex inference tasks is not straightforward.  
Early approaches to learning-aided multipath-based positioning extract specific features from the radio channel and apply model-agnostic supervised regression methods on these features \cite{MaranoJSAC2010,WymeerschIEEE2012}. While these approaches potentially provide high accuracy estimates at low computational demand (after training), they suffer from their dependence on a large representative database and can fail in scenarios that are not sufficiently represented by the training data. This is why recent algorithms use deep learning and auto encoding-based methods, directly operating on the received radio signal \cite{LiShenMILCOM2021,StahlkeSensors2021,HuangTMC2022} and hybrid, physics-informed learning models \cite{ContiProcIEEE2019, LiShenMILCOM2021,Kram2022GPR} to reduce the dependence on training data. 

Bayesian inference leveraging graphical models provides a powerful and flexible means that has been widely used in applications like %
multipath-based localization \cite{GentnerTWC2016,LeitingerTWC2019,LeitingerICC2019,KimTWC2020,Yu2020}, %
multiobject-tracking \cite{BarShalomTCS2009,MeyerProc2018,MeyWilJ21}, %
and parametric channel tracking \cite{LiTWC2022}. The underlying estimation problems pose common challenges such as uncertainties beyond Gaussian noise (missed detections and clutter), an uncertain origin of measurements, and unknown and time-varying numbers of objects to be localized and tracked. 
As the measurement models of these applications are non-linear, most methods typically rely on particle-based implementations or linearization \cite{ArulampalamTSP2002, DurrantWhyte2006}.
Similarly, the \ac{pda} algorithm \cite{BarShalomTCS2009,BarShalom1995} represents a low-complexity Bayesian method for robust localization and tracking with extension to multiple-sensors \ac{pda} \cite{JeoTugTAES2005} and \ac{pdaai} \cite{LerroACC1990,LeitingerICC2019}. 
All these methods can be categorized as ``two-step approaches'', in the sense that they do not operate on the received sampled radio signal, but use extracted measurements provided by a preprocessing step, providing a high level of flexibility and a significant reduction of computational complexity. In contrast, ``direct positioning approaches" such as \cite{ZhaStaJosWanGenDamWymHoeTAES2020,KropfreiterFUSION2021} directly exploit the received sampled signal, which can lead to a better detectability of low-\ac{snr} features, yet, they are computationally demanding.

\vspace{-3mm}
\subsection{Contributions} \label{sec:contributions}

The problem studied in this paper can be summarized as follows.

\vspace{0.5mm}
\noindent{\textit{Estimate the time-varying location of a mobile agent using \ac{los} propagation and NLOS propagation of radio signals.}}
\vspace{1.2mm}

We propose a particle-based algorithm for robust localization and tracking that estimates the state of a mobile agent by utilizing the position-related information contained in the \ac{los} component as well as in \acfp{mpc} of multiple sensors (anchors). 
{Similar to other ``two-step approaches", it uses \ac{mpc} delays and corresponding amplitude measurements provided by a snapshot-based parametric \acf{ceda}.}
The proposed algorithm performs joint probabilistic data association and sequential estimation \cite{WilliamsTAES2014,MeyerProc2018,LeitingerTWC2019} of a mobile agent state together with all parameters of a time-varying number of \acfp{pbo}\footnote{The introduced \ac{pbo} model is based on state-of-the art concepts of factor graph-based dynamic multi-object tracking \cite{MeyerProc2018,LeitingerTWC2019} and loopy joint probabilistic data association \cite{WilliamsTAES2014,MeyerProc2018} (see Section~\ref{sec:dynamic_mot})).},
using message passing by means of the \ac{spa} on a factor graph \cite{KschischangTIT2001}.
\acp{pbo} contain a state representation of ``delay biases", denoting the delay difference between the \ac{los} component and the respective \acp{mpc}, as well as a binary variable denoting the existence of the respective \acp{mpc}. %
This model enables the algorithm to utilize the position information contained in the \acp{mpc} without building an explicit representation\footnote{Unlike \ac{mpslam} methods \cite{GentnerTWC2016,LeitingerTWC2019,KimTWC2020}, map features such as \acp{va}\cite{PedersenJTAP2018,Meissner2015Diss} or scatter points \cite{GentnerTWC2016} are not explicitly modeled. Their geometric information is implicitly captured by sequential inference of delay biases.} of an environment map \cite{GentnerTWC2016,LeitingerTWC2019} in order to support the estimation of the agent state (see Fig.~\ref{fig:eye_catcher}). This allows the algorithm to operate reliably in challenging environments, characterized by strong multipath propagation and temporary \ac{olos} situations without using any prior information (no training data are needed). %
The key contributions of this paper are summarized as follows.
\begin{itemize}[leftmargin=1.5em]
	\item We introduce a Bayesian probabilistic model for \ac{mpc}-aided 
	localization and tracking of the position by sequential inference of a time-varying number of ``delay biases'' represented by \acp{pbo}. 
	\item We present an \ac{spa} based on the factor graph representation of the estimation problem where the \ac{pbo} states are estimated jointly and sequentially, demonstrating that the information contained in \acp{pbo} dramatically increases the performance in \ac{olos} situations. We also  demonstrate the capability of the proposed method to identify unreliable estimates using the existence probability of the \acp{pbo}.
	\item We compare the proposed \ac{spa} to other state-of-the-art algorithms for \ac{mpc}-aided localization and tracking as well as to the \ac{pcrlb} \cite{Tichavsky1998} using both synthetic and real radio measurements. Specifically, we compare to our robust positioning method from \cite{VenusTWC2023}, and to the \ac{mpslam} method presented in \cite{LeitingerTWC2019,LeitingerICC2019}. For synthetic measurements, we also compare to the channel SLAM algorithm from \cite{GentnerTWC2016}, and the learning-based methods presented in \cite{WymeerschIEEE2012} and \cite{Kram2022GPR}.
\end{itemize}

This work advances over the preliminary account of our %
conference publication \cite{VenusAsilomar2022} by (i) presenting a detailed derivation of the proposed \ac{spa} and its particle-based implementation, (ii) thoroughly analyzing the geometric relations underlying the proposed model, (iii) presenting a comprehensive numerical analysis of the algorithm performance, (iv) comparing the proposed \ac{spa} to the \ac{mpslam} algorithm presented in \cite{LeitingerTWC2019,LeitingerICC2019} and to the learning-based methods presented in \cite{WymeerschIEEE2012} and \cite{Kram2022GPR,StahlkeSensors2021}, (v) validating its performance using real radio measurements 
 and, (vi) demonstrating its implicit capability to identify unreliable measurements. %
\subsection{Notations and Definitions} \label{sec:notation}
Column vectors and matrices are denoted by boldface lowercase and uppercase letters. Random variables are displayed in san serif, upright font, e.g., $\rv{x}$ and $\RV{x}$ and their realizations in serif, italic font, e.g. $x$. %
$f({x})$ and $p({x})$ denote, respectively, the \ac{pdf} or \ac{pmf} of a continuous or discrete random variable $\rv{x}$ (these are short notations for $f_\rv{x}({x})$ or $p_\rv{x}({x})$). 
$(\cdot)^{\mathrm{T}}$, $(\cdot)^\ast$, and $(\cdot)^{\text{H}}$ denote matrix transpose, complex conjugation and Hermitian transpose, respectively. 
$ \norm{\cdot}{} $ is the Euclidean norm. $ \vert\cdot\vert $ represents the cardinality of a set. $ \mathrm{diag}\{\V{x}\} $ denotes a diagonal matrix with entries in $ \V{x} $. %
$\M{I}_{[\cdot]}$ is an identity matrix of dimension given in the subscript. $[\V{X}]_{n,n}$ denotes the $n$th diagonal entry of $ \V{X} $. %
Furthermore, %
${1}_{\mathbb{A}}(\V{x})$ denotes the indicator function that is ${1}_{\mathbb{A}}(\V{x}) = 1$ if $\V{x} \in \mathbb{A}$ and 0 otherwise, for $\mathbb{A}$ being an arbitrary set and $\mathbb{R}^{\text{+}}$ is the set of positive real numbers. %
We predefine the following \acp{pdf} \ac{wrt} $\rv{x}$: The truncated Gaussian \ac{pdf} is
\vspace*{-1mm}
\begin{align} \label{eq:eq:truncated_gaussian_pdf}
	f_\text{TN}(x; \mu , \sigma , \lambda) = \frac{1}{Q(\frac{\lambda - \mu}{\sigma}) \sqrt{2\pi} \sigma} e^{\frac{-(x-\mu)^2}{2\,\sigma^2}}  {1}_{\mathbb{R}^{\text{+}}}(x \minus \lambda)\\[-7mm]\nn
\end{align}
with mean $\mu$, standard deviation $\sigma$, truncation threshold $\lambda$ and $Q(\cdot)$ denoting the Q-function \cite{Kay1998}. Accordingly, the Gaussian \ac{pdf} is $f_\text{N}(x; \mu , \sigma) =f_\text{TN}(x; \mu , \sigma , \text{-}\s\infty)$. The truncated Rician \ac{pdf} is \cite[Ch. 1.6.7]{BarShalom2002EstimationTracking} 
\vspace*{-1mm}
\begin{align} \label{eq:truncated_rice_pdf}
	f_\text{TRice}(x;\rmv s ,\rmv u , \lambda) = \frac{1}{Q_1(\frac{u}{s}, \frac{\lambda}{s})}\frac{x}{s^2} e^{\frac{-(x^2+u^2)}{2\,s^2}} I_0(\frac{x\, u}{s^2}) {1}_{\mathbb{R}^{\text{+}}}(x \minus \lambda)\\[-7mm]\nn
\end{align}
with non-centrality parameter $u$, scale parameter $s$ and truncation threshold $\lambda$. $I_0(\cdot)$ is the 0th-order modified first-kind Bessel function and $Q_1(\cdot,\cdot)$ denotes the Marcum Q-function \cite{Kay1998}. The truncated Rayleigh \ac{pdf} is \cite[Ch. 1.6.7]{BarShalom2002EstimationTracking}
\vspace*{-1mm}
\begin{align} \label{eq:truncated_rayleigh_pdf}
	f_\text{TRayl}(x; s , \lambda) = \frac{x}{s^2}\, e^{\frac{-(x^2-\lambda^2)}{2\, s^2}}  {1}_{\mathbb{R}^{\text{+}}}(x - \lambda)\\[-7mm]\nn
\end{align}
with scale parameter $s$ and truncation threshold $\lambda$. This formula corresponds to the so-called Swirling I model\cite{BarShalom2002EstimationTracking}. %
Finally, we define the uniform \ac{pdf} $f_\mathrm{U}(x;a,b) = 1/(b-a) {1}_{[a,b]}(x)$ and the uniform \ac{pmf} $p_\mathrm{UD}(x;\mathcal{X}) = 1/|\mathcal{X}| {1}_{\mathcal{X}}(x)$.

\vspace{-2mm} 
\section{Geometrical Relations}\label{sec:geometric_model}
\begin{figure}[!t]
	
	\centering
	\setlength{\abovecaptionskip}{-0.5mm}
	\setlength{\belowcaptionskip}{0pt}
	
	\setlength{\figurewidth}{0.36\textwidth}
	\setlength{\figureheight}{0.12\textwidth}

	\tikzsetnextfilename{eye_catcher2}
	\scalebox{0.93}{\includegraphics{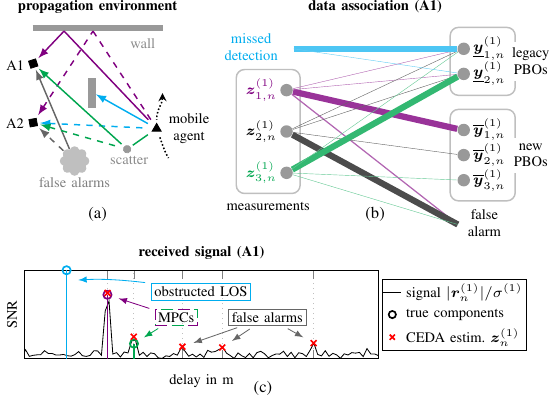}}
	\caption{Graphical overview of %
		(a) an exemplary radio propagation environment, (b) joint probabilistic data association using measurements of anchor A1, and (c) the received signal vector of anchor A1.
	}\label{fig:eye_catcher2}
	\vspace{-4.5mm}
\end{figure}

We consider a mobile agent equipped with a single antenna that moves along an unknown trajectory.
At each time $n$, the mobile agent at position $\bm{p}_{n} \triangleq [{p}_{\text{x}\s n}\;  {p}_{\text{y}\s n}]^\text{T}$ transmits a radio signal and each anchor $j \rmv\rmv \in \rmv\rmv \{1,\s...\s,J\}$ equipped with a single antenna at anchor position $\bm{p}_{\text{A}}^{(j)} \triangleq [p_{\text{Ax}}^{(j)} \; p_{\text{Ay}}^{(j)}]^\text{T}$ acts as a receiver\footnote{Note however that due to the reciprocity of wireless channels \cite{Bjornson2017MassiveMIMO}, both, agent and anchors, can equivalently act as signal transmitters or receivers.} (see Fig.~\ref{fig:eye_catcher2}a).
The Euclidean distance between mobile agent at time $n$ and anchor $j$ (i.e., the LOS path) is given as
$%
	\norm{\bm{p}_n - \bm{p}_\text{A}^{(j)}}{}
$. %
Specular reflections of radio signals on flat surfaces (planar walls, floor, ceiling,...) can be described by \acp{va} that are mirror images of the anchors \cite{PedersenJTAP2018,Meissner2015Diss}. Similarly, point scatters are described by the sum of their distance \ac{wrt} the agent ${\bm{p}}_n$ and their distance \ac{wrt} the anchor $\bm{p}_\text{A}^{(j)}$ \cite{GentnerTWC2016}, where the latter is constant over time $n$.
The ``distance bias" $b_{l,n}^{(j)}$ corresponding to the $l$-th \ac{mpc} at time $n$ and for anchor $j$, caused by one of the discussed phenomena, is given as
\begin{align} \label{eq:distance_bias_geometry}
	 b_{l,n}^{(j)}(\bm{p}_n) =  \norm{\bm{p}_{n} - \bm{p}_{\text{MPC}\s l}^{(j)}}{} -  \norm{\bm{p}_{n} - \bm{p}_\text{A}^{(j)}}{} + c_{\text{MPC}\s l}^{(j)}\\[-5mm]\nn
\end{align}
with $\bm{p}_{\text{MPC}\s l}^{(j)}$ being the position corresponding to the current \ac{va} or point scatter. The term $c_{\text{MPC}\s l}^{(j)}$ is an offset which equals zero for \acp{va} and $\norm{\bm{p}_\text{A}^{(j)} - \bm{p}_{\text{MPC}\s l}^{(j)}}{}$ for point scatters, respectively, and, thus, is constant over time $n$.  

\begin{figure}[!t]
	\centering
	\setlength{\abovecaptionskip}{0pt}
	\setlength{\belowcaptionskip}{0pt}
	\tikzsetnextfilename{plane_wave}
	\scalebox{0.9}{
		\includegraphics{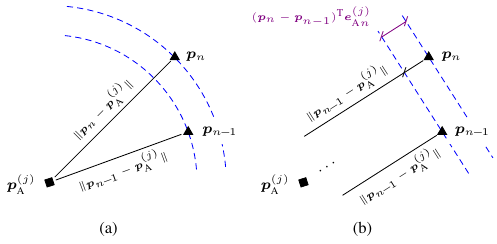}}
	\caption{Exact geometry (a) vs. far field assumption (b) \ac{wrt} $\bm{p}_{\text{A}\s n}^{(j)}$. %
	}\label{fig:plane_wave}
	\vspace*{-4mm}	
\end{figure}

In this work we are interested to model the temporal evolution of the distance bias. To this end, we consider the bias difference, denoted as
\begin{align} \label{eq:bias_difference}
	\Delta  b_{l,n}^{(j)}(\bm{p}_n, \bm{p}_{n\minus 1})  \triangleq  b_{l,n}^{(j)}(\bm{p}_n) -  b_{l,n-1}^{(j)}(\bm{p}_{n-1})\,\\[-5mm]\nn
\end{align}
which, in general, is a nonlinear function of $\bm{p}_{n}$ and $\bm{p}_{n\minus 1}$. 
However, if the distance between agent and anchor $\norm{\bm{p}_{n} - \bm{p}_\text{A}^{(j)}}{}$ is large compared to the agent movement %
$\norm{\bm{p}_{n} - \bm{p}_{n\minus 1}}{}$, the bias difference is well approximated as
\begin{align} 
	\label{eq:far_field}
		\Delta  b_{l,n}^{(j)}(\bm{p}_n, \bm{p}_{n\minus 1}) \approx  (\bm{p}_{n} -\bm{p}_{n-1})^\text{T} (\bm{e}_{\text{MPC}\s l,n}^{(j)} - \bm{e}_{\text{A}\s n}^{(j)} ) \,  \\[-5mm]\nn
\end{align}
where $\bm{e}_{\text{MPC}\s l,n}^{(j)} \triangleq ( \bm{p}_{n} - \bm{p}_{\text{MPC}\s l}^{(j)} ) / \norm{\bm{p}_{n} - \bm{p}_{\text{MPC}\s l}^{(j)}}{}$ and $\bm{e}_{\text{A}\s n}^{(j)} \triangleq  ( \bm{p}_{n} - \bm{p}_{\text{A}}^{(j)} ) / \norm{\bm{p}_{n} - \bm{p}_{\text{A}}^{(j)}}{}$ are unit vectors that point from $\bm{p}_{\text{MPC}\s l}^{(j)}$ and $\bm{p}_{\text{A}}^{(j)}$, respectively, in the direction of the mobile agent. 
Note that this implies the distances $\norm{\bm{p}_{n} - \bm{p}_{\text{MPC}\s l}^{(j)}}{}$ are also large \ac{wrt} the agent movement.
The approximation used in \eqref{eq:far_field} is known in literature as the far field assumption \cite{Thrun2005Affine,KuangICC2013} and is geometrically visualized in Fig.~\ref{fig:plane_wave} for $\bm{p}_{\text{A}\s n}^{(j)}$, and it applies equally for $\bm{p}_{\text{MPC}\s l}^{(j)}$. 
It is %
based on the observation that the unit vectors $\bm{e}_{\text{A}\s n\minus 1}^{(j)}$ and $\bm{e}_{\text{A}\s n}^{(j)}$ or $\bm{e}_{\text{MPC}\s l,n-1}^{(j)}$ and $\bm{e}_{\text{MPC}\s l,n}^{(j)}$, respectively, are similar. 
Analyzing \eqref{eq:far_field}, we observe that 
\begin{itemize}
\item[(i)] \eqref{eq:far_field} is a linear function \ac{wrt} the agent positions $\bm{p}_n$ and  $\bm{p}_{n-1}$, i.e., a locally linear agent movement leads to a locally linear change of the delay bias. 
\item[(ii)] \acp{va} or point scatters, which take a similar angle to the agent as the anchor, i.e., $ \bm{e}_{\text{A}\s n}^{(j)} \approx \bm{e}_{\text{MPC}\s l,n}^{(j)}$, lead to small bias differences $\Delta  b_{l,n}^{(j)}(\bm{p}_n, \bm{p}_{n\minus 1}) \approx 0$ even if the agent movement follows a non-linear path (e.g. sudden turns). 
\end{itemize}
Note that while observation (ii) is readily shown in \eqref{eq:far_field}, it does not require the far-field assumption. Thus, it applies for the bias differences in \eqref{eq:bias_difference} in general. 

The proposed method utilizes the above observations %
by tracking the distance biases  $b_{l,n}^{(j)}(\bm{p}_n)$ to each \ac{mpc} using a (locally linear) constant velocity model. Thus, the proposed method is exploiting ``local" map information without explicitly modeling map features such as \acp{va} or scatters. %

\begin{table*}[t!] \vspace*{-3mm}
	\renewcommand{\baselinestretch}{1}\small\normalsize
	\setlength{\tabcolsep}{3pt} %
	\renewcommand{\arraystretch}{1.1} %
	\footnotesize
	\centering
	
	\caption{Summary and description of all unobserved \acp{rv} of the system model.}\label{tbl:rvs}
	\begin{tabular}{|r||c|c||c|c|c|c||c|c|}
		\toprule
		\textbf{Description} & \textbf{agent position} & \textbf{agent velocity} & \textbf{bias} & \textbf{bias velocity} & \makecell{\textbf{normalized}\\\textbf{amplitude}} & \textbf{existence} & \multicolumn{2}{c|}{{\textbf{association} \textbf{variables}}} \\ 
		\textbf{Symbol} & $\RV{p}_n$ & $\RV{v}_n$ & $\rv{b}_{k,n}^{(j)}$ & $\rv{v}_{\text{b}\s k,n}^{(j)} $ & $\rv{u}_{k,n}^{(j)}$ & $\rv{r}^{(j)}_{k,n}$  & $\underline{\rv{a}}_{k,n}^{(j)}$ & $\overline{\rv{a}}_{m,n}^{(j)}$\\ 
		\textbf{Type} & continuous & continuous & continuous & continuous & continuous & Bernulli & \textit{discrete} & \textit{discrete} \\ %
		\textbf{State Transition} & Markov & Markov & Markov & Markov & Markov & Markov & independent & {independent} \\ 
		\textbf{\# per Time $n$} & $1$ & $1$ & $\sum_{j=1}^{J} \rv{K}_n^{(j)}$ &  $\sum_{j=1}^{J} \rv{K}_n^{(j)}$ &  $\sum_{j=1}^{J} \rv{K}_n^{(j)}$ &  $\sum_{j=1}^{J} \rv{K}_n^{(j)}$ & $\sum_{j=1}^{J} \rv{K}_n^{(j)}$ & $ \sum_{j=1}^{J} \rv{M}_n^{(j)}$ \\ \midrule
		\textbf{Description} & \multicolumn{2}{c||}{\textbf{agent state}} & \multicolumn{4}{c||}{\textbf{augmented \ac{pbo} state}} & \multicolumn{2}{c|}{} \\ 
		\textbf{Symbol} & \multicolumn{2}{c||}{$\RV{x}_n = [\RV{p}_n ^\text{T}\; \RV{v}_n^\text{T} ]^\text{T}$} & \multicolumn{4}{c||}{ $\RV{y}^{(j)}_{k,n} \triangleq [\RV{\psi}^{(j)\s \mathrm{T}}_{k,n} \iist \rv{r}^{(j)}_{k,n}]^{\mathrm{T}} \triangleq [\rv{b}_{k,n}^{(j)} \iist \rv{v}_{\text{b}\s k,n}^{(j)} \iist \rv{u}_{k,n}^{(j)} \iist \rv{r}^{(j)}_{k,n}]^{\mathrm{T}}$} & \multicolumn{2}{c|}{} \\ 
		\bottomrule
	\end{tabular} 
	\vspace{-2.5mm}
\end{table*}

\vspace{-1mm} 
\section{Radio Signal Model and Channel Estimation}\label{sec:signal_model}

The received complex baseband signal at the $j$th anchor is sampled $N_\text{s}$ times with sampling frequency $ f_{\text{s}} = 1/T_{\text{s}}$ yielding an observation period of $T =  N_{\text{s}} \, T_{\text{s}}$. 
By stacking the samples, we obtain the discrete-time received signal vector \cite{HanFleuRao:TSP2018,VenusTWC2023}
\vspace*{-1mm}
\begin{align}
\RV{r}_{n}^{(j)} = {\alpha}_{n,1}^{(j)} \V{s}({\tau}_{n,1}^{(j)}) + \sum_{l = 2}^{{L}_n^{(j)}}  {\alpha}_{n,l}^{(j)} \V{s}({\tau}_{n,l}^{(j)}) + \RV{\noise{}}_{n}^{(j)}
\label{eq:signal_model_sampled}\\[-7mm]\nn
\end{align}
where $\bm{s}(\tau) \triangleq [s(-(N_\text{s}-1)\ist T_\text{s}/2-\tau) \,\,\, \cdots \,\,\, s((N_\text{s}-1)\ist T_\text{s}/2  - \tau)]^\text{T}\in \mathbb{C}^{N_\text{s}\times 1}$ is the discrete-time transmit pulse with delay $\tau$.
The first and second terms describe the \ac{los} component and the sum of ${L}_n^{(j)} \minus 1$ specular \acp{mpc} with their corresponding complex amplitudes ${\alpha}_{n,l}^{(j)} \,\,\, \in \mathbb{C} $ and delays ${\tau}_{n,l}^{(j)} \,\,\, \in \mathbb{R}$, respectively. The delays are related to respective distances via the relation ${\tau}_{n,l}^{(j)} = {d}_{n,l}^{(j)}/ \s c$ with $c$ being the speed of light. 
The measurement noise vector $\RV{\noise{}}_{n}^{(j)} \in \mathbb{C}^{N_\text{s} \times 1} $ is a zero-mean, circularly-symmetric complex Gaussian random vector with covariance matrix %
${\sigma}^{(j)\s 2} \M{I}_{N_\text{s}}$ and noise variance is given by ${\sigma}^{(j)\s 2} = {N}_{0}^{(j)}/T_{\text{s}}$. The \acp{mpc} arise from reflection by unknown objects, since we assume that no map information is available. %
The component \ac{snr} of each \ac{mpc} is $ \mathrm{SNR}^{(j)}_{n,l} = |{\alpha}^{(j)}_{n,l}|^2 \norm{\V{s}({\tau}_{n,l}^{(j)})}{2} / {\sigma}^{(j)\s 2}$ and the corresponding normalized amplitude is ${u}^{(j)}_{n,l} \triangleq \mathrm{SNR}^{(j)\s\frac{1}{2}}_{n,l}$. 
We assume time %
and frequency synchronization between all anchors and the mobile agent. However, our model can be extended to an
unsynchronized system similarly as in \cite{GentnerTWC2016}. %

\vspace{-2mm} 
\subsection{Parametric Channel Estimation} \label{sec:channel_estimation}

We independently apply, at each time $ n $ and for each anchor $j$, a parametric \ac{ceda} \cite{ShutWanJos:CSTA2013,Hansen2014SAM,HanFleuRao:TSP2018} to the observed complex baseband signal vector $\bm{r}_n^{(j)}$. The \ac{ceda}  decomposes $\bm{r}_n^{(j)}$ into individual components that represent potential \ac{mpc} parameter estimates. It yields a number of $M_n^{(j)}$ measurements denoted by ${\V{z}^{(j)}_{m,n}}$ with $m \in \Set{M}_n^{(j)} \triangleq \{1,\,\dots\,,M_n^{(j)}\} $ that are collected by the vector $\vspace*{-0.2mm}\V{z}^{(j)}_{n} = [{\bm{z}^{(j)\text{T}}_{1,n}}  \rmv \cdots  {\V{z}^{(j)\text{T}}_{M_n^{(j)}\rmv\rmv\rmv\rmv,n}}]^\text{T}$. 
Each  ${\bm{z}^{(j)}_{m,n}} = [\zd~ \zu]^\text{T}$, contains a distance measurement $\zd \in [0, d_\text{max}]$ 
and a normalized amplitude measurement\footnote{Note that the normalized amplitude measurements are determined as $\zu  = |\mu^{(j)}_{\alpha,m,n}|/\sigma^{(j)}_{\alpha,m,n}$ with $\mu^{(j)}_{\alpha,m,n} \,\,\, \in \mathbb{C}$ and $\sigma^{(j)}_{\alpha,m,n} \,\,\, \in \mathbb{R}^{+}$, which denote the estimated mean and standard deviation of the complex amplitudes of a \ac{mpc} provided by the \ac{ceda}, respectively. The phases of the complex amplitudes are jointly estimated by the \ac{ceda} and are contained in $\mu^{(j)}_{\alpha,m,n} \in \mathbb{C}$.} $\zu \in [\gamma, \infty)$, where $d_\text{max}$ is the maximum possible distance and $\gamma$ is the detection threshold of the \ac{ceda}. 
Individual measurements $\zd$ and $\zu$ relate to true \ac{mpc} parameters ${d}^{(j)}_{n,l}$ and ${u}^{(j)}_{n,l}$, but it is unknown which measurement corresponds to which \ac{mpc}, or if a measurement is due to a false alarm (see Section~\ref{sec:DA}). 

The \ac{ceda} decomposes the discrete signal vector $\V{r}_n^{(j)}$ into individual, decorrelated components according to \eqref{eq:signal_model_sampled}, reducing the number of dimensions (as ${M}_n^{(j)}$ is usually much smaller than $N_\text{s}$). It thus can be said to compress the information\footnote{For real applications, the \ac{ceda} can be executed locally in the processing unit of each anchor in order to compress the signal samples into individual measurements, which are only then transmitted to a global processing unit.} contained in $\V{r}_n^{(j)}$ into $ \bm{z}^{(j)}_{n} = [{\bm{z}^{(j)\text{T}}_{1,n}}  ... \, {\bm{z}^{(j)\text{T}}_{M_n^{(j)},n}}]^\text{T}$. 
The stacked vector $\bm{z}_n = [\bm{z}^{(1)\, \text{T}}_{n} \rmv\rmv\rmv ... \, \bm{z}^{(J)\,\text{T}}_{n}]^\text{T}$ is used by the proposed algorithm as a noisy measurement.

\section{Dynamic Multi-Object Tracking} \label{sec:dynamic_mot}

When facing a time-varying and unknown number of objects, which in this work correspond to individual \acp{mpc}, the following challenges need to be addressed \cite{WilliamsTAES2014,BarShalomTCS2009,MeyerProc2018}, as outlined by Figure~\ref{fig:eye_catcher}. 
\begin{enumerate}
	\item \textit{Data association uncertainty:} It is unknown which measurement corresponds to which \ac{mpc}.
	\item \textit{Missed detections:} There can be \acp{mpc} that did not cause a measurement. 
	\item \textit{False alarms (clutter):} There can be measurements that are not caused by an \ac{mpc}, but by perturbations, such as measurement noise.
	\item \textit{Object birth:} New \acp{mpc} can appear.
	\item \textit{Object death:} Existing \acp{mpc} can vanish.
\end{enumerate}
All of the above challenges apply for each time $n$ and for each anchor $j$. 
Challenges 1 to 3 are related to the problem of \textit{joint data association}, which involves associating measurements to multiple objects. This is addressed in the system model using loopy probabilistic data association \cite{WilliamsTAES2014,MeyerProc2018} as described in Section~\ref{sec:DA}.  
Challenges 4 and 5 refer to the problem of \textit{dynamic multi-object tracking}, which involves the inference of an unknown and time-varying number of objects \cite{MeyerProc2018,LeitingerTWC2019}. This is addressed in the system model by the concept of \acp{pbo}, as discussed in in Sections~\ref{sec:pbo_states} and \ref{sec:npbo}.

\vspace{-1mm} 
\section{System Model} \label{sec:system_model}

The following section introduces the probabilistic system model used for the subsequent derivation of the proposed \ac{spa}. 
For the sake of clarity, Table~\ref{tbl:rvs} provides an overview of all the  \textit{unobserved} random variables introduced in the following Section. These variables are jointly inferred by the proposed \ac{spa}. 

\subsection{Agent State and \ac{pbo} States} \label{sec:pbo_states}
The current state of the mobile agent is described by the state vector $\RV{x}_n = [\RV{p}_n ^\text{T}\; \RV{v}_n^\text{T} ]^\text{T}$ containing the position $\RV{p}_n = [\rv{p}_{\text{x}\s n}\; \rv{p}_{\text{y}\s n}]^\text{T}$ and velocity $\RV{v}_n = [\rv{v}_{\text{x}\s n}\; \rv{v}_{\text{y}\s n}]^\text{T}$. 

Following \cite{LeitingerTWC2019, MeyerJTSP2017, MeyerProc2018}, we account for the time-varying and unknown number of \acp{mpc} by introducing \acp{pbo} indexed by $k \in \{1,\dots, K_{n}^{(j)}\} \triangleq \mathcal{K}_n^{(j)}$. {Thereby, we explicitly distinguish between the \ac{los} component at $k=1$ and \acp{mpc} $k \in \{2, \, ... \,, K_n^{(j)}\}$.} The number of \acp{pbo} $K_n^{(j)}$ corresponds to the maximum number of components that have produced measurements at anchor $j$ so far \cite{MeyerProc2018}. 

We define the augmented \ac{pbo} state, which is denoted as  $\RV{y}^{(j)}_{k,n} \triangleq [\RV{\psi}^{(j)\s \mathrm{T}}_{k,n} \iist \rv{r}^{(j)}_{k,n}]^{\mathrm{T}}$, where the \ac{pbo} state $\RV{\psi}^{(j)}_{k,n} \triangleq [\RV{x}_{\text{b}\s k,n}^{(j)}\iist \rv{u}_{k,n}^{(j)}]^{\mathrm{T}}$ with $\RV{x}_{\text{b}\s k,n}^{(j)} \triangleq [\rv{b}_{k,n}^{(j)} \iist \rv{v}_{\text{b}\s k,n}^{(j)}]$ consists of the bias $\rv{b}_{k,n}^{(j)}$, the respective distance bias velocity $\rv{v}_{\text{b}\s k,n}^{(j)}$, and the normalized amplitude $\rv{u}_{k,n}^{(j)}$. 
The existence / non-existence of \ac{pbo} $k$ is modeled by a binary random variable $ \rv{r}^{(j)}_{k,n} \in \{0,1\} $ in the sense that a \ac{pbo} exists if and only if $r^{(j)}_{k,n} = 1$. Although the \ac{los} existence $\rv{r}^{(j)}_{1,n}$ and amplitude $u^{(j)}_{1,n}$ are random, the distance bias is known and fixed to zero, i.e., $b_{1,n}^{(j)} \equiv 0,\iist v_{\text{b}\s 1,n}^{(j)} \equiv 0$. See Table~\ref{tbl:rvs} for an overview of all random variables of the system model. 

Formally, \ac{pbo} $k$ is also considered for $r_{k,n}^{(j)} = 0$, i.e., when it is non-existent. The states $\RV{\psi}_{k,n}^{(j)}$ of non-existent \acp{pbo} are obviously irrelevant and have no influence on the \ac{pbo} detection and state estimation. Therefore, all \acp{pdf} defined for \ac{pbo} states, $ f(\V{y}_{k,n}^{(j)}) = f(\V{\psi}_{k,n}^{(j)},r_{k,n}^{(j)}) $, are of the form $ f(\V{\psi}_{k,n}^{(j)},r_{k,n}^{(j)}=0) = f_{k,n}^{(j)} f_{\mathrm{d}}(\V{\psi}_{k,n}^{(j)}) $, where $ f_{\mathrm{d}}(\V{\psi}_{k,n}^{(j)}) $ is an arbitrary ``dummy \ac{pdf}'' and $ f_{k,n}^{(j)} \in [0,1]$ is a constant representing the probability of non-existence \cite{MeyerProc2018, MeyerJTSP2017, LeitingerTWC2019}. %

\vspace{-2mm} 
\subsection{Measurement Model} \label{sec:measurement_model} 

At each time $n$ and for each anchor $j$, the \ac{ceda} provides the currently observed measurement vector $\bm{z}_n^{(j)}$, with fixed ${M}^{(j)}_n$, according to Sec.~\ref{sec:channel_estimation}. Before the measurements %
are observed, they are random and represented by the vector  ${\RV{z}^{(j)}_{m,n}} = [\zdr~\zur]^\text{T}$. In line with Sec.~\ref{sec:channel_estimation} we define the nested random vectors $\RV{z}^{(j)}_{n} = [{\RV{z}^{(j)\text{T}}_{1,n}} \, ... \, {\RV{z}^{(j)\text{T}}_{\rv{M}_n^{(j)}\rmv\rmv\rmv\rmv,n}}]^\text{T}$ and $\RV{z}_n = [\RV{z}^{(1)\, \text{T}}_{n} ... \, \RV{z}^{(J)\,\text{T}}_{n}]^\text{T}$. The number of measurements $\rv{M}^{(j)}_n$ is also a random variable. The vector containing all numbers of measurements is defined as  $\RV{M}_n = [\rv{M}_n^{(1)}\, ... \, \rv{M}_n^{(J)}]^\text{T}$.

If ${\rv{z}^{(j)}_{m,n}}$ is generated by a \ac{pbo} $k$, i.e., by the \ac{los} component or an \ac{mpc}, we assume that the single-measurement \ac{lhf} $  f(\V{z}_{m,n}^{(j)}| \V{x}_n, \V{\psi}_{k,n}^{(j)}) $ is conditionally independent across $\zdr$, and $\zur$. %
Thus, it factorizes as
\begin{align}
	\rmv f(\V{z}_{m,n}^{(j)}| \V{x}_n, \V{\psi}_{k,n}^{(j)}) \rmv  = \rmv  f( \zd | {b}^{(j)}_{k,n},\V{p}_{n},u_{k,n}^{(j)}) \, f( \zu | u_{k,n}^{(j)}) . \nn
	\label{eq:single_measurement_likelihood}\\[-5mm]\nn
\end{align}
The \ac{lhf} of the distance measurement $\zdr$
is given by
\begin{equation} \label{eq:like_delay}
	f(\zd | {b}^{(j)}_{k,n}, \bm{p}_n , \rmv u_{k,n}^{(j)}\rmv
	) = f_\text{N}\big(\zd;\, d({b}^{(j)}_{k,n},\bm{p}_n),\, \sigma_{\text{d}}^{2} (u^{(j)}_{k,n}) \big)\ist .
\end{equation}
Its mean value is described by a distance function, which is assumed to be geometrically related to the agent position via 
\begin{align} \label{eq:distance_fun}
	d(\rv{b}^{(j)}_{k,n},\RV{p}_n) = \norm{\RV{p}_n - \bm{p}_\text{A}^{(j)}}{}  + \rv{b}_{k,n}^{(j)}\\[-5mm]\nn
\end{align}
where $\rv{b}_{k,n}^{(j)}$ represents the distance bias of \ac{mpc} $k$ from the \ac{los} component distance according to Sec.~\ref{sec:geometric_model}.  
The variance is determined from the Fisher information, given by
$ \sigma_{\text{d}}^{2} (u_{k,n}^{(j)}) =   c^2 / ( 8\,  \pi^2 \, \beta_\text{bw}^2 \,  %
 u_{k,n}^{(j)\s 2}) $ 
with $\beta_\text{bw}$ being the root mean squared bandwidth (see \cite{WitrisalSPM2016Copy,ShenTIT2010} for details) . %
\begin{figure}[t]
	\centering

	\setlength{\belowcaptionskip}{0pt}

	\tikzsetnextfilename{likelihoods}\includegraphics{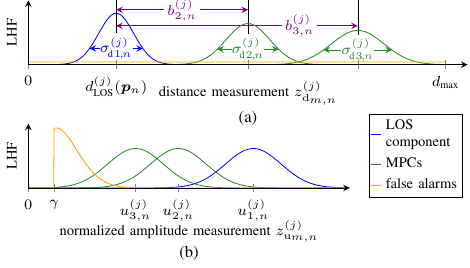}
	\vspace*{-3mm}
	\caption{Visualization of (a) the joint distance \ac{lhf} and (b) the joint amplitude \ac{lhf}. The \ac{los} distance is $d^{(j)}_\text{LOS}(\bm{p}_n) = \norm{\bm{p}_n - \bm{p}_\text{A}^{(j)}}{} $ and  $\sigma_{\rmv\text{d}\s {k\rmv,\rmv n}}^{(j)} = \sigma_{\text{d}} (u^{(j)}_{k,n}) $.
	}
	\label{fig:single_measurment_like}\vspace{-3mm}
	\vspace*{-2mm}	
\end{figure}
The \ac{lhf} of the normalized amplitude measurement $\zur$ is obtained\footnote{The proposed model describes the distribution of the amplitude estimates of the radio signal model given in \eqref{eq:signal_model_sampled}\cite{LerroACC1990,LiTWC2022,VenusTWC2023}.} 
as %
\begin{align} \label{eq:los_amplitude}
	f(\zu| u^{(j)}_{k,n}) \triangleq f_\text{TRice}\big(\zu; \sigma_\mathrm{u} (u^{(j)}_{k,n}) ,u^{(j)}_{k,n}, \gamma\big)\\[-6mm]\nn
\end{align}
 with non-centrality parameter corresponding to the normalized amplitude $\rv{u}^{(j)}_{k,n}$ and $\gamma$ being the detection threshold of the \ac{ceda}. Again, the scale parameter is determined from the Fisher information given as 
$\sigma_{\mathrm{u}}^{2} (\rv{u}^{(j)}_{k,n}) = 1/2 +\rv{u}^{(j)\s 2}_{k,n}\, /(4 N_{\text{s}})$ (see \cite{LiTWC2022} for a detailed derivation). Note that this expression reduces to $1/2$ if the \ac{awgn} noise variance $\sigma^{(j)\s 2}$ is assumed to be known or $N_{\text{s}}$ grows indefinitely. %
The probability of detection resulting from \eqref{eq:los_amplitude} is given by the Marcum Q-function %
\cite{Kay1998, LiTWC2022}
\begin{align}
	\pdrv =  Q_1\Bigg(\frac{\rv{u}^{(j)}_{n}}{\sigma_\mathrm{u} (\rv{u}^{(j)}_{n})}, \frac{\gamma}{\sigma_\mathrm{u} (\rv{u}^{(j)}_{n})}\Bigg)\, .\\[-6mm]\nonumber
\end{align}

False alarm measurements are assumed to be statistically independent of \ac{pbo} states and are modeled by a Poisson point process with mean $ {{\mu}_{\mathrm{fa}}} $ and \ac{pdf} $ f_{\mathrm{fa}}(\V{z}_{m,n}^{(j)}) $, which is assumed to factorize as $ f_{\mathrm{fa}}(\V{z}_{m,n}^{(j)}) = {f_{\mathrm{fa}}}({z_\mathrm{d}}_{m,n}^{(j)}) {f_{\mathrm{fa}}}({z_\mathrm{u}}_{m,n}^{(j)}) $. The false alarm \ac{lhf} of the distance measurement is uniformly distributed, i.e., $ {f_{\mathrm{fa}}}({z_\mathrm{d}}_{m,n}^{(j)})\rmv\rmv= f_\mathrm{U}({z_\mathrm{d}}_{m,n}^{(j)},0,d_\text{max})$. %
The false alarm \ac{lhf} of the normalized amplitude is given by 
\begin{align} \label{eq:nlos_amplitude_lhf}
	f_\mathrm{fa}(\zu)\rmv \triangleq \rmv f_\text{TRayl}(\zu \,; \sqrt{1/2} \,, \gamma)\\[-5mm]\nn
\end{align}
with the scale parameter given as $\sqrt{1/2}$ and detection threshold $\gamma$. See Fig.~\ref{fig:single_measurment_like} for a graphical representation of the joint likelihood function.
We approximate the mean number of false alarms as $\mu_\text{fa} = N_\text{s}\, e^{-\gamma^2}$, where the right-hand side expression corresponds to the false alarm probability $p_\text{fa}(u) = \int \rmv f_\text{TRayl}(u \,; \sqrt{1/2} \,, \gamma) \, \mathrm{d} u  = e^{-\gamma^2}$ according to \eqref{eq:nlos_amplitude_lhf}.

\vspace{-2mm} 
\subsection{State-Transition Model} \label{sec:state_transition_model}

For each \ac{pbo} with state $\RV{y}_{k,n-1}^{(j)} $ with $k \rmv\in\rmv \mathcal{K}_{n-1}^{(j)}$ at time $n-1$ and anchor $j$, there is one ``legacy'' \ac{pbo} with state $ \underline{\RV{y}}_{k,n}^{(j)} \rmv\rmv\triangleq\rmv\rmv [\underline{\RV{\psi}}_{k,n}^{(j)\s \mathrm{T}} \iist \underline{\rv{r}}_{k,n}^{(j)}]^{\mathrm{T}} $ with $k \rmv\rmv\in\rmv\rmv \mathcal{K}_{n-1}^{(j)}$ at time $n$, i.e., 
\begin{equation}\nonumber
	\big\{ \RV{y}_{k,n-1}^{(j)} ~\big|~ k \rmv\in\rmv \mathcal{K}_{n-1}^{(j)}  \big\}  \rightarrow   \big\{ \underline{\RV{y}}_{k,n}^{(j)} ~\big|~ k \rmv\in\rmv \mathcal{K}_{n-1}^{(j)}  \big\} ~~\forall~ j \in \{1 \, ... \, J \} \, .
\end{equation}
{We also define the joint states 
$\underline{\RV{y}}_n^{(j)} \!\triangleq [\underline{\RV{y}}_{1,n}^{(j)\text{T}} \rmv\cdots\ist \underline{\RV{y}}_{K_{n\minus 1}^{(j)}\rmv,n}^{(j)\text{T}} ]^{\mathrm{T}}$ and 
$\underline{\RV{y}}_n \!\triangleq [\underline{\RV{y}}_n^{(1)\text{T}} \rmv\cdots\ist \underline{\RV{y}}_n^{(J)\text{T}} ]^{\mathrm{T}}\rmv$ as well as 
${\RV{y}}_n^{(j)} \!\triangleq [{\RV{y}}_{1,n}^{(j)\text{T}} \rmv\cdots\ist {\RV{y}}_{K_{n}^{(j)}\rmv,n}^{(j)\text{T}} ]^{\mathrm{T}}$ and 
${\RV{y}}_n \!\triangleq [{\RV{y}}_n^{(1)\text{T}} \rmv\cdots\ist {\RV{y}}_n^{(J)\text{T}} ]^{\mathrm{T}}\rmv$.} 
Assuming that the agent state as well as the \ac{pbo} states evolve independently across $ k $ and $ n $ and $j$, the joint state-transition \ac{pdf} factorizes as \cite{MeyerProc2018}
\vspace*{-1mm}
\begin{align}
	&f(\V{x}_n ,\underline{\V{y}}_n|\V{x}_{n-1},\V{y}_{n-1}) \nonumber\\[-3mm] 
	&\hspace{20mm}= f(\V{x}_{n}|\V{x}_{n-1}) \prod_{j=1}^J\prod_{k=1}^{K_{n-1}^{(j)}} \!  f(\underline{\V{y}}_{k,n}^{(j)} | \V{y}_{k, n-1}^{(j)})
	\label{eq:state_transition}\\[-7mm]\nn
\end{align}
where $ f(\underline{\V{y}}_{k,n}^{(j)}|\V{y}_{k,n-1}^{(j)})\rmv\rmv=\rmv\rmv f(\underline{\V{\psi}}_{k,n}^{(j)}, \underline{r}_{k,n}^{(j)}|\V{\psi}_{k,n-1}^{(j)}, r_{k,n-1}^{(j)}) $ is the augmented state-transition \ac{pdf}\footnote{Note that for the variables $b_{1,n}^{(j)} $ and $\iist v_{\text{b}\s 1,n}^{(j)}$, which are constant for all $n$ and $j$ (see Sec.~\ref{sec:pbo_states}), there is no state transition. }. 

If a \ac{pbo} existed at time $ n-1 $, i.e., $ r_{k,n-1}^{(j)} = 1 $, at time $n$ it either dies, i.e., $ \underline{r}_{k,n}^{(j)} = 0 $, or it still exists, i.e., $ \underline{r}_{k,n}^{(j)} = 1 $, with the survival probability denoted as $ p_{\mathrm{s}} $. If it does survive, the \ac{pbo} state $ \underline{\RV{\psi}}_{k,n}^{(j)} $ is distributed according to the state-transition \ac{pdf} $ f(\underline{\V{\psi}}_{k,n}^{(j)}|\V{\psi}_{k,n-1}^{(j)}) $. Thus, 
\vspace*{-1mm}
\begin{align}
	\rmv\rmv\rmv\rmv\rmv\rmv\rmv f(\underline{\V{\psi}}_{k,n}^{(j)}, \underline{r}_{k,n}^{(j)}|\V{\psi}_{k,n\text{-}1}^{(j)}, 1) \rmv = \rmv\rmv
	\begin{cases}
		(1-p_{\mathrm{s}})f_{\mathrm{d}}(\underline{\V{\psi}}_{k,n}^{(j)}),  \rmv\rmv \rmv\rmv\rmv 			&\underline{r}_{k,n}^{(j)} = 0\\
		p_{\mathrm{s}}f(\underline{\V{\psi}}_{k,n}^{(j)}|\V{\psi}_{k,n\text{-}1}^{(j)}),	\rmv\rmv \rmv\rmv\rmv  &\underline{r}_{k,n}^{(j)} = 1 
	\end{cases}.\rmv\rmv\rmv\rmv\rmv
	\label{eq:state_transition_pdf2}\\[-6mm]\nn
\end{align}
If a \ac{pbo} did not exist at time $ n\rmv\rmv-\rmv\rmv1 $, i.e., $ r_{k,n-1}^{(j)}\rmv\rmv=\rmv\rmv 0 $, it cannot exist at time $ n $ as a legacy \ac{pbo}. This means that 
\vspace*{-2mm}
\begin{align}
	f(\underline{\V{\psi}}_{k,n}^{(j)}, \underline{r}_{k,n}^{(j)}|\V{\psi}_{k,n-1}^{(j)}, 0) =
	\begin{cases}
		f_{\mathrm{d}}(\underline{\V{\psi}}_{k,n}^{(j)}), 	&\underline{r}_{k,n}^{(j)} = 0\\
		0, 											&\underline{r}_{k,n}^{(j)} = 1 
	\end{cases}\,.
	\label{eq:state_transition_pdf1}\\[-6mm]\nn
\end{align}

To account for the smooth, continuous motion of the mobile agent, the agent state $\RV{x}_n$ is assumed to evolve in time according to a 2-dimensional, constant-velocity and stochastic-acceleration model \cite{BarShalom2002EstimationTracking} given as 
\begin{equation}
	\RV{x}_n = \bm{A}_2\, \RV{x}_{n\minus 1} + \bm{B}_2\, \RV{w}_{n},
\end{equation}
with the acceleration process $\RV{w}_n$ being i.i.d. across $n$, zero mean, Gaussian with covariance matrix ${\sigma_{\text{a}}^2}\, \bm{I}_2$; ${\sigma_{\text{a}}}$ %
is the acceleration standard deviation and  $\bm{A}_2 \in \mathbb{R}^{\text{4x4}}$ and $\bm{B}_2 \in \mathbb{R}^{\text{4x2}}$ are the model matrices. %
The model matrices for the constant-velocity and stochastic-acceleration model are constant over time $n$ and are given as \cite[p.~273]{BarShalom2002EstimationTracking}
	\begin{equation} \label{eq:constant_velocity}
		\bm{A}_{N_\text{D}}  =   \begin{bmatrix}
			1 &  \Delta T  \\
			0 &         1 \\
		\end{bmatrix}  \otimes \V{I}_{N_\text{D}} \, , ~~~~~~
		\bm{B}_{N_\text{D}}  = \begin{bmatrix}
			\frac{\Delta T^2}{2} \\
			\Delta T             \\
		\end{bmatrix}  \otimes \V{I}_{N_\text{D}} 
	\end{equation}
	where $N_\text{D}$ is the dimensionality of the problem, $\Delta T$ is the observation period and $\otimes$ denotes the Kronecker product of two matrices. 

The \ac{pbo} state-transition \ac{pdf} is factorized as  
\begin{equation}\nn
f(\underline{\V{\psi}}_{k,n}^{(j)}|\V{\psi}_{k,n-1}^{(j)}) = f(\underline{\V{x}}_{\text{b}\s k,n}^{(j)}|\V{x}_{\text{b}\s k,n-1}^{(j)})f(\underline{{u}}_{k,n}^{(j)}|{u}_{k,n-1}^{(j)})\, .
\end{equation}
According to the observations of Sec.~\ref{sec:geometric_model}, the legacy bias state $\underline{\RV{x}}_{\text{b}\s k,n}^{(j)}$ is assumed to evolve in time linearly, according to a 1-dimensional  constant-velocity and stochastic-acceleration model
\begin{equation} \label{eq:state_transition_bias}
\underline{\RV{x}}_{\text{b}\s k,n}^{(j)} = \bm{A}_1 \, {\RV{x}}_{\text{b}\s k,n\minus 1}^{(j)} + \bm{B}_1\,\,  \rv{w}_{\text{b}\s k,n}^{(j)}
\end{equation}
with the acceleration process $\rv{w}_{\text{b}\s k,n}^{(j)}$ being i.i.d. across $n$, $k$ and $j$, zero mean, Gaussian with standard deviation ${\sigma_{\text{b}}}$, and $\bm{A}_1 \in \mathbb{R}^{\text{2x2}}$ and $\bm{B}_1 \in \mathbb{R}^{\text{2x1}}$ given by \eqref{eq:constant_velocity} with $N_\text{D}=1$. %
The state-transition of the legacy normalized amplitude $\underline{\rv{u}}_{k,n}^{(j)}$%
, i.e., the state-transition \ac{pdf} $ f( \underline{u}_{k,n}^{(j)} | u_{k,n\minus 1}^{(j)})$, is {given by a random walk model} $\underline{\rv{u}}_{k,n}^{(j)} = \rv{u}_{k,n\minus 1}^{(j)} + \rv{w}_{\text{u}\s k,n}^{(j)}$, where the noise $\rv{w}_{\text{u}\s k,n}^{(j)}$ is i.i.d. across $n$, $k$, and $j$, zero mean, Gaussian, with variance $\sigma^2_\text{u}$.
{Note that the temporal evolution of the distance biases $\rv{b}_{k,n}^{(j)}$ is generally non-linear, leading to a model mismatch. In most scenarios, however, it is well approximated as linear over short periods (see  Sec.~\ref{sec:geometric_model}).}

\vspace{-2mm} 
\subsection{New \acp{pbo}}\label{sec:npbo}
Following \cite{MeyerProc2018, LeitingerTWC2019}, newly detected \acp{pbo} at time $n$ and anchor $j$, i.e., \acp{pbo} that generate measurements for the first time at time $n$ and anchor $j$,  are represented by new \ac{pbo} states $ \overline{\RV{y}}_{m,n}^{(j)} \triangleq [\overline{\RV{\psi}}_{m,n}^{(j)\s \mathrm{T}}\iist \overline{\rv{r}}_{m,n}^{(j)}]^{\mathrm{T}} $, $ m \in \mathcal{M}_n^{(j)} $. 
New \acp{pbo} are modeled by a Poisson point process with mean number of new \ac{pbo} ${\mu}_{\mathrm{n}}$ and \ac{pdf} $f_{\mathrm{n}}(\overline{\V{\psi}}_{m,n}^{(j)})$, where ${\mu}_{\mathrm{n}}$ is assumed to be a known constant.  
Each measurement $\RV{z}_{m,n}^{(j)}$ gives rise to a new \ac{pbo} $ \overline{\RV{y}}_{m,n}^{(j)} $. Thus, the number of new \acp{pbo} at time $n$ and anchor $j$ equals to the number of measurements $M_{n}^{(j)}$. Here, $\overline{r}_{m,n}^{(j)} = 1$ means that the measurement $\RV{z}_{m,n}^{(j)}$ was generated by a newly detected \ac{pbo}. The state vector of all new \acp{pbo} at time $n$ and anchor $j$ is given by  $ \overline{\RV{y}}_{n}^{(j)} \triangleq [\overline{\RV{y}}_{1,n}^{(j) \s \mathrm{T}} \ist\cdots\ist \overline{\RV{y}}_{\rv{M}_{n}^{(j)},n}^{(j)\s \mathrm{T}} ]^{\mathrm{T}} $. %
The new \acp{pbo} become legacy \acp{pbo} at time $n+1$. Accordingly, the number of legacy \acp{pbo} is updated as $K_{n}^{(j)} = K_{n-1}^{(j)} + M_{n}^{(j)}$. The vector containing all \ac{pbo} states at time $n$ is given by 
\begin{equation}
	\RV{y}_{n}^{(j)} \triangleq [\underline{\RV{y}}_{n}^{(j)\s \mathrm{T}} \iist \overline{\RV{y}}_{n}^{(j)\s \mathrm{T}} ]^{\mathrm{T}} = [{\RV{y}}_{1,n}^{(j)\text{T}} \rmv\cdots\ist \RV{y}_{k,n}^{(j)} \rmv\cdots\ist {\RV{y}}_{K_{n}^{(j)}\rmv,n}^{(j)\text{T}} ]^{\mathrm{T}}
\end{equation}
with $\RV{y}_{k,n}^{(j)}$ such that
$ k \in  \mathcal{K}_{n}^{(j)} $.
To avoid that the number of \acp{pbo} grows indefinitely, \ac{pbo} states with low existence probability are removed as detailed in Section~\ref{sec:factor_graph}. 

\vspace{-2mm} 
\subsection{Data Association Uncertainty}
\label{sec:DA}

Estimation of multiple \ac{pbo} states is complicated by the data association uncertainty, i.e., it is unknown which measurement $\RV{z}_{m,n}^{(j)} $ originated from which \ac{pbo} (see Fig.~\ref{fig:eye_catcher2}b). Furthermore, it is not known if a measurement did not originate from a \ac{pbo} (false alarm), or if a \ac{pbo} did not generate any measurement (missed detection). 

The associations between measurements and legacy \acp{pbo} are described by the \ac{pbo}-oriented association vector $ \underline{\RV{a}}_{n}^{(j)} \triangleq [\underline{\rv{a}}_{1,n}^{(j)} \ist \cdots \ist  \underline{\rv{a}}_{\rv{K}_{n-1},n}^{(j)}]^{\mathrm{T}} $. %
It contains $\rv{K}_{n-1}^{(j)}$ \ac{pbo}-oriented association variables, denoted as $	\underline{\rv{a}}^{(j)}_{k,n} $ where $k \rmv\in\rmv \mathcal{K}_{n-1}^{(j)}$, with entries  
\begin{equation}
	\underline{a}^{(j)}_{k,n} \rmv\rmv\triangleq \rmv\rmv
	\begin{cases}
		m \rmv\in\rmv  \mathcal{M}_n^{(j)} , \rmv\rmv & \text{legacy \ac{pbo} $ k $ causes measurement $ m $} \\[1mm]
		0,				     \rmv\rmv & \text{\parbox{\textwidth}{legacy \ac{pbo} $ k $ does not cause\\[-1mm] any measurement}} \label{eq:daoo} \\[-1mm]
	\end{cases} 
\end{equation}	
In accordance with \cite{WilliamsLauTAE2014,MeyerProc2018,LeitingerTWC2019}, the associations can be equivalently described by the measurement-oriented association vector $ \overline{\RV{a}}_{n}^{(j)} \triangleq [\overline{\rv{a}}_{1,n}^{(j)} \ist \cdots \ist \overline{\rv{a}}_{\rv{M}_{n},n}^{(j)}]^{\mathrm{T}} $. It contains  $\rv{M}_n^{(j)}$ measurement-oriented association variables, denoted as $\overline{\rv{a}}_{m,n}^{(j)}$ where $m \rmv\in\rmv  \mathcal{M}_n^{(j)}$, with entries 
\begin{equation}	
	\overline{a}^{(j)}_{m,n}  \rmv\rmv\triangleq \rmv\rmv
	\begin{cases}
		k \rmv\in\rmv \mathcal{K}_{n-1}^{(j)} , &\text{\parbox{\textwidth/4}{measurement $ m $ is caused by\\[-1mm] legacy \ac{pbo} $ k $}} \\[1mm]
		0,					     &\text{\parbox{\textwidth/4}{measurement $ m $ is not caused\\[-1mm] by any legacy \ac{pbo}}} 	\label{eq:damo}  \\[-1mm]
	\end{cases}
\end{equation}
Furthermore, we assume that at any time $ n $, each \ac{pbo} can generate at most one measurement, and each measurement can be generated by at most one \ac{pbo} (referred to in literature as point target assumption) \cite{WilliamsLauTAE2014, MeyerProc2018, LeitingerTWC2019}. This is enforced by the exclusion functions $ \Psi(\underline{\V{a}}_n^{(j)},\overline{\V{a}}_n^{(j)}) $ and \vspace{-1mm}{$\Gamma_{\underline{\V{a}}_{n}^{(j)}}(\overline{r}_{m,n}^{(j)}) $}. The exclusion function $ \Psi(\underline{\V{a}}_n^{(j)},\overline{\V{a}}_n^{(j)}) \triangleq \prod_{k = 1}^{K_{n-1}^{(j)}}\prod_{m = 1}^{M_{n}^{(j)}}\psi(\underline{a}_{k,n}^{(j)},\overline{a}_{m,n}^{(j)})$ is defined by its factors, given as
\begin{equation}
	\psi(\underline{a}_{k,n}^{(j)},\overline{a}_{m,n}^{(j)}) \rmv\rmv\triangleq \rmv\rmv
\begin{cases}
	0 , & \text{\parbox{\textwidth/4}{$ \underline{a}_{k,n}^{(j)} = m $ and $ \overline{a}_{m,n}^{(j)} \neq k $ or\\[0mm] $ \overline{a}_{m,n}^{(j)} = k $ and $ \underline{a}_{k,n}^{(j)} \neq m $}}\\[1mm]
	1,					     & \text{else}\label{eq:exclusion_psi}  \\[-1mm]
\end{cases}	%
\end{equation}
enforcing the facts that two legacy \acp{pbo} cannot be generated by the same measurement and two measurements cannot cause the same legacy \ac{pbo}. 
The function $ \Gamma_{\underline{\V{a}}_{n}^{(j)}}(\overline{r}_{m,n}^{(j)})$ is given as
\begin{equation}
\Gamma_{\underline{\V{a}}_{n}^{(j)}}(\overline{r}_{m,n}^{(j)})  \rmv\rmv\triangleq \rmv\rmv
	\begin{cases}
		0 , & \text{\parbox{\textwidth/4}{$\overline{r}_{m,n}^{(j)} = 1$ and $ \underline{a}_{k,n}^{(j)} = m $}} \\[1mm]
		1,					     &\text{else} \label{eq:exclusion_gamma} \\[-1mm]
	\end{cases}	 
\end{equation}
enforcing the fact that a measurement cannot be generated by a new \ac{pbo} and a legacy \ac{pbo}. 
The ``redundant'' formulation of using $ \underline{\RV{a}}_{n}^{(j)} $ together with $ \overline{\RV{a}}_{n}^{(j)} $ is the key to making the algorithm scalable for large numbers of \acp{pbo} and measurements (see also the supplementary material \mref{Sec.}{sec:iterative_da}). The joint vectors containing all association variables for time $n$ are given by $\underline{\RV{a}}_{n} \triangleq [\underline{\RV{a}}_{1}^{(j)\s \mathrm{T}} \, ... \, \underline{\RV{a}}_{n}^{(j)\s\mathrm{T}} ]^{\mathrm{T}}$, $\overline{\RV{a}}_{n} \triangleq [\overline{\RV{a}}_{1}^{(j)\s \mathrm{T}} \, ... \, \overline{\RV{a}}_{n}^{(j)\s \mathrm{T}} ]^{\mathrm{T}}$. 

Figure~\ref{fig:eye_catcher2} shows an exemplary propagation environment and conceptually illustrates the joint data association between measurements and \ac{pbo} states. The lines in Figure~\ref{fig:eye_catcher2} (b) represent the posterior association probabilities, where a thick line indicates a high probability. In the given example, the LOS component is blocked, thus there is no measurement that explains legacy PBO  $\underline{\RV{y}}_{1,n}^{(1)}$ and the probability of it being a missed detection is very high. Wall and scatter lead to \ac{mpc} that cause $\bm{z}_{1,n}^{(1)}$ and $\bm{z}_{3,n}^{(1)}$, respectively. As the mobile agent is moving upward, the scatter was visible at previous time $n-1$. Accordingly, its measurement has a high probability of corresponding to legacy PBO  $\underline{\RV{y}}_{2,n}^{(1)}$.  Due to the obstacle, the wall was not visible at previous time $n-1$ and its measurement has a high probability of corresponding to a new PBO $\overline{\RV{y}}_{1,n}^{(1)}$.

\vspace{-2mm} 
\section{Factor Graph and Sum-Product Algorithm} \label{sec:factor_graph}

The problem considered is the sequential estimation of the agent state ${\RV{x}}_n$ using all observed measurements $\V{z}_{1:n}\rmv$ from all anchors up to time $n$. This is done in a Bayesian sense by calculating the \ac{mmse} \cite{Kay1993} estimate of the extended agent state 
\vspace*{-1mm}
\begin{align}\label{eq:mmse} 
{	\hat{{\bm{x}}}^\text{MMSE}_{n} \,\triangleq \int \rmv {\bm{x}}_{n} \, f({\bm{x}}_{n} |\V{z}_{1:n} )\, \mathrm{d}{\bm{x}}_{n} }\\[-6mm]\nn
\end{align}
with $\hat{{\bm{x}}}^{\text{MMSE}}_{n} = [\hat{\bm{p}}^{\text{MMSE T}}_n $ $ \hat{\bm{v}}^{\text{MMSE {T}}}_n]^\text{T}$ and $\V{z}_{1:n} = [\V{z}^\text{T}_{{1}}\, ... \, \V{z}^\text{T}_{n}]^\text{T}$. We also calculate the states of all \textit{detected} \acp{pbo}\\
\vspace*{-4mm}
\begin{align} \label{eq:mmsepbo}
	\hat{\bm{\psi}}^{(j)\s\text{MMSE}}_{k,n}\triangleq \int \rmv \bm{\psi}_{k,n}^{(j)} \, f(\bm{\psi}_{k,n}^{(j)} | r_{k,n}^{(j)} = 1 , \V{z}_{1:n} )\, \mathrm{d} \bm{\psi}_{k,n}^{(j)} \\[-6mm]\nn
\end{align}
with $\hat{\bm{\psi}}_{k,n}^{(j)\s \text{MMSE}}  = [\hat{b}_{k,n}^{(j)\s \text{MMSE}}\, \hat{v}_{\text{b}\s k,n}^{(j)\s \text{MMSE}}\, \hat{u}_{k,n}^{(j)\s \text{MMSE}}]^\text{T}$.
A \ac{pbo} is detected if $ p(r_{k,n}^{(j)} = 1|\V{z}_{1:n}) > p_\mathrm{de} $ \cite{Kay1998}, where $ p_\mathrm{de} $ is the existence probability threshold not to be confused with $ \lambda $, the detection threshold of the \ac{ceda}. 
The existence probabilities $ p(r_{k,n}^{(j)} = 1|\V{z}_{1:n}) $ are obtained from the marginal posterior \acp{pdf} of the \ac{pbo} states, $ f(\V{y}_{k,n}^{(j)}| \V{z}_{1:n}) = f(\V{\psi}_{k,n}^{(j)}, r_{k,n}^{(j)} | \V{z}_{1:n}) $, according to 
\vspace*{-2mm}
\begin{align}
	p({r}_{k,n}^{(j)} \!=\!1 \big| \V{z}_{1:n}) =\rmv \int \rmv f({\V{\psi}}_{k,n}^{(j)}\ist, r_{k,n}^{(j)} \!=\! 1\big|\V{z}_{1:n}) \ist\mathrm{d}{\V{\psi}}_{k,n}^{(j)} 
	\label{eq:existProb}
	\\[-6mm]\nn
\end{align}
and the marginal posterior \acp{pdf} $ f(\V{\psi}_{k,n}^{(j)} | r_{k,n}^{(j)} = 1,\V{z}_{1:n}) $ are obtained from $ f(\V{\psi}_{k,n}^{(j)}, r_{k,n}^{(j)} | \V{z}_{1:n}) $ as \vspace*{-2mm}
\begin{align}
	f(\V{\psi}_{k,n}^{(j)} | r_{k,n}^{(j)} = 1,\V{z}_{1:n}) = \dfrac{f(\V{\psi}_{k,n}^{(j)}, r_{k,n}^{(j)} = 1 | \V{z}_{1:n})} {p(r_{k,n}^{(j)} = 1 | \V{z}_{1:n})}\ist.
	\label{eq:PSMC_existProb} \\[-6mm]\nn
\end{align}
We consider the estimates provided at time $n$ as ``reliable" when the LOS component, i.e., the \ac{pbo} at $k=1$, is detected by at least three anchors $j$, i.e., $|\mathcal{J}_{\text{LOS},n}| \geq 3$, where
\begin{equation} \label{eq:reliability}
	\mathcal{J}_{\text{LOS},n} = \{\, j \in \{1,\s...\s,J\} \, | \,  p(r_{1,n}^{(j)} = 1|\V{z}_{1:n}) > p_\mathrm{de} \,\} \ist .
\end{equation}
{As the number of \acp{pbo} grows with time $n$ (at each time by $ K_{n}^{(j)} = K_{n-1}^{(j)} + M_{n}^{(j)}$), \acp{pbo} with posterior existence probability $ p(r_{k,n}^{(j)} = 1 | \V{z}_{1:n})$ below a threshold $p_\mathrm{pr}$ are removed from the state space (``pruned'').} The \ac{los} \ac{pbo} at $k=1$ is not pruned, even if its existence probability $r_{1,n}^{(j)}$ falls below $p_\mathrm{pr}$.
\ifthenelse{0 = 1}{
	
\alex{Furthermore, we estimate the posterior variance of the agent state using
\begin{equation}\label{eq:post_var} 
	\hat{\V{\sigma}}^2_{\text{\textbf{x}}\s n} \,\triangleq \int \rmv ( {\bm{x}}_{n} - 	\hat{{\bm{x}}}^\text{MMSE}_{n})^2  \, f({\bm{x}}_{n} |\V{z}_{1:n} )\, \mathrm{d}{\bm{x}}_{n}
\end{equation}
with $\hat{\V{\sigma}}^2_{\text{\textbf{x}}\s n} = 
[\hat{\V{\sigma}}^{2}_{\text{\textbf{p}}\s n}\,
\hat{\V{\sigma}}^{2}_{\text{\textbf{v}}\s n}\,
]^\text{T}$.  Based on this estimates, we decide for the position measurement to be ``reliable" if
\vspace{-2mm}
\begin{equation}\label{eq:reliable_agent_detection} \vspace{-2mm}
\sum_{i=1}^{2}[\hat{\V{\sigma}}^{2}_{\text{\textbf{p}}\s n}]_{i}
< 2\, \text{tr}\{  \tilde{\bm{J}}_{\bm{p}\s \text{P} \s n}^{-1} \} %
\end{equation}
where $ \tilde{\bm{J}}_{\bm{p}\s \text{P} \s n} $ is at time $n$ the Fisher information matrix for joint positioning and tracking according to the supplementary material of  \cite[Sec. VI, Eq. 27]{VenusTWC2023}, given as
\begin{equation} \label{eq:pcrlb}
	\tilde{\bm{J}}_{\bm{p}\s \text{P} \s n} = ( \bm{A} \, \bm{J}_{\bm{p}\s \text{P} \s n\minus 1}^{-1} \, \bm{A}^\text{T} \, + \, {\sigma_{\text{a}}^2} \, \bm{B} \, \bm{B}^\text{T} )^{-1} + \tilde{\bm{J}}_{\bm{p}\s \text{S} \s n}
\end{equation}
where we replaced $ {\bm{J}}_{\bm{p} \s \text{S}\s n}  $ with  %

\begin{equation} \vspace{-1mm} \label{eq:spcrlb}
\tilde{\bm{J}}_{\bm{p} \s \text{S}\s n} = \frac{8 \pi^2 \beta_\text{bw}^2}{c^2} \sum_{j=1}^{J}  \tilde{u}_{n^{(j)}_\text{LOS}}^{(j)\s 2} \V{D}_{\text{r} \s n^{(j)}_\text{LOS}}^{(j)}
\end{equation}
where $n^{(j)}_\text{LOS}$ is the last time the \ac{pbo} corresponding to the \ac{los} component ($k=1$) of anchor $j$ was detected, i.e.,
\begin{equation}\label{eq:nlosj} 
n^{(j)}_\text{LOS} = \argmin_{n' \in \{1 ... n \}} n-n' ~|~  {p(r_{1,n'}^{(j)} = 1 | \V{z}_{1:n}) > p_\mathrm{de}} \ist .
\end{equation}
}

}{}

In order to obtain \eqref{eq:mmse}-\eqref{eq:PSMC_existProb}, the respective marginal posterior \acp{pdf} need to be calculated from the joint posterior \ac{pdf} $f( \V{x}_{0:n}, \V{y}_{1:n}, \underline{\V{a}}_{1:n}, \overline{\V{a}}_{1:n} ,\V{m}_{1:n} | \V{z}_{1:n} )$ representing the statistical model discussed in Sec.~\ref{sec:system_model}. Since direct marginalization of the joint posterior \ac{pdf} is computationally infeasible\cite{MeyerProc2018}, we perform message passing by means of the \ac{spa} rules on the factor graph that represents the factorization of the joint posterior \ac{pdf}.

\vspace{-4mm} 
\subsection{Joint Posterior and Factor Graph} \label{sec:joint_posterior}
\vspace{-0.5mm}
{%
The vectors containing all state variables for all times up to $n$ are given by
$\RV{z}_{1:n} = [\RV{z}^\text{T}_{{1}}\, ... \, \RV{z}^\text{T}_{n}]^\text{T}$, 
${\RV{x}_{0:n}} = [{\RV{x}}^\text{T}_{0}\, ... \, {\RV{x}}^\text{T}_{n}]^\text{T}$, 
$\underline{\RV{a}}_{1:n} \triangleq [\underline{\RV{a}}_{1}^{\text{T}} \, ... \, \underline{\RV{a}}_{n}^{\text{T}} ]^{\text{T}}$, 
$ \overline{\RV{a}}_{1:n} \triangleq [\overline{\RV{a}}_{1}^{\text{T}} \, ... \,\overline{\RV{a}}_{n}^{\text{T}} ]^{\text{T}}$, 
$\RV{y}_{1:n} = [\RV{y}^\text{T}_{1}\, ... \, \RV{y}^\text{T}_{n}]^\text{T}$, 
and $\RV{m}_{1:n} = [\RV{M}^\text{T}_{1}\, ... \, \RV{M}^\text{T}_{n}]^\text{T}$.
We now assume that the measurements $\bm{z}_{1:n}$ are observed and thus fixed.
Applying Bayes' rule {as well as some commonly used independence assumptions}\cite{MeyerProc2018,LeitingerTWC2019}, the joint posterior \ac{pdf} of all state variables $\RV{x}_{0:n}$, $\RV{y}_{1:n}$, $\underline{\RV{a}}_{1:n}$, $\overline{\RV{a}}_{1:n}$ ,$\RV{m}_{1:n}$ up to time $n$ can be derived up to a constant factor as 
\begin{align}
	&f( \V{x}_{0:n}, \V{y}_{1:n}, \underline{\V{a}}_{1:n}, \overline{\V{a}}_{1:n} ,\V{m}_{1:n} | \V{z}_{1:n} ) \nn \\[-1.3mm]
	&\hspace{4mm}\propto  f(\V{x}_{0}) \prod^{J}_{j'=1}\rmv\rmv\rmv f(\underline{\V{y}}^{(j')}_{1,0})   %
	\prod^{n}_{n'=1} \rmv \! \Phi_\mathrm{x}(\V{x}_{n'}|\V{x}_{n'-1}) \prod^{J}_{j=1} \rmv\rmv\Psi\big(\underline{\V{a}}^{(j)}_{n'} \rmv,\overline{\V{a}}^{(j)}_{n'}\big) \nn\\[0mm]
	&\hspace{4mm}\times \rmv %
	\prod^{K^{(j)}_{n'-1}}_{k=1}\rmv\rmv\rmv \Phi_k\big(\underline{\V{y}}^{(j)}_{k,n'} \big| \V{y}^{(j)}_{k,n'-1}\big) \underline{g}\big( \V{x}_{n'}, \underline{\V{\psi}}^{(j)}_{k,n'} , \underline{r}^{(j)}_{k,n'}, \underline{a}^{(j)}_{k,n'}; \V{z}^{(j)}_{n'} \big)\rmv
	\nn \\
	&\hspace{4mm}\times \prod^{M^{(j)}_{n'}}_{m=1} \overline{g}\big( \V{x}_{n'}, \overline{\V{\psi}}^{(j)}_{m,n'} , \overline{r}^{(j)}_{m,n'}, \overline{a}^{(j)}_{m,n'}; \V{z}^{(j)}_{n'} \big)  %
	\label{eq:factorization1}\\[-6mm]\nn
\end{align}
where we introduced the state-transition functions  $\Phi_\mathrm{x}({\bm{x}}_n|{\bm{x}}_{n-1}) \triangleq  f({\bm{x}}_n|{\bm{x}}_{n-1})$, and $\Phi_k(\underline{\bm{y}}_{k,n}^{(j)}|\bm{y}_{k,n-1}^{(j)}) \triangleq  f(\underline{\bm{y}}_{k,n}^{(j)}|\bm{y}_{k,n-1}^{(j)})$, as well as the pseudo \acp{lhf} $\underline{g}\big( \V{x}_n, \underline{\V{\psi}}^{(j)}_{k,n} , \underline{r}^{(j)}_{k,n}, \underline{a}^{(j)}_{k,n}; \V{z}^{(j)}_{n} \big)$ and  $\overline{g}\big( \V{x}_n, \overline{\V{\psi}}^{(j)}_{m,n} , \overline{r}^{(j)}_{m,n}, \overline{a}^{(j)}_{m,n}; \V{z}^{(j)}_{n} \big)$, for legacy \acp{pbo} and new \acp{pbo}, respectively.

For $\underline{g}\big( \V{x}_n, \underline{\V{\psi}}^{(j)}_{k,n} , \underline{r}^{(j)}_{k,n}, \underline{a}^{(j)}_{k,n}; \V{z}^{(j)}_{n} \big)$ one obtains
\vspace*{-2mm}
\begin{align}
	&\underline{g}\big( \V{x}_n, \underline{\V{\psi}}^{(j)}_{k,n} , 1, \underline{a}^{(j)}_{k,n}; \V{z}^{(j)}_{n} \big)  \nn \\[1.5mm]
	&\hspace{1mm}=\begin{cases}
		\displaystyle \frac{ \pd \ist f(\V{z}_{m,n}^{(j)}| \V{x}_n, \underline{\V{\psi}}_{k,n}^{(j)}) }  {\mu_\mathrm{fa} \ist f_{\mathrm{fa}}\big( \V{z}^{(j)}_{m,n} \big)} \ist, 
		&\rmv \underline{a}^{(j)}_{k,n}\!=\rmv m \in\! \Set{M}^{(j)}_n \\[4.5mm]
		1 \!-\rmv \pd \ist, &\rmv \underline{a}^{(j)}_{k,n} \!=\rmv 0 
	\end{cases}\label{eq:underlineg} \\[-6mm]\nn
\end{align}
and $\underline{g}\big( \V{x}_n, \underline{\V{\psi}}^{(j)}_{k,n} , 0, \underline{a}^{(j)}_{k,n}; \V{z}^{(j)}_{n} \big) = 1_{\{0\}}\big(\underline{a}^{(j)}_{k,n}\big)$. 
Similarly, for $\overline{g}\big( \V{x}_n, \overline{\V{\psi}}^{(j)}_{m,n} , \overline{r}^{(j)}_{m,n}, \overline{a}^{(j)}_{m,n}; \V{z}^{(j)}_{n} \big)$ one can write 
\vspace*{-2mm}
\begin{align}
	&\overline{g}\big( \V{x}_n, \overline{\V{\psi}}^{(j)}_{m,n} , 1, \overline{a}^{(j)}_{m,n}; \V{z}^{(j)}_{n} \big) \nn \\[1mm]
	&\triangleq \begin{cases}
		0 \ist,  &\rmv\rmv\rmv\rmv\rmv\rmv \overline{a}^{(j)}_{m,n} \!\in\rmv\Set{K}^{(j)}_{n-1} \\[1.5mm]
		\displaystyle \frac{\mu_{\mathrm{n}} \ist f_{\mathrm{n}}(\overline{\V{\psi}}_{m,n}^{(j)})  \ist f(\V{z}_{m,n}^{(j)}| \V{x}_n, \overline{\V{\psi}}_{m,n}^{(j)}) }{\mu_\mathrm{fa} \ist f_{\mathrm{fa}}\big( \V{z}^{(j)}_{m,n} \big)} 
		\ist, &\rmv\rmv\rmv\rmv\rmv\rmv \overline{a}^{(j)}_{m,n} \rmv=\rmv 0 
		\label{eq:overlineg}
	\end{cases}\\[-6mm]\nn
\end{align}
and $\overline{g}\big( \V{x}_n, \overline{\V{\psi}}^{(j)}_{m,n} , 0, \overline{a}^{(j)}_{m,n}; \V{z}^{(j)}_{n} \big) \rmv\triangleq\rmv f_{\mathrm{d}}\big(\overline{\V{\psi}}^{(j)}_{m,n}\big)$. 
The factor graph \cite{KschischangTIT2001,Loeliger2004SPM} representing the factorization in \eqref{eq:factorization1} is shown in Fig.~\ref{fig:factorGraph}. 
Note that $\RV{m}_{1:n}$ vanishes in \eqref{eq:factorization1} as it is fixed and thus constant, being implicitly defined by the measurements $\bm{z}_{1:n}$, and that the exclusion function $\Gamma_{\underline{\V{a}}_{n}^{(j)}}(\overline{r}_{m,n}^{(j)})$ %
has been considered in \eqref{eq:overlineg}. 
A detailed derivation of the joint posterior in \eqref{eq:factorization1} is given in \cite{MeyerProc2018,LeitingerTWC2019,LiTWC2022}.

\vspace{-2mm}
\subsection{Marginal Posterior and Sum-Product Algorithm (SPA)}  \label{sec:spa}
Since direct marginalization of the joint posterior \ac{pdf} in \eqref{eq:factorization1} is infeasible, we use loopy message passing (belief propagation) \cite{KschischangTIT2001} by means of the \acf{spa} rules \cite{KschischangTIT2001,Loeliger2004SPM} on the factor graph shown in Fig.~\ref{fig:factorGraph}. Due to the loops inside the factor graph, the resulting beliefs $ q(\V{x}_{n}) $, $ {q}(\underline{\V{y}}_{k,n}^{(j)}) = {q}(\underline{\V{\psi}}_{k,n}^{(j)}, \underline{r}_{k,n}^{(j)}) $, 
and
$ {q}(\overline{\V{y}}_{m,n}^{(j)}) = {q}(\overline{\V{\psi}}_{m,n}^{(j)}, \overline{r}_{m,n}^{(j)}) $ are only approximations of the respective posterior marginal \acp{pdf}. %
See the supplementary material \mref{Sec.}{sec:spa_messages} for a detailed derivation of the resulting \ac{spa}. 
Since the integrals involved in the calculations of the messages and beliefs cannot be obtained analytically, we use a computationally efficient sequential particle-based message passing implementation that performs approximate computations. As in \cite{MeyerPhd2015,MeyerJSIPN2016}, our implementation uses a ``stacked state", comprising the agent state as well as all \ac{pbo} states. %
A detailed derivation of the particle-based implementation is also given in the supplementary material \mref{Sec.}{sec:particle_based_implementation}. 
\begin{figure}[t]
	\centering
	\tikzsetnextfilename{factor_graph}
	\scalebox{0.96}{%
		\includegraphics{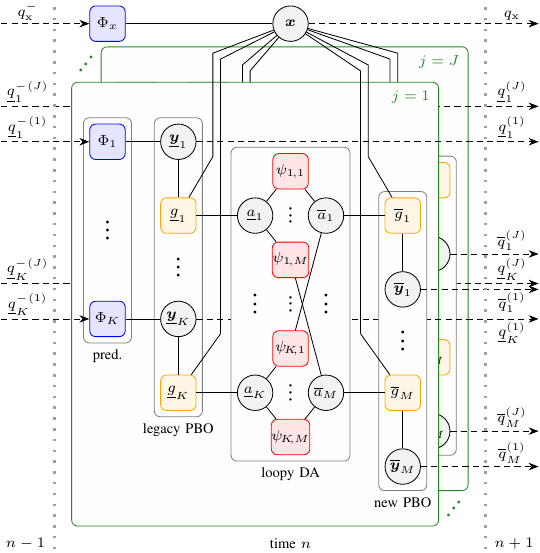}}
	\caption{Factor graph corresponding to the factorization shown in \eqref{eq:factorization1}. Dashed arrows represent messages that are only passed in one direction. The following short notations are used: %
		$ K \triangleq K_{n-1}^{(j)} $, 
		$ M \triangleq M_{n}^{(j)} $, 
		$ \underline{a}_{k} \triangleq \underline{a}_{k,n}^{(j)} $, 
		$ \overline{a}_{m} \triangleq \overline{a}_{m,n}^{(j)} $, 
		$ \V{x}\triangleq \V{x}_{n} $,
		$ \underline{\V{y}}_{k} \triangleq \underline{\V{y}}_{k,n}^{(j)} $, 
		$ \overline{\V{y}}_{m} \triangleq \overline{\V{y}}_{m,n}^{(j)} $, 
		$ \Phi_{x} \triangleq \Phi_x(\V{x}_{n} |\V{x}_{n-1} ) $, 
		$ \Phi_{k} \triangleq \Phi_k(\underline{\V{y}}_{k,n}^{(j)} |\V{y}_{k,n-1}^{(j)} ) $, 
		$ \underline{g}_{k} \triangleq \underline{g}( \V{x}_{n}, $ $ \underline{\V{\psi}}^{(j)}_{k,n} , \underline{r}^{(j)}_{k,n}, \underline{a}^{(j)}_{k,n}; \V{z}^{(j)}_{n} ) $, 
		$ \overline{g}_{m} \triangleq  \overline{g}( \V{x}_{n}, \overline{\V{\psi}}^{(j)}_{m,n} , \overline{r}^{(j)}_{m,n}, \overline{a}^{(j)}_{m,n}; \V{z}^{(j)}_{n} ) $, 
		$ \psi_{k,m} \triangleq \psi(\underline{a}_{k,n}^{(j)},\overline{a}_{m,n}^{(j)}) $, 
		$q_\mathrm{x} \triangleq q(\V{x}_{n})$,  
		$\underline{q}_{k}^{(j)} \triangleq \rmv  {q}(\underline{\V{y}}_{k,n}^{(j)})$, 
		$\overline{q}_{m}^{(j)} \triangleq	{q}(\overline{\V{y}}_{m,n}^{(j)})$, 
		$q_\mathrm{x}^{-} \triangleq q(\V{x}_{n\minus 1})$,  
		$\underline{q}_{k}^{- (j)} \triangleq {q}(\underline{\V{y}}_{k,n\minus 1}^{(j)})$. 
		\vspace{-4.5mm}}	 
	\label{fig:factorGraph}
\end{figure}
The computational complexity scales only linearly in the number of particles $I$. %
The initial distributions $f(\V{x}_{0})$ and $%
f(\underline{\V{y}}^{(j)}_{1,0})$ are determined heuristically, using an initial measurement vector $\V{z}_0$ containing $\V{M}_0$ measurements, as detailed in Section~\ref{sec:init_states}. %
For computational efficiency of the particle-based implementation, we approximate %
\eqref{eq:los_amplitude} by a truncated Gaussian \ac{pdf}, i.e., we set $f(\zu| u^{(j)}_{k,n}) \triangleq f_\text{TN}(\zu; \sigma_\mathrm{u} (u^{(j)}_{k,n}) ,u^{(j)}_{k,n}, \gamma)$.

\section{Initial States} \label{sec:init_states}
The initial distributions $f(\V{x}_{0})$ and $f(\underline{\V{y}}_{0,}) = \prod_{j=1}^{J}  
f(\underline{\V{y}}^{(j)}_{1,0})$ are determined heuristically, assuming an initial measurement vector $\V{z}_0$ containing $\V{M}_0$ measurements to be available. 
It is assumed to factorize as
${f}({\bm{x}}_{0}) = 
{f}({\bm{p}}_{0})
{f}({\bm{v}}_{0})$. 

For all anchors $j \rmv\in\rmv \mathcal{J}$, we assume that the joint \ac{pbo} state only contains the \ac{los} component, i.e., $\mathcal{K}_n^{(j)} = \{ 1 \}$, while \acp{pbo} corresponding to \acp{mpc} are initialized as new \acp{pbo} during filter operation (at times $n\geq 1$). As discussed in %
Sec.~\ref{sec:pbo_states} the bias of the \ac{los} component is zero, i.e., $b_{1,0}^{(j)} \equiv 0,\iist v_{\text{b}\s 1,0}^{(j)} \equiv 0$. We initialize the normalized amplitude \acp{pdf} as 
${f}(u_{1,0}^{(j)}) \sim f( \zuZero | u_{\s \text{init}}^{(j)})\, f(u_{\s \text{init}}^{(j)}) $ where $f(u_{\s \text{init}}^{(j)})$ is drawn from a uniform distribution given as $f(u_{\s \text{init}}^{(j)}) \triangleq f_\mathrm{U}(x;0,u_\text{max})$ and $u_\text{max}$ is the maximum amplitude. %
The existence variables are initialized uniformly distributed as $p({r}^{(j)}_{1,0} ) = p_\mathrm{UD}({r}^{(j)}_{1,0}; \{0.5,0.5\})$.

The agent position state is initialized as $f(\bm{p}_0) \sim \prod_{j=1}^{J}  \prod_{m=1}^{M^{(j)}_0} f( \zdZero | {b}^{(j)}_{1,0}~\!=~\!0,  \bm{p}_{\s \text{init}},\zuZeroMax ) \, f(\bm{p}_{\s \text{init}}) $, where $\zuZeroMax$ is the maximum normalized amplitude measurement in $\V{z}_0^{(j)}$. The proposal distribution $f(\bm{p}_{\s \text{init}})$ is drawn uniformly on two-dimensional discs around each anchor $j$, which are bounded by the maximum possible distance $d_\text{max}$ and a sample is drawn from each of the $J$ discs with equal probability. 
We assume the velocity vector ${\bm{v}}_{0}$ to be zero mean, Gaussian, with covariance matrix $\sigma_\text{v}^2\,  \bm{I}_2 $ and $\sigma_\text{v} = 6\, \text{m/s}$, as we do not know in which direction we are moving.
After drawing from the proposal distributions $ f(\bm{p}_{\s \text{init}})$ and $ f(u_{\s \text{init}}^{(j)}) $, we perform a resampling step (see the supplementary material  \mref{Sec.}{sec:particle_state_estimation}) that avoids particle degeneracy to obtain particle-based representations\footnote{Note that for numerical implementation this can also be realized by drawing samples directly from the measurement space.} of ${f}({\bm{p}}_{0})$ and ${f}(u_{1,0}^{(j)})$.

\vspace{-1mm}
\section{Results}\label{sec:results}

\newcommand{\shadesofgraysynthetic}{Different shades of gray represent different numbers of anchors in OLOS according to the legend in Fig.~\ref{fig:results_va}.}
\newcommand{\shadesofgrayreal}{Different shades of gray represent different numbers of anchors in OLOS according to the legend in Fig.~\ref{fig:track_nxp}.}
We validate the proposed algorithm by analyzing its performance %
using both synthetic data obtained using numerical simulation of different propagation scenarios and real radio measurements. The performance is compared with state-of-the-art multipath-aided positioning and tracking methods, including the \ac{mpslam} algorithm presented in \cite{LeitingerTWC2019,LeitingerICC2019}, the channel SLAM algorithm from \cite{GentnerTWC2016}, and the multipath ``cluster''%
-based robust positioning algorithm from \cite{VenusTWC2023}. 
For synthetic measurements, we also compare to the learning-based method introduced in \cite{WymeerschIEEE2012}, as well as to the hybrid approach from \cite{Kram2022GPR}.
As a performance benchmark we provide the \ac{pcrlb} on the agent position error and compare to a particle-based variant of the multi-sensor \ac{pdaai} \cite{JeoTugTAES2005}, which, in contrast to the other methods, does not facilitate multipath and, thus, acts as an additional baseline. 
Table~\ref{tbl:reference_methods} provides a summary of all reference methods along with their respective abbreviations. In the remainder of the paper, we use these abbreviations to refer to the respective methods. 

\begin{table}[htb]
	\vspace{-2mm}
	\renewcommand{\baselinestretch}{1}\small\normalsize
	\setlength{\tabcolsep}{3pt} %
	\renewcommand{\arraystretch}{1.1} %
	\centering
	\footnotesize
	\caption{Reference methods and respective abbreviations}\label{tbl:reference_methods}
	\vspace{-1mm}
	\begin{tabular}{ |r p{6.8cm}| }  
		\toprule
		abbreviation & description  \\
		\midrule
		\acs{pdaai} &  particle-based variant of the multi-sensor \ac{pdaai} \cite{JeoTugTAES2005} \\
		\acs{mpslam} &  \acl{mpslam} algorithm presented in \cite{LeitingerTWC2019,LeitingerICC2019} \\
		\acs{chslam}&  channel SLAM algorithm presented in \cite{GentnerTWC2016} \\
		\acs{cluster} &  multipath cluster-based robust tracking algorithm from \cite{VenusTWC2023} \\
		\acs{delaybias} &  learning-based bias mitigation algorithm presented in \cite{WymeerschIEEE2012} \\
		\acs{gptrack} &  learning-based robust positioning method from \cite{Kram2022GPR} \\
		\bottomrule
	\end{tabular}
	\vspace{-2.5mm}
\end{table}

\vspace{-2mm}
\subsection{Common Analysis Setup} \label{sec:common_setup}
The following setup and parameters are commonly used for all analyses presented unless noted otherwise.

{To obtain the measurements $\bm{z}_{n,m}^{(j)}$ for each anchor $j$ at each time $n$ we used the \ac{ceda} from the supplementary material of \cite{VenusTWC2023}.} 
The state transition variances are set as 
$\sigma_a=2~\mathrm{m/s^2}$, 
$\sigma_\text{u} = 0.05\,\hat{u}_{n\minus 1}^{(j)\s \text{MMSE}}$, 
$\sigma_\text{b} = 0.05\, \hat{b}_{n\minus 1}^{(j)\s \text{MMSE}} $. While $\sigma_a$ is set according to the maximum agent acceleration \cite{BarShalom_AlgorithmHandbook}, for the state transition variances of all other parameters we use values proportional to the \ac{rmse} estimate of the previous time step $n \minus 1$ as a heuristic. 
Note that this choice allows no tuning of the state transition variances to be required for all experiments presented, even though the propagation environments are considerably different.
The particles for the initial state of a new \ac{pbo} $ {\overline{\RV{\psi}}}_{m,n}^{(j)} $ are drawn from independent uniform distributions in the respective observation space, according to the joint \ac{pdf} $f_{\mathrm{n}}\big(\overline{\V{\psi}}_{m,n}^{(j)}\big) \triangleq f_\mathrm{U}(\overline{b}_{k,n}^{(j)};0,d_\text{max})\, f_\mathrm{U}( \overline {v}_{\text{b}\s k,n}^{(j)}; -v_{\text{b}\s \text{max}} , v_{\text{b}\s \text{max}} )\, f_\mathrm{U}(\overline{u}_{k,n}^{(j)} ;0 , u_\text{max}) $, where the maximum normalized amplitude and bias velocity are assumed to be $u_\text{max} = 40~\mathrm{dB}$ and $v_{\text{b}\s \text{max}} = 4~\mathrm{m/s}$, respectively.
The other simulation parameters are as follows: the survival probability is $ p_{\mathrm{s}} = 0.99 $, the existence probability threshold is $ p_{\mathrm{de}} = 0.99 $,
 the pruning threshold is $ p_{\mathrm{pr}} = 10^{-2} $, the mean number of newly detected \acp{mpc} is $ \mu_{\mathrm{n}} = 0.05 $, the maximum number of message passing iterations for the loopy DA is {$P = 5000 $} and the \acp{pdf} of the states are represented by {$ I = 30^4 $} particles each. 
We set the detection threshold to $\gamma = 2$ ($6\,\mathrm{dB}$) for all simulations, which allows the algorithm to facilitate low-energy \acp{mpc}. %
For numerical stability, we reduced the root mean squared bandwidth $\beta_\text{bw}$ in \eqref{eq:like_delay} for \acp{mpc} (i.e., $ k \in {2, \, ... \, , K_n^{(j)}}$) by a factor of $1/3$. 
To prevent the algorithm after an \ac{olos} situation from initializing a new \ac{pbo} which competes with the explicit \ac{los} component ($k=1$), we introduce a gate region according to \cite[p. 95]{BarShalom1995}. %
Measurements inside the gate region do not create new \acp{pbo}. The corresponding gate threshold is chosen such that the probability that the \ac{los} measurement is inside the gate is 0.999.
\begin{figure}[t!]
	
	\centering
	\setlength{\abovecaptionskip}{0pt}
	\setlength{\belowcaptionskip}{0pt}

	\tikzsetnextfilename{track_va}
	\includegraphics{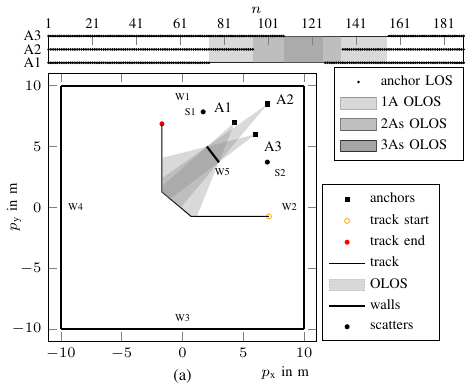}
	\vspace{-1mm}
	\tikzsetnextfilename{training_data_ml}
	\scalebox{0.98}{\includegraphics{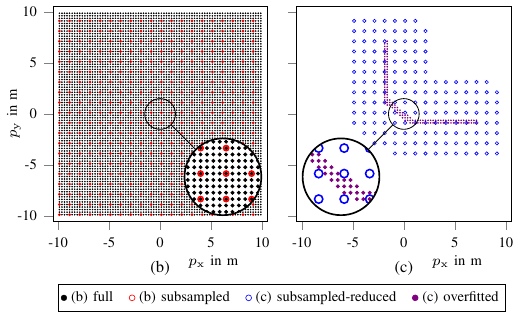}}
	\vspace{-0.5mm}
	\caption{Graphical representation of the investigated synthetic experiments. Fig. (a) shows the simulated trajectory, anchor positions, walls, and OLOS intervals. Figs. (b) and (c) show the positions of all simulated training datasets.
	}\label{fig:track_geometric}
	\vspace{-4.5mm}
\end{figure}
\begin{figure*}[t]
	\centering
	\setlength{\abovecaptionskip}{0pt}
	\setlength{\belowcaptionskip}{0pt}
	\tikzsetnextfilename{single_realization}
	\includegraphics{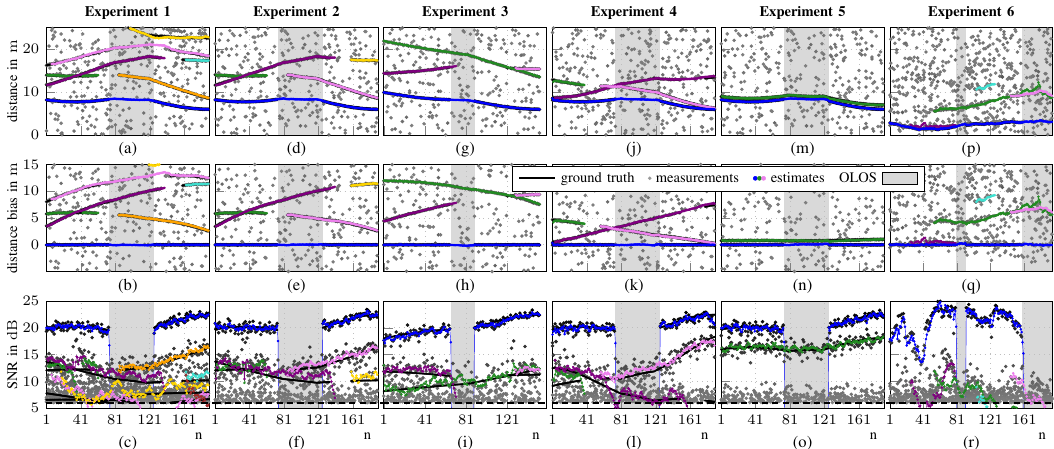}
	\vspace{-1mm}
	\caption{Single measurement realization in terms of distance, distance bias to the \ac{los} component and \ac{snr} of anchor A1 for all six experiments investigated. We show the \ac{ceda} measurements together with the true values (for synthetic experiments) and the \ac{mmse} estimates of the proposed algorithm. 
	}\label{fig:single_realization}
	\vspace{-4.5mm}
\end{figure*}

\subsubsection{Implementation of Reference Methods} \label{sec:reference_methods}

For consistency, the state-transition \acp{pdf} and initial state distributions of the agent state of all reference methods (and the normalized amplitude state of \ac{mpslam}, \acs{cluster} and \acs{pdaai}) are set as described in Sec.~\ref{sec:common_setup} and {Sec.}~\ref{sec:init_states}. All reference methods rely on particle-based state-space representations. In line with the proposed method, we used $30^4$ particles for \ac{mpslam}. For \acs{chslam}, we used $10^4$ particles for the agent state and $300$ particles for each \ac{va} state. As recommended in \cite{Kram2022GPR} and \cite{VenusTWC2023}, we used $5000$ particles for state-space representation of all other reference methods. 
\ac{mpslam} is implemented according to \cite{LeitingerICC2019, LeitingerTWC2019} using the measurements $\bm{z}_{m,n}^{(j)}$, i.e., distance and amplitude measurements, determined by applying the selected \ac{ceda} to the individual radio signal vectors $\bm{r}_n^{(j)}$ as an input%
. 
 The parameters of the dynamic object model (mean number of false alarms, mean number of potential \acp{va}, probability of survival, pruning threshold) as well as detection threshold and the number of particles are set in accordance with Sec.~\ref{sec:common_setup}. %
The anchor driving noise was set to $\sigma_\text{An} = 0.02~\mathrm{m}$. 
In line with the proposed algorithm, we introduced a gate region as described in Sec.~\ref{sec:common_setup} to prevent \ac{mpslam} from initializing new \acp{va} after an \ac{olos} situation, which compete with the physical anchors. For stability, we increased the distance measurement variances of all virtual anchors (not the physical anchors) by a factor of $3$ \ac{wrt} the Fisher information-based value. 
\acs{cluster} is implemented according to \cite{VenusTWC2023}. Again, we use the measurements $\bm{z}_{m,n}^{(j)}$ provided by the selected \ac{ceda} as an input. 
\acs{chslam} is implemented according to \cite{GentnerTWC2016}. We have only implemented the channel SLAM algorithm proposed in \cite{GentnerTWC2016}, not the full two-stage method that includes a channel estimator/-tracker. For consistency, we again use the measurements $\bm{z}_{m,n}^{(j)}$ obtained by applying the selected \ac{ceda} to the individual radio signal vectors $\bm{r}_n^{(j)}$, where \acs{chslam} uses only the distances $\zd$ as measurement inputs. To ensure a fair comparison, we use Fisher information-based variance values, consistent with the other methods, which are determined using the corresponding amplitude measurements $\zu$. Since the data association is unknown, we perform Monte-Carlo data association for each particle separately, following the conventional approach in classical Rao-Blackwellized SLAM \cite{DurrantWhyte2006,MonThrKolWeg:AAAI2002}. 
\acs{pdaai} is implemented identically to \acs{cluster} assuming an uninformative \ac{nlos} distribution (conventional uniformly distributed clutter model\cite{BarShalom1995}). 
Since \acs{delaybias} and \acs{gptrack} do not perform data association, we estimate the LOS component distance using a search-forward method\cite{DardariProcIEEE2009}. On the interpolated Bartlett spectrum\cite{KrimVibergSPM1996}, we search in a super-resolution manner for the first maximum that exceeds a relative threshold, which we chose as six times the noise variance. 
The search-forward approach enables correctly identifying the LOS component (i.e., the first visible signal component), even when there are \acp{mpc} with amplitudes higher than that of the \ac{los} component. 
\acs{gptrack} additionally applies to the received baseband signal vector $\bm{r}_n^{(j)}$ an \ac{aednn} compressing it into a small number of feature measurements, as well as a variational \ac{aednn} used for ``anomaly detection"\cite{StahlkeSensors2021}, i.e. data-driven identification of \ac{olos} situations. Also a \ac{gpr}-based \ac{lhf} is learned for representing the fingerprint of \ac{nlos} measurements. We set up the \acp{aednn} as well as the \ac{gpr} using the configurations reported to yield the best performance in \cite{Kram2022GPR}. See the supplementary material \mref{Sec.}{sec:reference_methods_details} for details. 
\acs{gptrack} models the \ac{los} component using a delay \ac{lhf} with heuristically set variance values. To ensure a fair comparison, 
we instead use Fisher information-based variance values.  
For the \acs{delaybias} method, we provide results using the setup referred to as ``GP", which learns a bias correction term using \ac{gpr} for the six parametric features suggested by the authors.\footnote{Note that the approach based on support vector machines (termed ``SVM" in \cite{WymeerschIEEE2012}) did not yield stable results for the investigated experiment. Using logarithmic features (``log-GP") also did not improve the results, while this variant is prohibitive when negative bias values occur.} After error correction according to \cite{WymeerschIEEE2012} of the distance measurements (provided by the search-forward method), we applied a particle filter with %
Fisher information-based likelihood variances in line with the other compared methods. %

\vspace{-2mm}
\subsection{Synthetic Radio Measurements (Experiments 1-5)} \label{sec:measurent_setup_synthetic}

We evaluate the proposed algorithm using synthetic radio measurements, %
where the agent moves along a trajectory with two distinct direction changes as shown in Fig.~\ref{fig:track_geometric}. The agent is observed at {$N=190$} discrete time steps $n \in \{1, \,...\, , N\}$ at a constant observation rate\footnote{A state-of-the-art \ac{uwb} ranging device, such as NXP SR040/SR150, can provide more than $10$ measurements per second.} of $\Delta T = 100\,\mathrm{ms}$, resulting in a continuous observation time of $19\,\mathrm{s}$. %
We simulate three anchors, labeled A1-A3 in Fig.~\ref{fig:track_geometric}, which are placed in close vicinity to each other. The limited directional diversity of the anchors (corresponding to a poor geometric dilution of precision (GDOP) \cite{GodHaiBlu:TIT2010}) poses a challenging setup for delay-measurement-based position estimation. %
We choose the transmitted signal to be of root-raised-cosine shape with a roll-off factor of $0.6$ and a %
$3$-dB bandwidth of $500\,\mathrm{MHz}$%
. The received baseband signal is critically sampled, i.e., $T_\text{s} = 1.25\,\mathrm{ns}$, with a total number of $N_\text{s} =81$ samples, amounting to a maximum distance $d_\text{max} = 30\,\mathrm{m}$. 
The normalized amplitudes (SNRs) of the LOS component as well as the \acp{mpc} are set to $38~\mathrm{dB}$ at an \ac{los} distance of $1~\mathrm{m}$. Unless otherwise noted, the normalized amplitudes of the individual \acp{mpc} are assumed to follow free-space path loss and are additionally attenuated by $3~\mathrm{dB}$ per reflection (e.g. $6~\mathrm{dB}$ for double-bounce reflections).
We show results of 5 synthetic experiments, referred to as Experiment 1-5. {The environment setup (i.e., walls and scatters) differs for the individual experiments as detailed in the following.} 
For all experiments investigated, the anchors are obstructed by an obstacle, labeled W5 in Fig.~\ref{fig:track_geometric}, which leads to partial and full \ac{olos} situations in the center of the track. 
Figs.~\ref{fig:single_realization}a-o provide a graphical representation of the observation space of anchor A1 for all experiments. It shows the measurements obtained by the \ac{ceda} in distance, distance bias and SNR domain together with the ground truth values, and the respective \ac{mmse} estimates of the proposed algorithm. The distance bias is obtained by subtracting the LOS distance for the true agent position from the respective \ac{pbo} distances values. For Fig.~\ref{fig:single_realization}, we used $ p_{\mathrm{de}} = 0.011 $ to visualize all \acp{pbo} available. 
\subsubsection*{Training of Reference Methods} \label{sec:training}

Training of \acs{gptrack} involves a two step procedure. In the first training step, both \acp{aednn} are trained using $6400$ unlabeled samples of baseband signal vectors that cover the entire floorplan, constituting a two dimensional grid from $-10$~m to $10$~m in $p_\text{x}$ and $p_\text{y}$ directions with $0.25$~m spacing. The \ac{aednn} used for ``anomaly detection" additionally requires LOS-only data (i.e., no OLOS situations). Thus, we created two training datasets, once deactivating the obstacle (W5). In the second training step \ac{gpr} is used to learn a feature-based measurement model using samples of baseband signal vectors (preprocessed by the feature extraction \acp{aednn}) labeled with their respective positions. We use a similar grid that also covers the entire floorplan, but with $1$~m grid spacing. The respective positions of both training datasets are depicted in Fig.~\ref{fig:track_geometric}b. 
Training of \acs{delaybias} requires only one set of baseband signal data labeled with their respective positions. Here, we instead provide results using the two different training datasets depicted in Fig.~\ref{fig:track_geometric}c. The ``reduced" dataset consists of positions where the overall received signal power remains within a moderate range. We found a low received signal to be detrimental for this method leading to strong fluctuations of the distance error for adjacent positions. Additionally, we used an ``overfitted" dataset, which contains only data located around the trajectory. 
In line with the proposed method, \acs{pdaai}, \ac{mpslam}, \acs{chslam} and \acs{cluster} require no training. 

\subsubsection*{Experiment 1 -- High Information} \label{sec:performance_synthetic_1}

In this experiment, the ground truth \ac{mpc} positions and corresponding distances are calculated based on the \ac{va} model (single-bounce and double-bounce reflections only), assuming the walls to act as large, flat surfaces. We use all walls (W1 to W4) of the floor plan shown in Fig.~\ref{fig:track_geometric}. Note that the image sources (\acp{va}) caused by walls W1-W4 are also obstructed by the obstacle (W5).
This experiment represents a ``high information" environment as several high-\ac{snr} \acp{mpc} are caused by walls arranged convexly all around the trajectory.
\pgfplotsset{
	resultStyle3/.style={mark=none ,line width=0.5pt, mycolor12, decmark={+}{0}},
	resultStyle9/.style={mark=none, line width=0.5pt, mycolor03, decmark={x}{0}}, 
	compareStyle3/.style={mark=none ,line width=0.5pt, mycolor12},
	compareStyle9/.style={mark=none, line width=0.5pt, mycolor03}, 
}
\begin{figure*}[t]	
	\centering
	\setlength{\belowcaptionskip}{0pt}	
	\tikzsetnextfilename{mse_all_va}
	\includegraphics{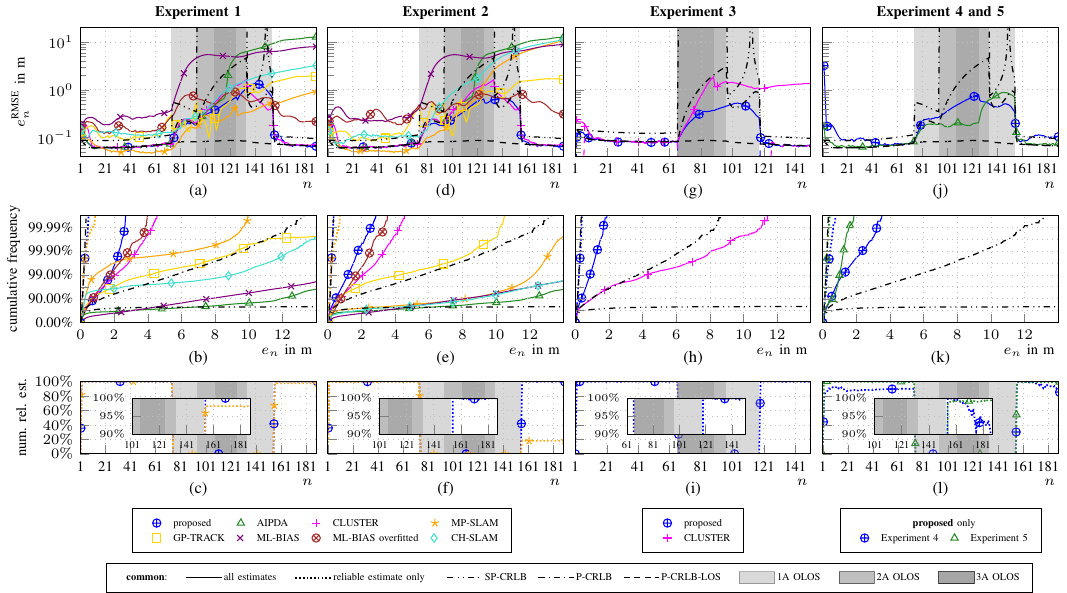}
	\vspace{-6mm}
	\caption{Performance in terms of the \ac{rmse} of the estimated agent position over time $n$ in (a),(d),(g),(j), in terms of the cumulative frequency of the estimated agent position error in inverse logarithmic scale in (b),(e),(h),(k) %
	, and the relative number of reliable estimates per time $n$ in (c),(f),(i),(l) for Experiment 1 in (a)-(c), for Experiment 2 in (d)-(f), for Experiment 3 in (g)-(i), and for Experiment 4 and 5 in (j)-(l). Note that legends vary between the experiments. The performance baselines (see Sec.~\ref{sec:benchmarks}) for Experiment 4 and 5 are identical. Thus, the results are presented in a single plot for the sake of brevity.
	}\label{fig:results_va}
	\vspace{-4.5mm}	
\end{figure*}

\subsubsection*{Experiment 2 -- Low Information} \label{sec:performance_synthetic_2}

In line with Experiment 1, the ground truth \ac{mpc} positions and corresponding distances in this experiment are calculated based on the \ac{va} model. However, for this experiment we only use walls W1 to W2 of the floor plan shown in Fig.~\ref{fig:track_geometric} (single-bounce reflections only).
This experiment represents a ``low-information" environment with few \acp{mpc} that are caused by walls whose \acp{va} take similar directions \ac{wrt} the agent as the physical anchors. This leads to %
all image sources (created by W1 and W2) being temporarily obstructed by the obstacle W5 as the agent moves along the trajectory (see Figs.~\ref{fig:single_realization}d-f). 

\subsubsection*{Experiment 3 -- Appearing Obstruction} \label{sec:performance_synthetic_3}

In line with Experiment 2, the ground truth \ac{mpc} positions and corresponding distances are calculated based on the \ac{va} model using only walls W1 and W2 of the floor plan shown in Fig.~\ref{fig:track_geometric}. However, we assume the obstacle (W5) to \textit{appear at time $n=67$}. Additionally, we have shifted wall W1 by three meters in y direction, i.e., it reaches from $[-10~13]$ to $[10~13]$. This leads to several image sources (\acp{va}) of all anchors disappearing simultaneously with the \ac{los} component, i.e., the \ac{mpc} visibility changes when the \ac{olos} situation occurs (see Figs.~\ref{fig:single_realization}g-i). 
Note that for this experiment, we shortened the trajectory, covering only the second turn during full \ac{olos}. 

\subsubsection*{Experiment 4 -- Scatter} \label{sec:performance_synthetic_4}

In this experiment, the ground truth \ac{mpc} positions and corresponding distances are calculated based on the scatters S1 and S2 of the floor plan shown in Fig.~\ref{fig:track_geometric}, which are the only source of multipath propagation. The \ac{mpc} distances are calculated as the sum of the respective scatter-anchor distances and the scatter-agent distances. %
The ground truth amplitudes are obtained assuming free-space path loss for both, the scatter-anchor distance and the scatter-agent distance \cite{GentnerTWC2016}, and lossless re-scattering. %
In this experiment, the resulting \acp{mpc} interfere strongly with the LOS component (see Figs.~\ref{fig:single_realization}j-l), when a scatter is near the path between agent and anchor. %
Thus, to obtain measurements $\bm{z}_{n,m}^{(j)}$, we used the \ac{ceda} from \cite{Hansen2014SAM} with adaptive initialization for new components\cite{ShutWanJos:CSTA2013}%
, which provides increased reliability considering the correlations between individual signal components (at the cost of increased computational complexity).  

\subsubsection*{Experiment 5 -- Ground Reflection} \label{sec:performance_synthetic_5}

In this experiment, the ground truth \ac{mpc} positions and corresponding distances are again calculated from the \ac{va} model. However, we assume multipath propagation to be caused by ground reflection assuming the agent as well as all anchors to be at a height of $1$ m \ac{wrt} the ground. For demonstration, we assume that the corresponding \acp{va} are not obstructed by the obstacle (W5). 

\vspace{-3mm} 
\subsection{Real Radio Measurements (Experiment 6)} \label{sec:performance_real}
\vspace{-1mm} 

In this experiment, we use real radio measurements collected in a laboratory hall of {NXP Semiconductors, Gratkorn, Austria}. The hall features a wide, open space and includes a demonstration car (Lancia Thema 2011), furniture, and metallic surfaces, thereby representing a typical multipath-prone industrial environment. 
An agent is assumed to move along a pseudo-random trajectory (selected out of a grid of agent positions), obtained in a static measurement setup. We selected $N=195$ measurements, assuming a sampling rate of $\Delta T = 170\,\mathrm{ms}$. The agent velocity is set to vary around a magnitude of $0.35\,\mathrm{m/s}$. This leads to a corresponding continuous observation time $33.15\,\mathrm{s}$.
At each selected position, a radio signal was transmitted from the assumed agent position, which was received by 4 anchors. Fig.~\ref{fig:car_ant} shows the measurement setup.
The agent was represented by a polystyrene build, while the anchor antennas were mounted on the demonstration car. The agent as well as the anchors were equipped with a dipole antenna with an approximately uniform radiation pattern in the azimuth plane and zeros in the floor and ceiling directions. The radio signal was recorded by an M-sequence correlative channel sounder with frequency range $3-10\,\mathrm{GHz}$. Within the measured band, the actual signal band was selected by a filter with raised-cosine impulse response $s(t)$, with a roll-off factor of $0.6$, a two-sided 3-dB bandwidth of $499.2\,\mathrm{MHz}$ and a center frequency of $7.9872\,\mathrm{GHz}$, corresponding to channel 9 of IEEE 802.15.4a. We used $N_\text{s} =81$ samples, amounting to a $d_\text{max}$ slightly below $30\,\mathrm{m}$. 
We created two full \ac{olos} situations at $n \in [80 , 92]$ and $n \in [159 , 170]$ using an obstacle consisting of a metal plate covered with attenuators, as depicted in Fig.~\ref{fig:absorber}. 
A floor plan showing the track, the environment (i.e, the car, other reflecting objects and walls), the antenna positions, and the \ac{olos} conditions \ac{wrt} all antennas is given in Fig.~\ref{fig:measurement_setup_nxp}. 
The metal surface of the car strongly reflected the radio signal, leading to a radiation pattern of $270^\circ$ for A1 and A2 and $180^\circ$ for A3 and A4. Thus, during large parts of the trajectory, the \ac{los} of 2 or 3 out of 4 anchors was not available. Moreover, the pulse reflected by the car surface strongly interfered with the \ac{los} pulse, leading to significant fluctuations of the amplitudes. Also, this leads to the channel estimator being prone to produce a high-\ac{snr} component just after the \ac{los} component. %
\begin{figure}[t]
	
	\centering
	\setlength{\belowcaptionskip}{1pt}	
	\captionsetup[subfloat]{farskip=8pt,captionskip=5pt} 
	\subfloat[\label{fig:car_ant}]{\includegraphics[height=0.20\textwidth]{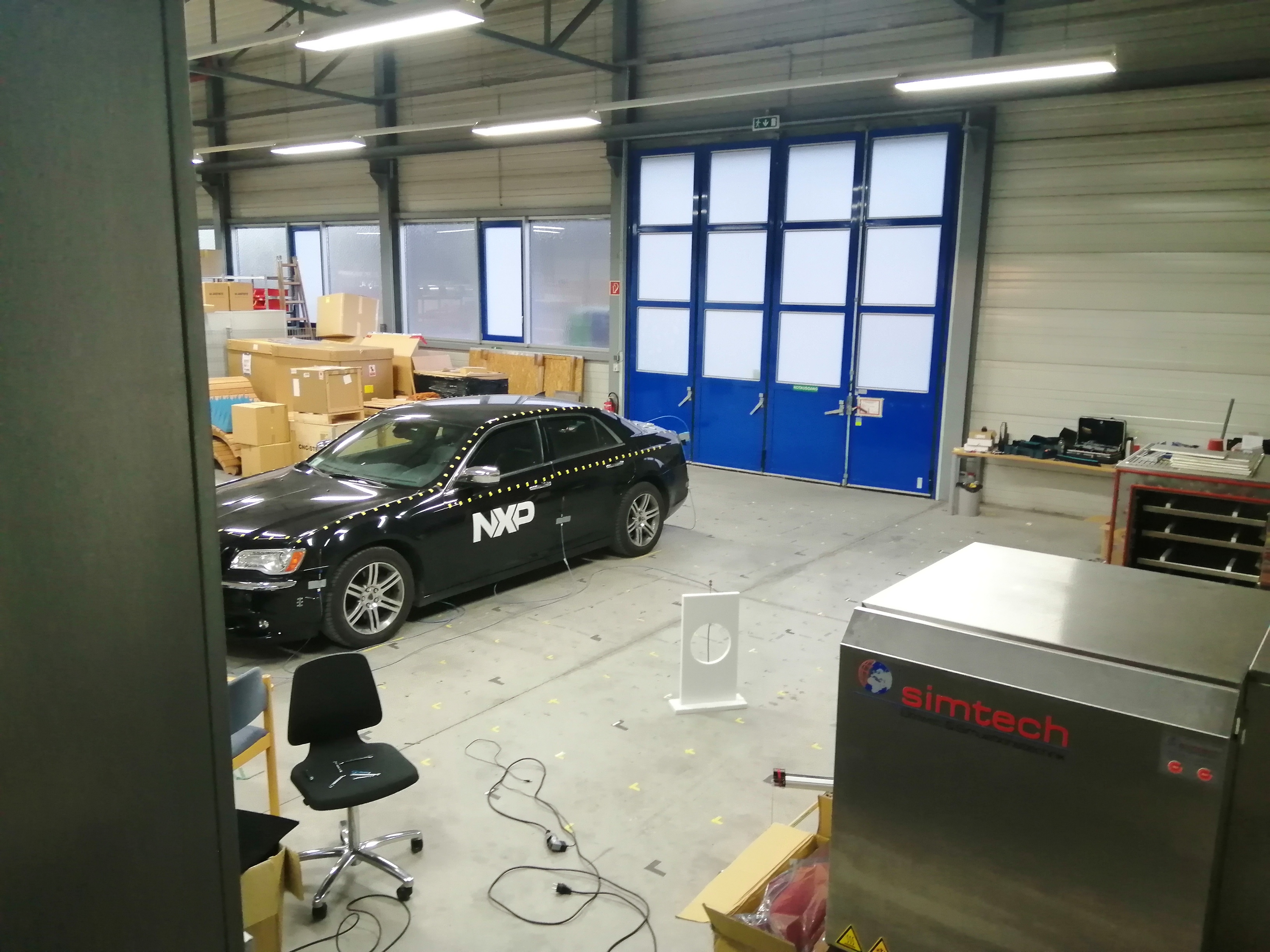}}\vspace{2mm}
	\subfloat[\label{fig:absorber}]{\includegraphics[height=0.20\textwidth]{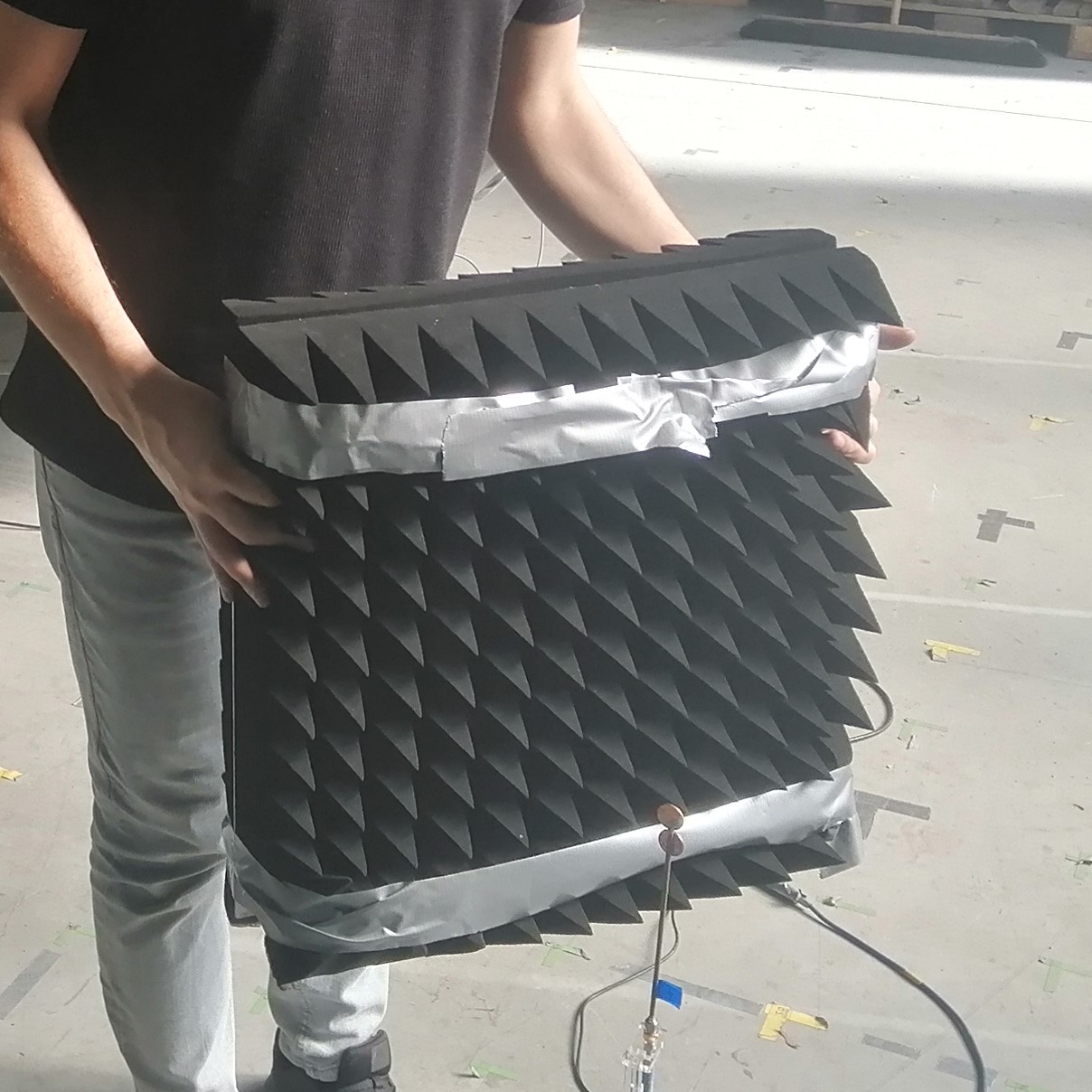}}
	\vspace{-3mm}
	\setlength{\figurewidth}{0.28\textwidth}
	\setlength{\figureheight}{0.28\textwidth}
	\tikzsetnextfilename{track_nxp}
	\captionsetup[subfloat]{captionskip=-3pt} 
	\subfloat[\label{fig:measurement_setup_nxp}]{\includegraphics{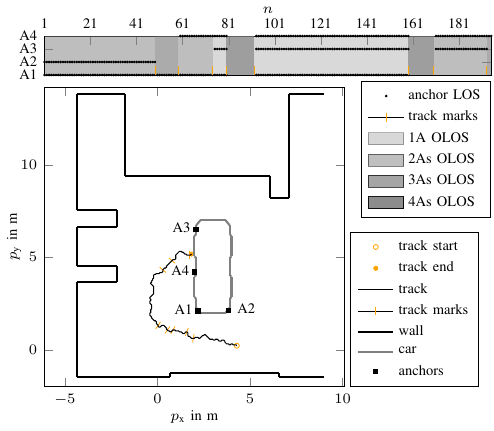}}
	\caption{Measurement setup for Experiment 6 using real radio-signals. We show pictures of (a) the overall scenario and (b) the OLOS setup used, as well as (c) the abstracted floorplan and trajectory.
	}\label{fig:measurement_setup_nxp_all}
	\vspace{-5mm}
\end{figure}
As only two antennas (A1 and A2) are visible at the track starting point, the position estimate obtained by trilateration is ambiguous. In the scenario presented, the relative antenna position \ac{wrt} the car can be assumed to be known. Thus, for this experiment, we used the antenna pattern as prior information for initialization of the position state. %
For the numerical evaluation presented, we added \ac{awgn} to the real radio signal obtained. %
We set $\norm{\bar{\V{r}}_{\mathrm{raw}}^{(j)}}{2}/\sigma^{(j)\s 2} = 20 \text{dB}$, where $\norm{\bar{\V{r}}_{\mathrm{raw}}^{(j)}}{2}$ is the average energy of the real measured signal per anchor $j$.
Figs.~\ref{fig:single_realization}q-r show the observation space of anchor A1 for this experiment. 
\vspace{-4mm} 
\subsection{Joint Performance Evaluation} \label{sec:performance_analysis}
\vspace{-1mm} 
We provide the performance of the proposed algorithm and all applicable\footnote{Experiment 3 highlights the advantages of the proposed method \ac{wrt} the \acs{cluster} method, while Experiment 4 and 5 demonstrate the applicability of the proposed method for alternative sources of multipath. Thus, we do not compare to other methods as this would provide no additional insights.} 
reference methods for all investigated experiments in Figs.~\ref{fig:results_va}, and \ref{fig:results_nxp}. 
\subsubsection{Performance Metrics and Baseline} \label{sec:benchmarks}
For each of the experiments investigated, we analyze the performance %
in terms of both, the \ac{rmse} of the estimated agent position over time $n$ given as  $e_{n}^{\text{RMSE}}~=~\sqrt{\E{\norm{\hat{\bm{p}}^{\text{MMSE}}_n -\bm{p}_n}{2}}}$ and the cumulative frequency of the magnitude error of the estimated agent position $e_{n} \triangleq  \norm{\hat{\bm{p}}^{\text{MMSE}}_n -\bm{p}_n}{}$, and are evaluated using a numerical simulation with 500 realizations.

As a performance benchmark, we provide the \ac{crlb} on the position error variance considering \textit{all visible \ac{los} measurements} of all $J$ anchors a single time step $n$, which we refer to as the snapshot-based positioning CRLB (SP-CRLB). 
Using the results from \cite{Jourdan2008,ShenTIT2010,WitrisalJWCOML2016}, we get the Fisher information matrix 
\vspace{-1mm}
\begin{equation} \vspace{-1mm} \label{eq:spcrlb}
	\bm{J}_{\bm{p} \s \text{SP}\s n} = \frac{8 \pi^2 \beta_\text{bw}^2}{c^2} \sum_{j=1}^{J}  {u}_{1,n}^{(j)\s 2} \V{D}_{\text{r} \s n}^{(j)} 1_{\mathbb{V}_n^{(j)}}
\end{equation}
where $\V{D}_{\text{r} \s n}^{(j)} = [\cos({\phi}_n^{(j)})\, \sin({\phi}_n^{(j)})]\, [\cos({\phi}_n^{(j)})\,  \sin({\phi}_n^{(j)})]^\text{T} $ is the ranging direction matrix \cite{ShenTIT2010}, with the (true) angle of arrival ${\phi}_n^{(j)} = \mathrm{atan2}(p_{\text{Ax}}^{(j)}-{p}_{\text{x}\s n}, p_{\text{Ay}}^{(j)}-{p}_{\text{y}\s n} )$, and $\mathbb{V}_n^{(j)}$ being the set containing the time indices $n$ of all times where the \ac{los} component is visible. 

Furthermore, we provide the corresponding \acf{pcrlb} \cite{Tichavsky1998} that additionally considers the information provided by the state transition model of the \textit{agent state} $\RV{x}_n$. %
Following the derivation in \cite[Sec. III]{Tichavsky1998}, we get the Fisher information matrix 
\begin{equation} \label{eq:pcrlb}
	\bm{J}_{\bm{p}\s \text{P} \s n} = ( \bm{A}_2 \, \bm{J}_{\bm{p}\s \text{P} \s n\minus 1}^{-1} \, \bm{A}_2^\text{T} \, + \, {\sigma_{\text{a}}^2} \, \bm{B}_2 \, \bm{B}_2^\text{T} )^{-1} + \bm{J}_{\bm{p}\s \text{S} \s n}
\end{equation}
which is a recursive equation corresponding to the covariance update equations of the Kalman filter \cite{ArulampalamTSP2002}. %
Since we initialize the {agent state} $\RV{x}_n$ using an initial measurement $\bm{z}_0$ (see {Sec.}~\ref{sec:init_states} for details), we accordingly calculate $\bm{J}_{\bm{p}\s \text{P} \s 0}$ using \eqref{eq:spcrlb} with the corresponding true values ${u}_{0}^{(j)\s 2}$ and ${\bm{p}}_0$. 
Note that 
$
	\sqrt{\text{tr}\{  \bm{J}_{\bm{p}\s \text{S} \s  n}^{-1}  \}} \geq \sqrt{\text{tr}\{  \bm{J}_{\bm{p}\s \text{P} \s n}^{-1}  \}}  %
$ %
since the \ac{pcrlb} uses additional information. 

\begin{figure}[t]	
	\centering	
	\setlength{\belowcaptionskip}{0pt}
	\tikzsetnextfilename{mse_all_nxp}
	\includegraphics{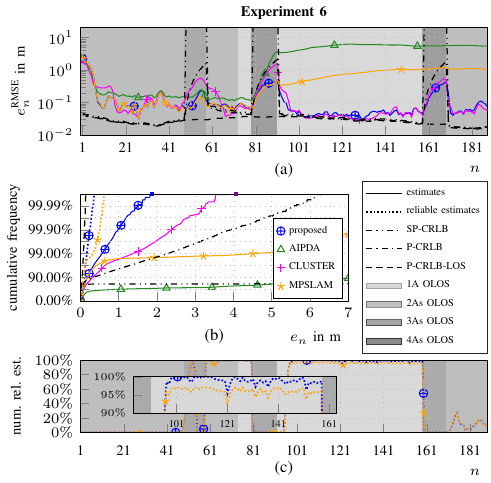}
	\vspace{-3mm}
	\caption{Performance for Experiment 6 in terms of the \ac{rmse} of the estimated agent position over time $n$ in (a), the cumulative frequency of the agent position error in inverse logarithmic scale in (b) %
		, and the relative number of reliable estimates per time $n$ in (c).}
	\label{fig:results_nxp}
	\vspace{-5mm}
\end{figure}

We note that \eqref{eq:spcrlb} does {not model} the additional information provided by coupling the \acp{mpc} with the \ac{los} object via the distance biases $\rv{b}_{k,n}^{(j)}$ in \eqref{eq:distance_fun}. This allows the RMSE of the proposed algorithm to \textit{fall below} the provided SP-CRLB and P-CRLB, demonstrating the additional information leveraged using the proposed \ac{mpc}-aided model. 
	
However, in contrast to mapping approaches \cite{LeitingerTWC2019,GentnerTWC2016,KimTWC2020}, which can facilitate multipath information via estimated map features (\acp{va}), the proposed method just allows to mitigate the distance bias between MPC-related distance measurements and the LOS component distance. Thus, a tight bound on the proposed method can be obtained by assuming the LOS component to be available at all times $n$, i.e., by setting $1_{\mathbb{V}_n^{(j)}}\vspace{0.23mm} \triangleq 1$ in \eqref{eq:spcrlb}. We refer to the corresponding P-CRLB as ``P-CRLB-LOS'' in the following performance analysis.

\subsubsection{Overall performance}  \label{sec:overall_performance}
Figs.~\ref{fig:results_va}, and \ref{fig:results_nxp} show in solid lines (termed ``all estimates") the performance considering the estimates of every realization of the performed numerical simulations of at all times $n \in \{1, \,...\, , N\}$ (c.f. Sec.~\ref{sec:performance_reliable}). 
Analyzing the performance of the proposed method across all experiments, it can not only utilize the position information contained in \acp{mpc} caused by flat surfaces (\ac{va} model), as shown in Experiment 1, 2, and 3, but it can also leverage \acp{mpc} caused by scatters, as demonstrated in Experiment 4, and three dimensional structures, such as the ground reflections in Experiment 5 (that lead to distance biases that evolve approximately linear over time $n$). 
Its \ac{rmse} attains the \ac{pcrlb} in \ac{los} conditions and outperforms it during full \ac{olos} situations due to the additional information provided by the \acp{mpc}. 
However, we observe a slightly reduced performance for Experiment 4, which is due to the small distance between \ac{los} and \acp{mpc} \ac{wrt} the bandwidth of the simulated \ac{uwb} system, in particular during initialization (see Fig.~\ref{fig:single_realization}j). 
While \acs{cluster} also manages to maintain the track in every single realization, it shows reduced performance in Experiment 1, 2, and 6 during the \ac{olos} situation, which is due to the approximate nature and resulting reduced curvature of its measurement model. Furthermore, in Experiment 3 the method diverged in about $50\%$ of the simulation runs. This is due to the simultaneous disappearance of the LOS component (i.e, the start of the \ac{olos} situation) and the first \ac{mpc} as illustrated in Fig.~\ref{fig:single_realization}g, which significantly alters the shape of the observed ``multipath cluster" and, thus, violates the  system model of this method. 
\acs{mpslam} can achieve a significantly reduced \ac{rmse} in high-information scenarios with many reflecting surfaces. This is shown in Experiment 1, where \acs{mpslam} clearly outperforms the proposed algorithm in terms of accuracy (see e.g. Fig.~\ref{fig:results_va}b). Examining Fig.~\ref{fig:results_va}a, we even observe the \ac{rmse} of \acs{mpslam}  to fall below the \ac{pcrlb} in the first part of the \ac{olos} situation. This is possible due to the additional position information provided by the \acp{mpc} measurements (associated to the jointly inferred \acp{va}) as investigated in \cite{LeitingerJSAC2015, Mendrzik2019}. 
However, as visible in Figs.~\ref{fig:results_va}a and b, in 6 simulation runs \acs{mpslam} loses the track after the full \ac{olos} situation%
\footnote{The full OLOS situations ends at time $n=56$, yielding approx. $6\cdot (190-155)/(190\cdot 500) = 0.3\%$ of outliers with errors larger than $3\;\mathrm{m}$ in Fig.~\ref{fig:results_va}b.}, while the proposed algorithm keeps the track for every realization. A possible explanation is the reduced number of dimensions of the \ac{pbo} model (1-D distance bias) \ac{wrt} \ac{mpslam} (2-D \ac{va} position) and the associated reduced chance of all particles converging to a wrong mode. 
\acs{chslam} shows a significantly higher \ac{rmse} in the first part of the trajectory. A possible explanation for this behavior is the insufficient number of particles\footnote{For an even larger number of \acs{va} particles (see  Section~\ref{sec:reference_methods} for details about the configuration used), the runtime and memory requirements of \acs{chslam} become prohibitive.} representing the \ac{va} state in connection with the high number of \acp{va} visible in this Experiment. Also it looses the track in about $8\%$ of simulation runs after the full \ac{olos} situation. 
In the low-information scenario investigated in Experiment 2, \ac{mpslam} and \acs{chslam} diverge in about $50\%$ and $80\%$ of the simulation runs, respectively. This is due to the geometric ambiguity of the scenario %
leading to the estimated agent distribution collapsing to an ambiguous mode.\footnote{Geometric ambiguity occurs in mapping based on delay measurements, since the non-linear relationship between delay and position can lead to multiple modes.} Ambiguities can be resolved over time when there is sufficient directional change in the agent movement \cite{KrekovicTSP2020}. However, in Experiment 2 this is not possible despite the significant directional change at point [0.5,-0.5] before the full \ac{olos} situation (see Fig.~\ref{fig:track_geometric}). This is because most of the \acp{va} are obstructed by the obstacle (W5) \textit{after} the directional change at [0.5,-0.5] leading to \ac{mpslam} discarding these \acp{va}, since they do not cause measurements for many times $n$ and thus there probability of existence approaches zero. When a \ac{va} becomes visible again, a new \ac{va} is initialized, and the information of the directional change is lost. 
The \acs{pdaai} method does not facilitate multipath at all, acting as a baseline that emphasizes the challenging nature of the investigated experiments. While it shows excellent performance in \ac{los} conditions,  it follows ambiguous paths (i.e., it loses the track) after the full \ac{olos} situation for many realizations, leading to a significantly reduced performance. 
While \acs{gptrack} significantly outperforms the \acs{pdaai} method, which demonstrates the effectiveness of the learned multipath representation, it shows reduced overall accuracy by not attaining the \ac{pcrlb} in \ac{los} conditions, as well as reduced robustness by losing the track in many realizations. The reduced accuracy of \acs{gptrack} is caused by the insufficient precision of the learned geometric imprint\footnote{Note that we chose the data grid to be at a $1\,\mathrm{m}$ spacing as otherwise the execution time of the method would be prohibitive (see also Sec.~\ref{sec:execution_time}).} that interferes with the LOS model. Possible explanations for the reduced robustness are both, the significant number of false alarms of the anomaly detection method and the reduced robustness of the conventional particle filter \ac{wrt} \ac{pda}-type filters \cite{BarShalom1995} (see also the supplementary material {\mref{Sec.}{sec:message_interpretation}}). 
Finally, the \acs{delaybias} method performs robustly, not showing any lost tracks when using the ``overfitted" training dataset, which confirms the validity of our implementation. Yet, it still shows a low overall accuracy not attaining the \ac{pcrlb} in \ac{los} condition. In contrast, for the ``subsampled-reduced" dataset, the method failed to produce consistent estimates, leading to  divergence of the subsequently applied particle filter. This is due to the fact, that the bias representations are learned per anchor, i.e., the features that lead to the bias estimate cannot offer angular information. Furthermore, jumps in the estimated delay can lead to instability of the method as we can only apply the particle filter to the corrected estimate, which is in contrast to the ``soft" information fusion offered by \ac{pda}-type methods including the proposed algorithm.

The results using real radio measurements in Experiment 6 confirm the validity of the numerical results presented. We observe a consistent performance gain of the proposed algorithm \ac{wrt} the reference methods.
However, different to Experiment 1 to 5, in Experiment 6 all presented algorithms fail to reach the \ac{pcrlb} over parts of the track, as can be observed from Fig.~\ref{fig:results_nxp}. The exact consistency in progression of the \ac{rmse} curves suggests unmodeled effects (e.g. diffraction at the vehicle body) as well as inaccuracies in the reference as a probable reason. 

\subsubsection{Identification of Unreliable Measurements} \label{sec:performance_reliable}
For the proposed algorithm and the \ac{mpslam} method, Figs.~\ref{fig:results_va} and \ref{fig:results_nxp} show results considering estimates identified as reliable according to \eqref{eq:reliability}, termed ``reliable estimates only" (dotted lines). %
 Additionally, Figs.~\ref{fig:results_va} and \ref{fig:results_nxp} show the relative number of reliable estimates over time $n$.
Analyzing the performance of both methods across all experiments investigated, they consistently identify outliers and, thus, they almost attain the P-CRLB-LOS. The \ac{mpslam} method falls even slightly below the P-CRLB-LOS in Experiment 1 and 2, due to the additional information provided by the \acp{mpc}. However, while the number of reliable measurements for \ac{mpslam} is reduced after the full \ac{olos} situation, especially for Experiment 2, the proposed method provides $100 \%$ reliable estimates with all runs converging after the full OLOS situation. 
Consistent with the observations from Sec.~\ref{sec:overall_performance}, we notice a slightly reduced number of reliable estimates of the proposed method for Experiment 4. However, it still identifies unreliable estimates, only slightly exceeding the P-CRLB-LOS.

\section{Runtime} \label{sec:execution_time}
Table~\ref{tbl:execution_times} shows the average runtime of the proposed algorithm and compares it to the runtime of all reference methods (see Table~\ref{tbl:reference_methods}). All runtimes are estimated using Matlab implementations executed on an AMD Ryzen Threadripper 1900X 8-Core Processor with up to $4\,\text{GHz}$ for all scenarios investigated. We also show the average number of measurements (over all anchors and time steps) ${M}_\text{mean}$ and the number of anchors $J$, which together with then number of particles used (see Sec.~\ref{sec:reference_methods}) determine the algorithm complexity per time step. 
Note that we used the ``subsampled-reduced" dataset (see Fig.~\ref{fig:track_geometric}) to determine the runtime of the \acs{delaybias} method. 
The runtime of the proposed method has the same order of magnitude as that of \ac{mpslam}, slightly outperforming it due to the higher number of objects (\acp{pbo} vs. potential \acp{va}) initialized by \ac{mpslam}. 
The runtime of \acs{delaybias} is the lowest, as it only requires one \ac{gpr}-based bias calculation per measurement. Therefore, the time-limiting component is the particle filter. \acs{pdaai} and \acs{cluster} offer a significantly lower runtime than the proposed method due to the lower complexity of the inference model as well as the lower number of particles required for inference. Finally, \acs{gptrack} and \acs{chslam} show the highest runtime. The particle-based representation in \acs{chslam} is very computationally demanding, since its posterior representation explicitly considers the dependency of map features on the agent state. \acs{gptrack} requires to evaluate the \ac{gpr}-based mapping once per particle, anchor and feature. Its computational complexity is given as $\mathcal{O}(N_\text{T}^{\s 2} J I F)$ \cite{Rasmussen2006GP}, where $N_\text{T}$ is the number of training samples and $F$ is the size of the feature space of the \ac{aednn}. $J$ and $I$ are respectively, the 
Note that we did not include the runtime of the pre-processing algorithms (\ac{ceda} and feature extraction \acp{aednn} of \acs{gptrack}) in this comparison.

\begin{table}[thb]
	\renewcommand{\baselinestretch}{1}\small\normalsize
	\setlength{\tabcolsep}{3pt} %
	\renewcommand{\arraystretch}{1} %
	\centering
	\footnotesize
	\caption{Algorithm runtimes and characteristic values of all investigated scenarios.}\label{tbl:execution_times}
	\begin{tabular}{ | r c c c c c c |} 
		\toprule
		\textbf{method}  &  \textbf{Ex. 1} &   \textbf{Ex. 2} &   \textbf{Ex. 3} &   \textbf{Ex. 4}  &   \textbf{Ex. 5}  &   \textbf{Ex. 6}  \\ %
		\midrule
		\textbf{proposed}  &$ 460\,\mathrm{ms}$  & $ 244\,\mathrm{ms}$ & $231\,\mathrm{ms}$ & $224\,\mathrm{ms}$ & $170\,\mathrm{ms}$ & $312\,\mathrm{ms}$ \\
		\textbf{\acs{pdaai}} & $ 32\,\mathrm{ms}$  & $ 28\,\mathrm{ms}$ & $-$ & $-$ & $-$ & $27\,\mathrm{ms}$  \\
		\textbf{\acs{mpslam}}  & $750\,\mathrm{ms}$ 	& $283\,\mathrm{ms}$ &   $-$ & $-$ & $-$ & $340\,\mathrm{ms}$  \\
		\textbf{\acs{cluster}} & $ 54\,\mathrm{ms}$  & $ 42\,\mathrm{ms}$ & $ 41\,\mathrm{ms}$ & $-$ & $-$ & $46\,\mathrm{ms}$  \\
		\textbf{\acs{delaybias}} & $ 6\,\mathrm{ms}$  & $ 5\,\mathrm{ms}$ & $-$  & $-$ & $-$ & $-$ \\
		\textbf{\acs{gptrack}} & $ 1340\,\mathrm{ms}$  & $ 1330\,\mathrm{ms}$ &  $-$ & $-$ & $-$ & $-$\\
		\textbf{\acs{chslam}} & $ 11.2\,\mathrm{s}$  & $ 3.5\,\mathrm{s}$ &  $-$ & $-$ & $-$ & $-$\\
		\midrule 
		${M}_\text{mean}\times J$ & $8.6 \times 3 $ & $4.2 \times 3$ & $4.1 \times 3$ & $4 \times 3$  & $3.7 \times 4$  & $4.3 \times 4$  \\
		\bottomrule
	\end{tabular}
	\vspace{-2.5mm}
\end{table}

\ifthenelse{1=0}{

\section{\rd{Snapshot-Based \acl{crlb} (SP-CRLB) and Posterior \acs{crlb} (P-CRLB)}}\label{sec:app_crlb}
\acused{pcrlb}
\rd{Here, we provide the expressions for what we refer to as the ``snapshot-based positioning CRLB (SP-CRLB)" and the corresponding posterior CRLB (P-CRLB) \cite{Tichavsky1998}, which are used as a performance benchmark in the main text \mref{Sec.}{sec:results}). They provide lower bounds on the \ac{rmse} of the position estimate, given as \cite{ShenTIT2010}
\begin{equation}
	\sqrt{\text{tr}\{  \bm{J}_{\bm{p}\s \text{S} \s  n}^{-1}  \}} \geq \sqrt{\text{tr}\{  \bm{J}_{\bm{p}\s \text{P} \s n}^{-1}  \}}  %
\end{equation}
with $\bm{J}_{\bm{p}\s \text{S} \s  n}$ and  $\bm{J}_{\bm{p}\s \text{P} \s  n}$ being the respective Fisher information matrices.
In particular, the SP-CRLB considers the information contained in the signal waveforms recorded by all $J$ anchors at a \textit{single} time step $n$. %
We use the results from \cite[Eq. 14]{WitrisalJWCOML2016} as the model used fits our signal model in \meqref{eq:signal_model_sampled_stochastic}. We get
\vspace{-2mm}
\begin{equation} \vspace{-1mm} \label{eq:spcrlb}
	\bm{J}_{\bm{p} \s \text{S}\s n} = \frac{8 \pi^2 \beta_\text{bw}^2}{c^2} \sum_{j=1}^{J}  \tilde{u}_{n}^{(j)\s 2} \V{D}_{\text{r} \s n}^{(j)} 1_{\mathbb{V}_n^{(j)}}
\end{equation}
where $\V{D}_{\text{r} \s n}^{(j)} = [\cos(\tilde{\phi}_n^{(j)})\, \sin(\tilde{\phi}_n^{(j)})]\, [\cos(\tilde{\phi}_n^{(j)})\,  \sin(\tilde{\phi}_n^{(j)})]^\text{T} $ is the ranging direction matrix \cite{ShenTIT2010}, with the (true) angle of arrival $\tilde{\phi}_n^{(j)} = \mathrm{atan2}(p_{\text{Ax}}^{(j)}-\tilde{p}_{\text{x}\s n}, p_{\text{Ay}}^{(j)}-\tilde{p}_{\text{y}\s n} )$, and $\mathbb{V}_n^{(j)}$ is the set containing all time step indices $n$ %
where the \ac{los} component is visible. $c$, $\beta_\text{bw}$, $\tilde{u}_{n}^{(j)}$, $p_{\text{Ax}}^{(j)}$, $\tilde{p}_{\text{x}\s n}$, $p_{\text{Ay}}^{(j)}$, $\tilde{p}_{\text{y}\s n}$ and $1_{\mathbb{V}_n^{(j)}}$ are defined in accordance to the main text (see \mref{Sec.}{sec:signal_model} and \mref{Sec.}{sec:system_model}). 
The P-CRLB additionally considers the information provided by the state transition model of the {agent state} $\RV{x}_n$. Following \cite[Sec. III]{Tichavsky1998}, we get
\begin{equation} \label{eq:pcrlb}
	\bm{J}_{\bm{p}\s \text{P} \s n} = ( \bm{A} \, \bm{J}_{\bm{p}\s \text{P} \s n\minus 1}^{-1} \, \bm{A}^\text{T} \, + \, {\sigma_{\text{a}}^2} \, \bm{B} \, \bm{B}^\text{T} )^{-1} + \bm{J}_{\bm{p}\s \text{S} \s n}
\end{equation}
which is a recursive equation corresponding to the covariance update equations of the Kalman filter \cite{ArulampalamTSP2002}. $\bm{A}$, $\bm{B}$ and ${\sigma_{\text{a}}}$ are defined in accordance to the main text (see \mref{Sec.}{sec:simulation_model}). Since we initialize the {agent state} $\RV{x}_n$ using an initial measurement $\bm{z}_0$ (see \mref{Sec.}{sec:init}), we accordingly calculate $\bm{J}_{\bm{p}\s \text{P} \s 0}$ using \eqref{eq:spcrlb} with the corresponding true values $\tilde{u}_{0}^{(j)\s 2}$ and $\tilde{\bm{p}}_0$.}

\rd{While \eqref{eq:spcrlb} considers the (inevitable asymptotic) loss of SNR related to the stochastic process  $\rv{\nu}_{\text{D}\s n}^{(j)}(\tau)$ in \meqref{eq:signal_model_sampled_stochastic}, it does {not model} the additional information provided by coupling the \acp{mpc} with the \ac{los} object via the NLOS bias $\tilde{b}_n^{(j)}$ in \meqref{eq:delta_fun}. This allows the RMSE of the proposed algorithm to \textit{fall below} the provided CRLB, demonstrating the additional information leveraged using the proposed NLOS model. However, in contrast to mapping approaches \cite{LeitingerTWC2019,GentnerTWC2016,KimTWC2020}, which can facilitate multipath information via estimated map features (virtual anchors \cite{PedersenJTAP2018,Meissner2015Diss}), our model just allows to mitigate the NLOS bias between MPC-related distance measurements and the LOS component distance. Thus, a strict lower bound can be obtained by assuming the LOS component to be available at all times $n$, i.e., setting $1_{\mathbb{V}_n^{(j)}}\vspace{0.23mm} \triangleq 1$ in \eqref{eq:spcrlb}. We refer to the corresponding P-CRLB as P-CRLB-LOS in the main text.}
}{}

\vspace{-1mm} 
\section{Conclusion}\label{sec:conclusion}
\vspace{-1mm} 

\acresetall
We presented a particle-based \ac{spa} that sequentially estimates the position of a mobile agent %
using range and amplitude measurements provided by a snapshot-based \acf{ceda}.
We analyzed the performance of the proposed algorithm using numerically simulated radio signals and real radio measurements in different propagation environments, comprising flat surfaces (e.g., walls and floor) and scatters. We showed that the additional information provided by the \acp{pbo} can support the estimation of the agent position. Furthermore, we demonstrated the capability of the proposed method to identify unreliable measurements and, thus, to identify lost tracks. 
Our algorithm outperforms state-of-the-art methods for \ac{mpc}-aided robust positioning and tracking and consistently attains the \ac{pcrlb} in partial \acf{olos} situations. 
While \acf{mpslam} can naturally provide high-accuracy results in environments with flat surfaces that offer high geometric diversity, we have shown that the proposed method consistently provides a lower number of lost tracks. 
Possible directions for future research include extending the model to diffuse \ac{mpc} that lead to multiple measurements (e.g. caused by rough walls) by using data association with extended objects \cite{MeyWilJ21}.

 \acrodef{mimo}[MIMO]{multiple input multiple output}
 \acrodef{awgn}[AWGN]{additive white Gaussian noise}
 \acrodef{bw}[BW]{bandwidth}
 \acrodef{blt}[BLT]{bluetooth}
 \acrodef{cdf}[CDF]{cumulative distribution function}
 \acrodef{crlb}[CRLB]{Cram\'er-Rao lower bound}
 \acrodef{dmc}[DMC]{dense multipath component}
 \acrodef{dut}[DUT]{device under test}
 \acrodef{eirp}[EIRP]{equivalent isotropic radiated power}
 \acrodefplural{esl}[ESLs]{electronic shelf labels} 
 \acrodef{los}[LOS]{line-of-sight}
 \acrodef{mf}[MF]{matched filter}
 \acrodef{ml}[ML]{maximum likelihood}
 \acrodef{mpc}[MPC]{multipath component}
 \acrodef{nlos}[NLOS]{non-line-of-sight}
 \acrodef{pcb}[PCB]{printed circuit board}
 \acrodef{pdf}[PDF]{probability density function}
 \acrodef{reb}[REB]{ranging error bound}
 \acrodef{rss}[RSS]{received signal strength}
 \acrodef{smc}[SMC]{specular multipath component}
 \acrodef{snr}[SNR]{signal-to-noise-ratio}
 \acrodef{sinr}[SINR]{signal-to-interference-plus-noise-ratio}
 \acrodef{tdoa}[TDOA]{time difference of arrival}
 \acrodef{tka}[TKA]{trusted keyless access}
 \acrodef{toa}[TOA]{time-of-arrival}
 \acrodef{aoa}[AOA]{angle-of-arrival}
 \acrodef{uwb}[UWB]{ultra wide band}
 \acrodef{mie}[MIE]{method of interval estimation}
 \acrodef{mc}[MC]{Monte Carlo}
 \acrodef{mse}[MSE]{mean squared error}
 \acrodef{ci}[CI]{confidence interval}
 \acrodef{cl}[CL]{confidence level}
 \acrodef{pdp}[PDP]{power delay profile}
 \acrodef{dps}[DPS]{delay power spectrum}
 \acrodef{dm}[DM]{dense multipath}
 \acrodef{nlike}[NLIKE]{normalized likelihood}
 \acrodef{zzb}[ZZB]{Ziv-Zakai bound}
 \acrodef{ut}[UT]{unscented transform}
 \acrodef{glrt}[GLRT]{generalized likelihood ratio test}
 \acrodef{mse}[MSE]{mean squared error}
 \acrodef{rmse}[RMSE]{root mean squared error}
 \acrodef{nnlike}[NNLIKE]{normalized noise-free likelihood}
 \acrodef{stdv}[STDV]{standard deviation}
 \acrodef{rv}[RV]{random variable}
 \acrodef{bp}[BP]{belief propagation}
 \acrodef{pda}[PDA]{probabilistic data association}
 \acrodef{mp}[MP]{multipath}
 \acrodef{pmf}[PMF]{probability mass function}
 \acrodef{pdaf}[PDAF]{probabilistic data association filter}
 \acrodef{pdaai}[AIPDA]{amplitude-information \ac{pda}}
 \acrodef{olos}[OLOS]{obstructed line-of-sight}
 \acrodef{spa}[SPA]{sum-product algorithm}
 \acrodef{mmse}[MMSE]{minimum mean-square error}
 \acrodef{lhf}[LHF]{likelihood function}
 \acrodef{fa}[FA]{false alarm}
 \acrodef{ceda}[CEDA]{channel estimation and detection algorithm} 
 \acrodef{pcrlb}[P-CRLB]{posterior Cram\'er-Rao lower bound}
 \acrodef{mpslam}[MP-SLAM]{multipath-based SLAM}
 \acrodef{va}[VA]{virtual anchor}
 \acrodef{dnr}[DNR]{dense-to-noise ratio}
 \acrodef{pbo}[PBO]{potential bias object}
 \acrodef{npbo}[NPBO]{new \ac{pbo}}
 \acrodef{lpbo}[LPBO]{legacy \ac{pbo}}
 \acrodef{aednn}[AE-DNN]{autoencoder deep neural network}   
 \acrodef{gpr}[GPR]{Gaussian process regression}  
 \acrodef{cluster}[CLUSTER]{{\color{red}error}}  
 \acrodef{delaybias}[ML-BIAS]{{\color{red}error}}  
 \acrodef{gptrack}[GP-TRACK]{{\color{red}error}}  
 \acrodef{chslam}[CH-SLAM]{{\color{red}error}}  
 \acrodef{wrt}[w.r.t.]{with respect to} 

\renewcommand{\baselinestretch}{0.93}\small\normalsize %

\vspace{-2mm} 
\bibliographystyle{IEEEtran}
\bibliography{IEEEabrv, References, TempRefs}

\begin{thebibliography}{10}
\providecommand{\url}[1]{#1}
\csname url@samestyle\endcsname
\providecommand{\newblock}{\relax}
\providecommand{\bibinfo}[2]{#2}
\providecommand{\BIBentrySTDinterwordspacing}{\spaceskip=0pt\relax}
\providecommand{\BIBentryALTinterwordstretchfactor}{4}
\providecommand{\BIBentryALTinterwordspacing}{\spaceskip=\fontdimen2\font plus
\BIBentryALTinterwordstretchfactor\fontdimen3\font minus
  \fontdimen4\font\relax}
\providecommand{\BIBforeignlanguage}[2]{{%
\expandafter\ifx\csname l@#1\endcsname\relax
\typeout{** WARNING: IEEEtran.bst: No hyphenation pattern has been}%
\typeout{** loaded for the language `#1'. Using the pattern for}%
\typeout{** the default language instead.}%
\else
\language=\csname l@#1\endcsname
\fi
#2}}
\providecommand{\BIBdecl}{\relax}
\BIBdecl

\bibitem{VenusAsilomar2022}
A.~Venus, E.~Leitinger, S.~Tertinek, F.~Meyer, and K.~Witrisal, ``Graph-based
  simultaneous localization and bias tracking for robust positioning in
  obstructed los situations,'' in \emph{Proc. Asilomar-22}, 2022, pp. 1--8.

\bibitem{WitrisalSPM2016Copy}
K.~Witrisal, P.~Meissner \emph{et~al.}, ``High-accuracy localization for
  assisted living: {5G} systems will turn multipath channels from foe to
  friend,'' \emph{{IEEE} Signal Process. Mag.}, vol.~33, no.~2, pp. 59--70,
  Mar. 2016.

\bibitem{Mendrzik2019}
R.~Mendrzik, H.~Wymeersch, G.~Bauch, and Z.~Abu-Shaban, ``Harnessing {NLOS}
  components for position and orientation estimation in {5G} {M}illimeter
  {W}ave {MIMO},'' \emph{{IEEE} Trans. Wireless Commun.}, vol.~18, no.~1, pp.
  93--107, 2019.

\bibitem{Karlsson2017}
R.~{Karlsson} and F.~{Gustafsson}, ``{T}he future of automotive localization
  algorithms: {A}vailable, reliable, and scalable localization: {A}nywhere and
  anytime,'' \emph{{IEEE} Signal Process. Mag.}, vol.~34, no.~2, pp. 60--69,
  2017.

\bibitem{KoEMBMag2010}
J.~Ko, T.~Gao, R.~Rothman, and A.~Terzis, ``Wireless sensing systems in
  clinical environments: Improving the efficiency of the patient monitoring
  process,'' \emph{{IEEE} Eng. Med. Biol. Mag.}, vol.~29, pp. 103--9, 05 2010.

\bibitem{Chaccour2022_6G}
C.~Chaccour, M.~N. Soorki, W.~Saad, M.~Bennis, P.~Popovski, and M.~Debbah,
  ``Seven defining features of terahertz ({THz}) wireless systems: {A}
  fellowship of communication and sensing,'' \emph{{IEEE} Commun. Surveys
  Tuts.}, vol.~24, no.~2, pp. 967--993, 2022.

\bibitem{Wymeersch2022_LocSense1}
H.~Wymeersch and G.~Seco-Granados, ``Radio localization and sensing-{P}art i:
  {F}undamentals,'' \emph{{IEEE} Commun. Lett.}, vol.~26, no.~12, pp.
  2816--2820, 2022.

\bibitem{DardariProcIEEE2009}
D.~{Dardari}, A.~{Conti}, U.~{Ferner}, A.~{Giorgetti}, and M.~Z. {Win},
  ``Ranging with ultrawide bandwidth signals in multipath environments,''
  \emph{Proc. {IEEE}}, vol.~97, no.~2, pp. 404--426, Feb. 2009.

\bibitem{TaponeccoTWC2011}
L.~Taponecco, A.~D'Amico, and U.~Mengali, ``Joint {TOA} and {AOA} estimation
  for {UWB} localization applications,'' \emph{{IEEE} Trans. Wireless Commun.},
  vol.~10, no.~7, pp. 2207--2217, 2011.

\bibitem{RusekSPM2013}
F.~Rusek, D.~Persson, B.~K. Lau, E.~G. Larsson, T.~L. Marzetta, O.~Edfors, and
  F.~Tufvesson, ``Scaling up {MIMO}: {O}pportunities and challenges with very
  large arrays,'' \emph{{IEEE} Signal Process. Mag.}, vol.~30, no.~1, pp.
  40--60, Jan. 2013.

\bibitem{ShenTIT2010}
Y.~Shen and M.~Z. Win, ``Fundamental limits of wideband localizationpart {I}: A
  general framework,'' \emph{{IEEE} Trans. Inf. Theory}, vol.~56, no.~10, pp.
  4956--4980, 2010.

\bibitem{AdityaProc2018}
S.~Aditya, A.~F. Molisch, and H.~M. Behairy, ``A survey on the impact of
  multipath on wideband time-of-arrival based localization,'' \emph{Proc.
  {IEEE}}, vol. 106, no.~7, pp. 1183--1203, 2018.

\bibitem{GiffordTSP2022}
W.~M. Gifford, D.~Dardari, and M.~Z. Win, ``The impact of multipath information
  on time-of-arrival estimation,'' \emph{{IEEE} Trans. Signal Process.},
  vol.~70, pp. 31--46, 2022.

\bibitem{WymeerschIEEE2012}
H.~Wymeersch, S.~Maran{\`o}, W.~M. Gifford, and M.~Z. Win, ``{A} machine
  learning approach to ranging error mitigation for {UWB} localization,''
  \emph{{IEEE} Trans. Wireless Commun.}, vol.~60, no.~6, pp. 1719--1728, 2012.

\bibitem{GentnerTWC2016}
C.~Gentner, T.~Jost, W.~Wang, S.~Zhang, A.~Dammann, and U.~C. Fiebig,
  ``Multipath assisted positioning with simultaneous localization and
  mapping,'' \emph{{IEEE} Trans. Wireless Commun.}, vol.~15, no.~9, pp.
  6104--6117, Sep. 2016.

\bibitem{ShahmansooriTWC2018}
A.~Shahmansoori, G.~E. Garcia, G.~Destino, G.~Seco-Granados, and H.~Wymeersch,
  ``Position and orientation estimation through mm {W}ave {MIMO} in {5G}
  systems,'' \emph{{IEEE} Trans. Wireless Commun.}, vol.~17, no.~3, pp.
  1822--1835, Mar. 2018.

\bibitem{LeitingerTWC2019}
E.~{Leitinger}, F.~{Meyer}, F.~{Hlawatsch}, K.~{Witrisal}, F.~{Tufvesson}, and
  M.~Z. {Win}, ``A belief propagation algorithm for multipath-based {SLAM},''
  \emph{{IEEE} Trans. Wireless Commun.}, vol.~18, no.~12, pp. 5613--5629, 2019.

\bibitem{LeitingerICC2019}
E.~{Leitinger}, S.~{Grebien}, and K.~{Witrisal}, ``Multipath-based {SLAM}
  exploiting {AoA} and amplitude information,'' in \emph{Proc. IEEE ICCW-19},
  Shanghai, China, May 2019, pp. 1--7.

\bibitem{KimTWC2020}
H.~{Kim}, K.~{Granstr{\"o}m}, L.~{Gao}, G.~{Battistelli}, S.~{Kim}, and
  H.~{Wymeersch}, ``{5G} {mmWave} cooperative positioning and mapping using
  multi-model {PHD} filter and map fusion,'' \emph{{IEEE} Trans. Wireless
  Commun.}, vol.~19, no.~6, pp. 3782--3795, Mar. 2020.

\bibitem{PedersenJTAP2018}
T.~Pedersen, ``Modeling of path arrival rate for in-room radio channels with
  directive antennas,'' \emph{{IEEE} Trans. Antennas Propag.}, vol.~66, no.~9,
  pp. 4791--4805, 2018.

\bibitem{VenusTWC2023}
A.~{Venus}, E.~{Leitinger}, S.~{Tertinek}, and K.~{Witrisal}, ``A graph-based
  algorithm for robust sequential localization exploiting multipath for
  obstructed-{LOS}-bias mitigation,'' \emph{{IEEE} Trans. Wireless Commun.},
  2023.

\bibitem{WieVenWilLeiArxiv2023}
\BIBentryALTinterwordspacing
L.~Wielandner, A.~Venus, T.~Wilding, and E.~Leitinger, ``Multipath-based {SLAM}
  for non-ideal reflective surfaces exploiting multiple-measurement data
  association,'' \emph{ArXiv e-prints}, 2023. [Online]. Available:
  \url{http://arxiv.org/abs/2304.05680}
\BIBentrySTDinterwordspacing

\bibitem{Yu2020}
Z.~{Yu}, Z.~{Liu}, F.~{Meyer}, A.~{Conti}, and M.~Z. {Win}, ``Localization
  based on channel impulse response estimates,'' in \emph{Proc. IEEE/ION
  PLANS-20}, 2020, pp. 1014--1021.

\bibitem{KimWym:TVT2022}
H.~Kim, K.~Granstrom, L.~Svensson, S.~Kim, and H.~Wymeersch, ``{PMBM}-based
  {SLAM} filters in {5G} {mmWave} vehicular networks,'' \emph{{IEEE} Trans.
  Veh. Technol.}, pp. 1--1, May 2022.

\bibitem{KrekovicTSP2020}
M.~Krekovic, I.~Dokmanic, and M.~Vetterli, ``Shapes from echoes: Uniqueness
  from point-to-plane distance matrices,'' \emph{{IEEE} Trans. Signal
  Process.}, vol.~68, pp. 2480--2498, 2020.

\bibitem{MaranoJSAC2010}
S.~Marano~and, W.~Gifford, H.~Wymeersch, and M.~Win, ``{NLOS} identification
  and mitigation for localization based on {UWB} experimental data,''
  \emph{{IEEE} J. Sel. Areas Commun.}, vol.~28, no.~7, pp. 1026 --1035, Sept.
  2010.

\bibitem{LiShenMILCOM2021}
Y.~Li, S.~Mazuelas, and Y.~Shen, ``A semi-supervised learning approach for
  ranging error mitigation based on {UWB} waveform,'' in \emph{Proc. IEEE
  MILCOM-21}, 2021, pp. 533--537.

\bibitem{StahlkeSensors2021}
M.~Stahlke, S.~Kram, F.~Ott, T.~Feigl, and C.~Mutschler, ``Estimating {TOA}
  reliability with variational autoencoders,'' \emph{{IEEE} Sensors J.}, pp.
  1--1, 2021.

\bibitem{HuangTMC2022}
Y.~Huang, S.~Mazuelas, F.~Ge, and Y.~Shen, ``Indoor localization system with
  {NLOS} mitigation based on self-training,'' \emph{{IEEE} Trans. Mobile
  Comput.}, pp. 1--1, 2022.

\bibitem{ContiProcIEEE2019}
A.~{Conti}, S.~{Mazuelas}, S.~{Bartoletti}, W.~C. {Lindsey}, and M.~Z. {Win},
  ``Soft information for localization-of-things,'' \emph{Proc. {IEEE}}, vol.
  107, no.~11, pp. 2240--2264, Nov. 2019.

\bibitem{Kram2022GPR}
\BIBentryALTinterwordspacing
S.~Kram, C.~Kraus, T.~Feigl, M.~Stahlke, J.~Robert, and C.~Mutschler,
  ``Position tracking using likelihood modeling of channel features with
  {G}aussian processes,'' \emph{ArXiv e-prints}, vol. abs/2203.13110, 2022.
  [Online]. Available: \url{http://arxiv.org/abs/2203.13110}
\BIBentrySTDinterwordspacing

\bibitem{BarShalomTCS2009}
Y.~Bar-Shalom, F.~Daum, and J.~Huang, ``The probabilistic data association
  filter,'' \emph{{IEEE} Control Syst. Mag.}, vol.~29, no.~6, pp. 82--100, Dec
  2009.

\bibitem{MeyerProc2018}
F.~Meyer, T.~Kropfreiter, J.~L. Williams, R.~Lau, F.~Hlawatsch, P.~Braca, and
  M.~Z. Win, ``Message passing algorithms for scalable multitarget tracking,''
  \emph{Proc. {IEEE}}, vol. 106, no.~2, pp. 221--259, Feb. 2018.

\bibitem{MeyWilJ21}
F.~Meyer and J.~L. Williams, ``Scalable detection and tracking of geometric
  extended objects,'' \emph{{IEEE} Trans. Signal Process.}, vol.~69, pp.
  6283--6298, Oct. 2021.

\bibitem{LiTWC2022}
X.~Li, E.~Leitinger, A.~Venus, and F.~Tufvesson, ``Sequential detection and
  estimation of multipath channel parameters using belief propagation,''
  \emph{{IEEE} Trans. Wireless Commun.}, vol.~21, no.~10, pp. 8385--8402, Apr.
  2022.

\bibitem{ArulampalamTSP2002}
M.~S. Arulampalam, S.~Maskell, N.~Gordon, and T.~Clapp, ``A tutorial on
  particle filters for online nonlinear/non-{Gaussian} {Bayesian} tracking,''
  \emph{{IEEE} Trans. Signal Process.}, vol.~50, no.~2, pp. 174--188, Feb.
  2002.

\bibitem{DurrantWhyte2006}
H.~Durrant-Whyte and T.~Bailey, ``Simultaneous localization and mapping: {Part
  I},'' \emph{IEEE Robot. Autom. Mag.}, vol.~13, no.~2, pp. 99--110, Jun. 2006.

\bibitem{BarShalom1995}
Y.~Bar-Shalom and X.-R. Li, \emph{{Multitarget-Multisensor Tracking: Principles
  and Techniques}}.\hskip 1em plus 0.5em minus 0.4em\relax Storrs, CT, USA:
  Yaakov Bar-Shalom, 1995.

\bibitem{JeoTugTAES2005}
S.~Jeong and J.~Tugnait, ``Multisensor tracking of a maneuvering target in
  clutter using {IMMPDA} filtering with simultaneous measurement update,''
  \emph{{IEEE} Trans. Aerosp. Electron. Syst.}, vol.~41, no.~3, pp. 1122--1131,
  Nov. 2005.

\bibitem{LerroACC1990}
D.~Lerro and Y.~Bar-Shalom, ``Automated tracking with target amplitude
  information,'' in \emph{1990 American Control Conference}, May 1990, pp.
  2875--2880.

\bibitem{ZhaStaJosWanGenDamWymHoeTAES2020}
S.~Zhang, E.~Staudinger, T.~Jost, W.~Wang, C.~Gentner, A.~Dammann,
  H.~Wymeersch, and P.~A. Hoeher, ``Distributed direct localization suitable
  for dense networks,'' \emph{{IEEE} Trans. Aerosp. Electron. Syst.}, vol.~56,
  no.~2, pp. 1209--1227, July 2020.

\bibitem{KropfreiterFUSION2021}
T.~Kropfreiter, J.~L. Williams, and F.~Meyer, ``A scalable track-before-detect
  method with {P}oisson/multi-{B}ernoulli model,'' in \emph{Proc. IEEE
  FUSION-21}, 2021.

\bibitem{WilliamsTAES2014}
J.~Williams and R.~Lau, ``Approximate evaluation of marginal association
  probabilities with belief propagation,'' \emph{{IEEE} Trans. Aerosp.
  Electron. Syst.}, vol.~50, no.~4, pp. 2942--2959, 2014.

\bibitem{KschischangTIT2001}
F.~Kschischang, B.~Frey, and H.-A. Loeliger, ``Factor graphs and the
  sum-product algorithm,'' \emph{{IEEE} Trans. Inf. Theory}, vol.~47, no.~2,
  pp. 498--519, Feb. 2001.

\bibitem{Meissner2015Diss}
P.~Meissner, ``{Multipath-Assisted Indoor Positioning},'' Ph.D. dissertation,
  Graz University of Technology, 2014.

\bibitem{Tichavsky1998}
P.~Tichavsky, C.~Muravchik, and A.~Nehorai, ``Posterior {Cramer-Rao} bounds for
  discrete-time nonlinear filtering,'' \emph{{IEEE} Trans. Signal Process.},
  vol.~46, no.~5, pp. 1386--1396, May 1998.

\bibitem{Kay1998}
S.~Kay, \emph{Fundamentals of Statistical Signal Processing: Detection
  Theory}.\hskip 1em plus 0.5em minus 0.4em\relax Upper Saddle River, NJ, USA:
  Prentice Hall, 1998.

\bibitem{BarShalom2002EstimationTracking}
Y.~Bar-Shalom, T.~Kirubarajan, and X.-R. Li, \emph{Estimation with Applications
  to Tracking and Navigation}.\hskip 1em plus 0.5em minus 0.4em\relax New York,
  NY, USA: John Wiley \& Sons, Inc., 2002.

\bibitem{Bjornson2017MassiveMIMO}
E.~Bj{\"o}rnson, J.~Hoydis, L.~Sanguinetti \emph{et~al.}, ``Massive mimo
  networks: Spectral, energy, and hardware efficiency,'' \emph{Foundations and
  Trends{\textregistered} in Signal Processing}, vol.~11, no. 3-4, pp.
  154--655, 2017.

\bibitem{Thrun2005Affine}
S.~Thrun, ``Affine structure from sound,'' \emph{Advances in Neural Information
  Processing Systems}, vol.~18, 2005.

\bibitem{KuangICC2013}
Y.~Kuang, K.~Astroem, and F.~Tufvesson, ``Single antenna anchor-free {UWB}
  positioning based on multipath propagation,'' in \emph{Proc. IEEE ICCW-13},
  2013, pp. 5814--5818.

\bibitem{HanFleuRao:TSP2018}
T.~L. Hansen, B.~H. Fleury, and B.~D. Rao, ``Superfast line spectral
  estimation,'' \emph{{IEEE} Trans. Signal Process.}, vol.~PP, no.~99, pp.
  1--1, Feb. 2018.

\bibitem{ShutWanJos:CSTA2013}
D.~Shutin, W.~Wang, and T.~Jost, ``Incremental sparse {B}ayesian learning for
  parameter estimation of superimposed signals,'' in \emph{Proc. SAMPTA-2013},
  no.~1, Sept. 2013, pp. 6--9.

\bibitem{Hansen2014SAM}
T.~L. Hansen, M.~A. Badiu, B.~H. Fleury, and B.~D. Rao, ``A sparse {B}ayesian
  learning algorithm with dictionary parameter estimation,'' in \emph{Proc.
  IEEE SAM-14}, 2014, pp. 385--388.

\bibitem{MeyerJTSP2017}
F.~{Meyer}, P.~{Braca}, P.~{Willett}, and F.~{Hlawatsch}, ``A scalable
  algorithm for tracking an unknown number of targets using multiple sensors,''
  \emph{{IEEE} Trans. Signal Process.}, vol.~65, no.~13, pp. 3478--3493, 2017.

\bibitem{WilliamsLauTAE2014}
J.~Williams and R.~Lau, ``Approximate evaluation of marginal association
  probabilities with belief propagation,'' \emph{IEEE Trans. Aerosp. Electron.
  Syst.}, vol.~50, no.~4, pp. 2942--2959, Oct. 2014.

\bibitem{Supplement}
A.~{Venus}, E.~{Leitinger}, S.~{Tertinek}, and K.~{Witrisal}, ``Graph-based
  simultaneous localization and bias tracking: Supplementary material,''
  \emph{arXiv:2310.02814}, 2023.

\bibitem{Kay1993}
S.~Kay, \emph{Fundamentals of Statistical Signal Processing: Estimation
  Theory}.\hskip 1em plus 0.5em minus 0.4em\relax Upper Saddle River, NJ, USA:
  Prentice Hall, 1993.

\bibitem{Loeliger2004SPM}
H.-A. Loeliger, ``An introduction to factor graphs,'' \emph{{IEEE} Signal
  Process. Mag.}, vol.~21, no.~1, pp. 28--41, Feb. 2004.

\bibitem{MeyerPhd2015}
F.~Meyer, ``{Navigation and Tracking in Networks: Distributed Algorithms for
  Cooperative Estimation and Information-seeking Control},'' Ph.D.
  dissertation, TU Wien, Vienna, Austria, 2015.

\bibitem{MeyerJSIPN2016}
F.~Meyer, O.~Hlinka, H.~Wymeersch, E.~Riegler, and F.~Hlawatsch, ``Distributed
  localization and tracking of mobile networks including noncooperative
  objects,'' \emph{{IEEE} Trans. Signal Inf. Process. Netw.}, vol.~2, no.~1,
  pp. 57--71, 2016.

\bibitem{BarShalom_AlgorithmHandbook}
Y.~{Bar-Shalom}, P.~K. {Willett}, and X.~{Tian}, \emph{Tracking and data
  fusion: a handbook of algorithms}.\hskip 1em plus 0.5em minus 0.4em\relax
  Storrs, CT, USA: Yaakov Bar-Shalom, 2011.

\bibitem{MonThrKolWeg:AAAI2002}
M.~Montemerlo, S.~Thrun, D.~Koller, and B.~Wegbreit, ``{FastSLAM: A factored
  solution to the simultaneous localization and mapping problem},'' in
  \emph{Proc. AAAI-02}, Edmonton, Canda, Jul. 2002, pp. 593--598.

\bibitem{KrimVibergSPM1996}
H.~Krim and M.~Viberg, ``Two decades of array signal processing research: {T}he
  parametric approach,'' \emph{{IEEE} Signal Process. Mag.}, vol.~13, no.~4,
  pp. 67--94, 1996.

\bibitem{GodHaiBlu:TIT2010}
H.~Godrich, A.~Haimovich, and R.~Blum, ``Target localization accuracy gain in
  {MIMO} radar-based systems,'' \emph{{IEEE} Trans. Inf. Theory}, vol.~56,
  no.~6, pp. 2783 --2803, {J}une 2010.

\bibitem{Jourdan2008}
D.~B. {Jourdan}, D.~{Dardari}, and M.~Z. {Win}, ``Position error bound for
  {UWB} localization in dense cluttered environments,'' \emph{{IEEE} Trans.
  Aerosp. Electron. Syst.}, vol.~44, no.~2, pp. 613--628, 2008.

\bibitem{WitrisalJWCOML2016}
K.~Witrisal, E.~Leitinger, S.~Hinteregger, and P.~Meissner, ``Bandwidth scaling
  and diversity gain for ranging and positioning in dense multipath channels,''
  \emph{{IEEE} Wireless Commun. Lett.}, vol.~5, no.~4, pp. 396--399, Aug. 2016.

\bibitem{LeitingerJSAC2015}
E.~Leitinger, P.~Meissner, C.~Ruedisser, G.~Dumphart, and K.~Witrisal,
  ``Evaluation of position-related information in multipath components for
  indoor positioning,'' \emph{{IEEE} J. Sel. Areas Commun.}, vol.~33, no.~11,
  pp. 2313--2328, Nov. 2015.

\bibitem{Rasmussen2006GP}
C.~E. Rasmussen and C.~K.~I. Williams, \emph{Gaussian processes for machine
  learning}.\hskip 1em plus 0.5em minus 0.4em\relax {MIT} Press, 2006.

\end{thebibliography}


\begin{thebibliography}{10}
\providecommand{\url}[1]{#1}
\csname url@samestyle\endcsname
\providecommand{\newblock}{\relax}
\providecommand{\bibinfo}[2]{#2}
\providecommand{\BIBentrySTDinterwordspacing}{\spaceskip=0pt\relax}
\providecommand{\BIBentryALTinterwordstretchfactor}{4}
\providecommand{\BIBentryALTinterwordspacing}{\spaceskip=\fontdimen2\font plus
\BIBentryALTinterwordstretchfactor\fontdimen3\font minus
  \fontdimen4\font\relax}
\providecommand{\BIBforeignlanguage}[2]{{%
\expandafter\ifx\csname l@#1\endcsname\relax
\typeout{** WARNING: IEEEtran.bst: No hyphenation pattern has been}%
\typeout{** loaded for the language `#1'. Using the pattern for}%
\typeout{** the default language instead.}%
\else
\language=\csname l@#1\endcsname
\fi
#2}}
\providecommand{\BIBdecl}{\relax}
\BIBdecl

\bibitem{Main}
A.~{Venus}, E.~{Leitinger}, S.~{Tertinek}, and K.~{Witrisal}, ``Graph-based
  simultaneous localization and bias tracking,'' \emph{arXiv:2310.02814}, 2023.

\bibitem{MeyerJTSP2017}
F.~{Meyer}, P.~{Braca}, P.~{Willett}, and F.~{Hlawatsch}, ``A scalable
  algorithm for tracking an unknown number of targets using multiple sensors,''
  \emph{{IEEE} Trans. Signal Process.}, vol.~65, no.~13, pp. 3478--3493, 2017.

\bibitem{MeyerProc2018}
F.~Meyer, T.~Kropfreiter, J.~L. Williams, R.~Lau, F.~Hlawatsch, P.~Braca, and
  M.~Z. Win, ``Message passing algorithms for scalable multitarget tracking,''
  \emph{Proc. {IEEE}}, vol. 106, no.~2, pp. 221--259, Feb. 2018.

\bibitem{LeitingerICC2017}
E.~Leitinger, F.~Meyer, F.~Tufvesson, and K.~Witrisal, ``Factor graph based
  simultaneous localization and mapping using multipath channel information,''
  in \emph{Proc. IEEE ICCW-17}, Paris, France, May 2017, pp. 652--658.

\bibitem{LeitingerTWC2019}
E.~{Leitinger}, F.~{Meyer}, F.~{Hlawatsch}, K.~{Witrisal}, F.~{Tufvesson}, and
  M.~Z. {Win}, ``A belief propagation algorithm for multipath-based {SLAM},''
  \emph{{IEEE} Trans. Wireless Commun.}, vol.~18, no.~12, pp. 5613--5629, 2019.

\bibitem{KschischangTIT2001}
F.~Kschischang, B.~Frey, and H.-A. Loeliger, ``Factor graphs and the
  sum-product algorithm,'' \emph{{IEEE} Trans. Inf. Theory}, vol.~47, no.~2,
  pp. 498--519, Feb. 2001.

\bibitem{Loeliger2004SPM}
H.-A. Loeliger, ``An introduction to factor graphs,'' \emph{{IEEE} Signal
  Process. Mag.}, vol.~21, no.~1, pp. 28--41, Feb. 2004.

\bibitem{WilliamsLauTAE2014}
J.~Williams and R.~Lau, ``Approximate evaluation of marginal association
  probabilities with belief propagation,'' \emph{IEEE Trans. Aerosp. Electron.
  Syst.}, vol.~50, no.~4, pp. 2942--2959, Oct. 2014.

\bibitem{WilliamsTAES2014}
------, ``Approximate evaluation of marginal association probabilities with
  belief propagation,'' \emph{{IEEE} Trans. Aerosp. Electron. Syst.}, vol.~50,
  no.~4, pp. 2942--2959, 2014.

\bibitem{MeyerJSIPN2016}
F.~Meyer, O.~Hlinka, H.~Wymeersch, E.~Riegler, and F.~Hlawatsch, ``Distributed
  localization and tracking of mobile networks including noncooperative
  objects,'' \emph{{IEEE} Trans. Signal Inf. Process. Netw.}, vol.~2, no.~1,
  pp. 57--71, 2016.

\bibitem{ArulampalamTSP2002}
M.~S. Arulampalam, S.~Maskell, N.~Gordon, and T.~Clapp, ``A tutorial on
  particle filters for online nonlinear/non-{Gaussian} {Bayesian} tracking,''
  \emph{{IEEE} Trans. Signal Process.}, vol.~50, no.~2, pp. 174--188, Feb.
  2002.

\bibitem{Doucet2005}
A.~Doucet and X.~Wang, ``Monte {Carlo} methods for signal processing: {A}
  review in the statistical signal processing context,'' \emph{{IEEE} Signal
  Process. Mag.}, vol.~22, no.~6, pp. 152--170, Nov. 2005.

\bibitem{MeyerPhd2015}
F.~Meyer, ``{Navigation and Tracking in Networks: Distributed Algorithms for
  Cooperative Estimation and Information-seeking Control},'' Ph.D.
  dissertation, TU Wien, Vienna, Austria, 2015.

\bibitem{ContiProcIEEE2019}
A.~{Conti}, S.~{Mazuelas}, S.~{Bartoletti}, W.~C. {Lindsey}, and M.~Z. {Win},
  ``Soft information for localization-of-things,'' \emph{Proc. {IEEE}}, vol.
  107, no.~11, pp. 2240--2264, Nov. 2019.

\bibitem{Bishop2006}
C.~M. Bishop, \emph{Pattern recognition and machine learning}.\hskip 1em plus
  0.5em minus 0.4em\relax Springer, vol.~4, no.~4.

\bibitem{Kingma2019}
\BIBentryALTinterwordspacing
D.~P. Kingma and M.~Welling, ``An introduction to variational autoencoders,''
  \emph{Foundations and Trends{\textregistered} in Machine Learning}, vol.~12,
  no.~4, pp. 307--392, 2019. [Online]. Available:
  \url{https://doi.org/10.1561%2F2200000056}
\BIBentrySTDinterwordspacing

\bibitem{Kram2022GPR}
\BIBentryALTinterwordspacing
S.~Kram, C.~Kraus, T.~Feigl, M.~Stahlke, J.~Robert, and C.~Mutschler,
  ``Position tracking using likelihood modeling of channel features with
  {G}aussian processes,'' \emph{ArXiv e-prints}, vol. abs/2203.13110, 2022.
  [Online]. Available: \url{http://arxiv.org/abs/2203.13110}
\BIBentrySTDinterwordspacing

\end{thebibliography}

 \begin{IEEEbiography}[{\includegraphics[width=25mm,height=32.15mm,clip,keepaspectratio]{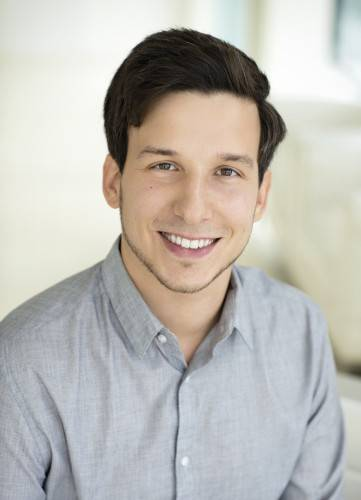}}]{Alexander~Venus} (Member, IEEE) received his BSc, MSc and Ph.D. degrees (all with highest honors) in Biomedical Engineering, Information and Computer Engineering and Information and Communications Engineering from Graz University of Technology, in 2012, 2015 and 2024, respectively. From 2014 to 2019 he was a research and development engineer at Anton Paar GmbH, Graz, and from 2020 to 2023 a research scientist at the Christian Doppler Laboratory for Location Aware Electronic Systems, Graz University of Technology. In 2022 he was a guest researcher at the University of California San Diego. He is currently a postdoctoral researcher at the Institute of Communication Networks and Satellite Communications, Graz University of Technology.
	
His research interests include inference on graphs, radio-based localization, navigation and sensing, statistical signal processing, nonlinear estimation, artificial intelligence and performance bounds. 
\end{IEEEbiography}

\begin{IEEEbiography}[{\includegraphics[width=25mm,height=32.15mm,clip,keepaspectratio]{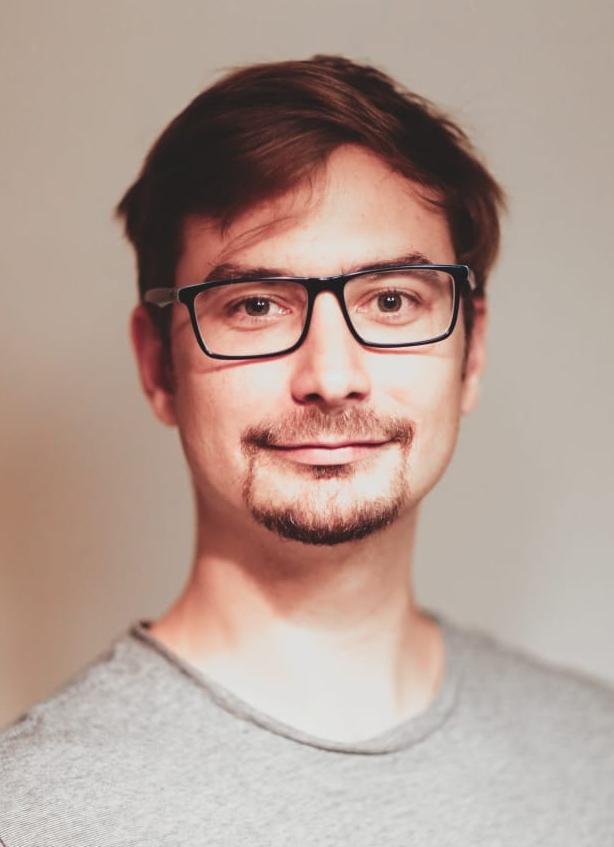}}]{Erik~Leitinger} (Member, IEEE) received his MSc and PhD degrees (with highest honors) in electrical engineering from Graz University of Technology, Austria in 2012 and 2016, respectively. He was postdoctoral researcher at the department of Electrical and Information Technology at Lund University from 2016 to 2018. He is currently a senior researcher at Graz University of Technology. His research interests include inference on graphs, statistical signal processing, high-dimensional and nonlinear estimation, localization and navigation, machine learning, stochastic modeling and estimation of radio channels, and estimation/detection theory. 
		
		Dr.\ Leitinger is an Associate Editor with the IEEE Transactions on Wireless Communications and the ISIF Journal of Advances in Information Fusion. He served as co-chair of a special session ``Positioning Energy Constraint Devices'' at the Asilomar Conference on Signals, Systems, and Computers 2020 and of the special session ``Synergistic Radar Signal Processing and Tracking'' at the IEEE Radar Conference in 2021. He is an Erwin Schr\"odinger Fellow. 
	\end{IEEEbiography}

\begin{IEEEbiography}[{\includegraphics[width=25mm,height=32.15mm,clip,keepaspectratio]{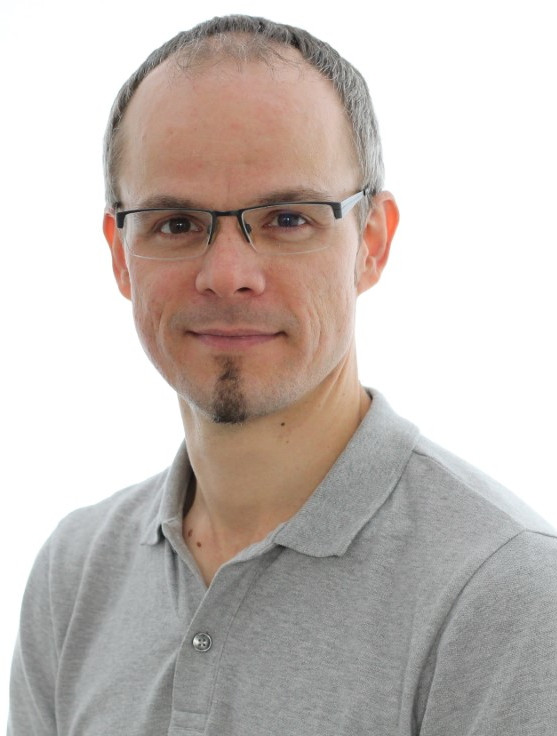}}]{Stefan~Tertinek} received the Dipl.-Ing. degree in electrical engineering from Graz University of Technology, Austria, in 2007, and the Ph.D. degree in electrical engineering from University College Dublin, Dublin, Ireland, in 2011. From 2011 to 2018 he was with Danube Mobile Communications Engineering GmbH \& Co KG (majority owned by Intel Austria GmbH), Linz, Austria, as a RF System Engineer involved in research and product development of multiple generations of cellular RF transceiver and modem platforms. In 2018 he joined NXP Semiconductors Austria GmbH \& Co KG as a RF System Architect in the Product Line Secure Car Access, where he works on ultra-wideband (UWB) and Bluetooth radio technologies with a focus on localization, radar and machine learning.
\end{IEEEbiography}

\begin{IEEEbiography}[{\includegraphics[width=25mm,height=32.15mm,clip,keepaspectratio]{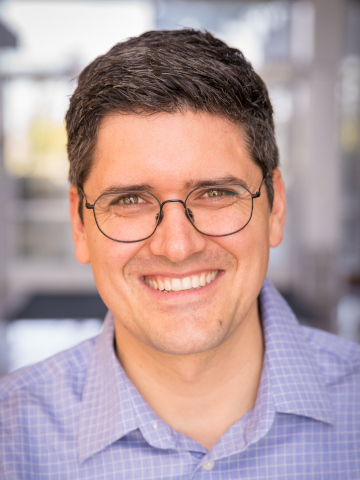}}]{Florian~Meyer}(Member, IEEE) received the MSc and PhD degrees (with highest honors) in electrical engineering from TU Wien, Vienna, Austria in 2011 and 2015, respectively. He is an Assistant Professor with the University of California San Diego, La Jolla, CA, jointly between the Scripps Institution of Oceanography and the Electrical and Computer Engineering Department. From 2017 to 2019 he was a Postdoctoral Fellow and Associate with the Laboratory for Information \& Decision Systems at the Massachusetts Institute of Technology, Cambridge, MA, and from 2016 to 2017 he was a Research Scientist with the NATO Centre for Maritime Research and Experimentation, La Spezia, Italy. Prof. Meyer is the recipient of the 2021 ISIF Young Investigator Award, a 2022 NSF CAREER Award, a 2022 DARPA Young Faculty Award, and a 2023 ONR Young Investigator Award. He is currently an Associate Editor for the \textit{IEEE Transactions on Signal Processing} and also served as an Associate Editor for the \textit{IEEE Transactions on Aerospace and Electronic Systems} from 2021 to 2023 and the \textit{ISIF Journal of Advances in Information Fusion} from 2019 to 2022. His research interests include statistical signal processing, high-dimensional and nonlinear estimation, inference on graphs, machine perception, and graph neural networks.
\end{IEEEbiography}

\begin{IEEEbiography}[{\includegraphics[width=1in,height=1.25in,clip,keepaspectratio]{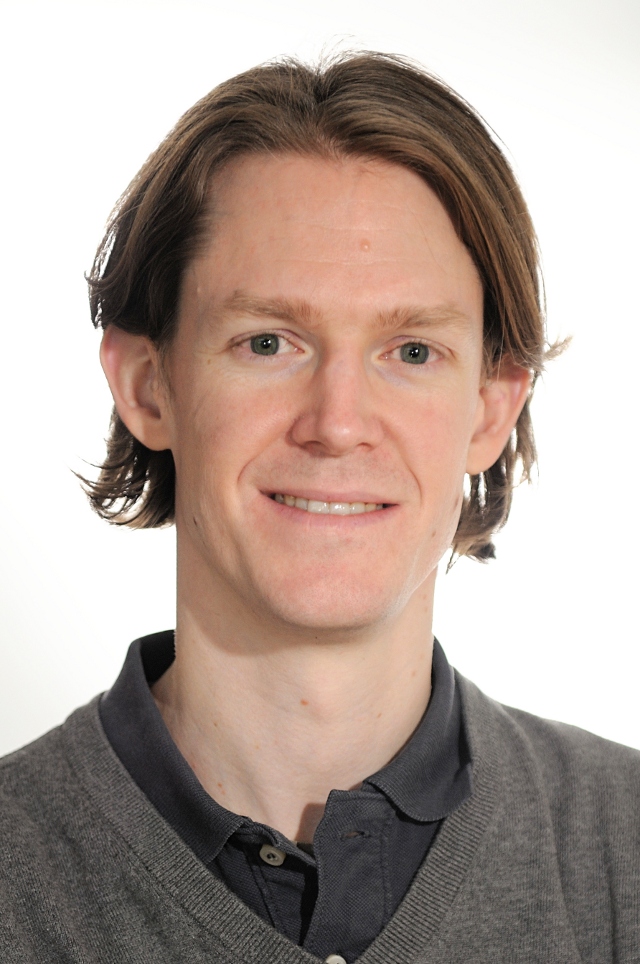}}]{Klaus~Witrisal} (Member, IEEE) received the Ph.D. degree (cum laude) from Delft University of Technology, Delft, The Netherlands, in 2002, and the Habilitation from Graz University of Technology in 2009. He is currently a Full Professor at Graz University of Technology and head of the Christian Doppler Laboratory for Location-aware Electronic Systems, where he participates in a range of national and international research projects with industrial and academic partners. His research interests are in signal processing for wireless communications, propagation channel modeling, and positioning. Klaus Witrisal served as an associate editor of IEEE Communications Letters and of the KICS/IEEE Journal of Communications and Networks, co-chair of the TWG ``Indoor'' of the COST Action IC1004, co-chair of the EWG ``Localisation and Tracking'' of the COST Action CA15104, leading chair of the IEEE Workshop on Advances in Network Localization and Navigation (ANLN), and TPC (co)-chair of the Workshop on Positioning, Navigation and Communication (WPNC). 
	\end{IEEEbiography}
\clearpage

\end{document}


\title{\huge{Graph-based Simultaneous Localization and Bias Tracking:\\ Supplementary Material}}
\author{\large{Alexander Venus, Erik Leitinger, Stefan Tertinek, Florian Meyer, and  Klaus Witrisal}\\[2mm]\small{September 2023}}
\maketitle
\frenchspacing

\noindent This manuscript provides additional analysis for the publication ``Graph-based Simultaneous Localization and Bias Tracking'' by the same authors \cite{Main}.
\vspace{-2mm}

\ifthenelse{0=1}
{
	
\section{Derivation of Likelihood and Joint Posterior} \label{sec:derivation_posterior}

\subsection{Joint Prior \ac{pdf}}\label{app:JointPrior}

Before presenting derivations, we first define a few sets as follows: $ \mathcal{D}_{\V{a}_n,\underline{\V{r}}_n} \triangleq \{k \in \{1, \dots, K_{n-1}\}: \underline{r}_{k,n} = 1, a_{k,n} \neq 0 \} $ denotes the existing legacy \acp{pbo} set, and $ \mathcal{N}_{\overline{\V{r}}_n} \triangleq \{m \in \{1,\dots,M_{n}\}: \overline{r}_{m,n} = 1, b_{m,n} = 0 \} $ denotes the existing new \acp{pbo} set. Correspondingly, the sets of non-existing legacy \acp{pbo} are given by $ \overline{\mathcal{D}}_{\V{a}_n, \underline{\V{r}}_n} \triangleq \{1, \dots, K_{n-1}\} \setminus \mathcal{D}_{\V{a}_n,\underline{\V{r}}_n}$, and the sets of non-existing new \acp{pbo} are given as $ \overline{\mathcal{N}}_{\overline{\V{r}}_n} \triangleq \{1,\dots,M_{n}\} \setminus \mathcal{N}_{\overline{\V{r}}_n}$. Hence, the number of false alarm measurements can be represented with the sets as $ {k_\mathrm{fa}}_{n} = M_{n} - |\mathcal{D}_{\V{a}_n,\underline{\V{r}}_n}| - |\mathcal{N}_{\overline{\V{r}}_n}| $, and the number of \acp{pbo} states is given as $ K_{n-1} + M_{n} = |\mathcal{D}_{\V{a}_n,\underline{\V{r}}_n} | + |\overline{\mathcal{D}}_{\V{a}_n,\underline{\V{r}}_n}| + |\mathcal{N}_{\overline{\V{r}}_n}| + |\overline{\mathcal{N}}_{\overline{\V{r}}_n}| $.

Assuming that the new \acp{pbo} states $ \overline{\RV{y}}_{n} $ and the \ac{pbo}-oriented association variables $ \RV{a}_{n} $ are conditionally independent given the legacy \acp{pbo} state $ \underline{\RV{y}}_{n} $, the joint prior \ac{pdf} of $\underline{\RV{y}}_{1:n}$, $\overline{\RV{y}}_{1:n}$, $ \RV{a}_{1:n} $, $ \RV{b}_{1:n} $, ${\rv{\mu}_{\mathrm{fa}}}_{1:n} $, and the number of the measurements $\RV{m}_{1:n}$ factorizes as 
\begin{align}
	&f(\V{y}_{1:n}, \V{a}_{1:n}, \V{b}_{1:n}, {\V{\mu}_{\mathrm{fa}}}_{1:n}, \V{m}_{1:n}) \nn \\ 
	& \hspace{2mm} = f(\underline{\V{y}}_{1:n}, \overline{\V{y}}_{1:n}, \V{a}_{1:n}, \V{b}_{1:n}, {\V{\mu}_{\mathrm{fa}}}_{1:n}, \V{m}_{1:n}) \nn \\ 
	& \hspace{2mm} = f({\mu_{\mathrm{fa}}}_{1}) f(\overline{\V{x}}_{1} | \overline{\V{r}}_{1}, M_{1}) p(\overline{\V{r}}_{1}, \V{a}_{1}, \V{b}_{1}, M_{1}|{\mu_{\mathrm{fa}}}_{1}) \nn \\
	& \hspace{5mm} \times \prod_{ n' = 2 }^{n} f({\mu_{\mathrm{fa}}}_{n'}|{\mu_{\mathrm{fa}}}_{n'-1}) \left(\prod_{k = 1}^{K_{n'-1}} f(\underline{\V{y}}_{k,n'}|\V{y}_{k,n'-1})\right) \nn \\
	& \hspace{5mm} \times f(\overline{\V{x}}_{n'} | \overline{\V{r}}_{n'}, M_{n'}) p(\overline{\V{r}}_{n'}, \V{a}_{n'}, \V{b}_{n'}, M_{n'}|{\mu_{\mathrm{fa}}}_{n'},\underline{\V{y}}_{n'})
	\label{eq:jointPriorPDF_global} 
\end{align}
where $ p(\overline{\V{r}}_{1}, \V{a}_{1}, \V{b}_{1}, M_{1}|{\mu_{\mathrm{fa}}}_{1}, \underline{\V{y}}_{1}) = p(\overline{\V{r}}_{1}, \V{a}_{1}, \V{b}_{1}, M_{1}|{\mu_{\mathrm{fa}}}_{1}) $ since no legacy \acp{pbo} exist at time $ n=1 $. We determine the prior \ac{pdf} of new \acp{pbo} $ f(\overline{\V{x}}_{n} | \overline{\V{r}}_{n}, M_{n}) $ and the joint conditional prior \ac{pmf} $ p(\overline{\V{r}}_{n}, \V{a}_{n}, \V{b}_{n}, M_{n}|{\mu_{\mathrm{fa}}}_{n}, \underline{\V{y}}_{n}) $ as follows.

Before the current measurements are observed, the number of measurements $\rv{M}_{n}$ is random. The Poisson \ac{pmf} of the number of existing new \acp{pbo} evaluated at $ |\mathcal{N}_{\overline{\V{r}}_n}| $ is given by $ p(|\mathcal{N}_{\overline{\V{r}}_n}|) = \mu_{\mathrm{n}}^{|\mathcal{N}_{\overline{\V{r}}_n}|}/|\mathcal{N}_{\overline{\V{r}}_n}|!\mathrm{e}^{\mu_{\mathrm{n}}} $. The prior \ac{pdf} of the new \ac{pbo} state $ \overline{\RV{x}}_{n} $ conditioned on $\overline{\RV{r}}_{n}$ and $\rv{M}_{n}$ is expressed as 
\begin{align}
	& f(\overline{\V{x}}_{n} | \overline{\V{r}}_{n}, M_{n}) = \prod_{m \in \mathcal{N}_{\overline{\V{r}}_n}} f_{\mathrm{n}}(\overline{\V{x}}_{m,n}) \prod_{m' \in \overline{\mathcal{N}}_{\overline{\V{r}}_n} } f_{\mathrm{D}}(\overline{\V{x}}_{m',n}).
	\label{eq:priorPDF_newPSMC} 
\end{align}
The \ac{pmf} for the number of false alarm measurements is given by $ p({k_\mathrm{fa}}_{n}) = {\mu_{\mathrm{fa}}}_{n}^{{k_\mathrm{fa}}_{n}}/ {k_\mathrm{fa}}_{n}!\mathrm{e}^{{\mu_{\mathrm{fa}}}_{n}} $. The joint conditional prior \ac{pmf} of the binary existence variables of new \acp{pbo} $\overline{\RV{r}}_{n} \triangleq [\overline{\rv{r}}_{1,n}\ist\cdots$ $\overline{\rv{r}}_{\rv{M}_n,n}]$, the \ac{da} vectors $ \RV{a}_{n} $ and $ \RV{b}_{n} $ and the number of the measurements $ \rv{M}_{n} $ conditioned on $ {\rv{\mu}_{\mathrm{fa}}}_{n} $ and $ \underline{\RV{y}}_{n} $ is obtained as \cite{BarShalom_AlgorithmHandbook, Florian_Proceeding2018, Erik_SLAM_TWC2019} 
\begin{align} 
	& p(\overline{\V{r}}_{n}, \V{a}_{n}, \V{b}_{n}, M_{n}|{\mu_{\mathrm{fa}}}_{n},\underline{\V{y}}_{n}) \nn \\ 
	& = \chi_{\overline{\V{r}}_{n}, \V{a}_{n}, M_{n}} \left(\prod_{ m\in \mathcal{N}_{\overline{\V{r}}_n} } \Gamma_{\V{a}_{n}}(\overline{r}_{m,n}) \right) \hspace*{-1mm} \left(\prod_{ k \in \mathcal{D}_{\V{a}_n,\underline{\V{r}}_n} } p_{\mathrm{d}}(\underline{u}_{k,n}) \right) \nn \\
	& \hspace*{2mm} \times \Psi(\V{a}_n,\V{b}_n) \left(\prod_{k' \in \overline{\mathcal{D}}_{\V{a}_n,\underline{\V{r}}_n} } \hspace*{-3mm} \big( 1(a_{k',n}) - \underline{r}_{k',n} p_{\mathrm{d}}(\underline{u}_{k',n}) \big)\right)\ist .
	\label{eq:priorPDF_DA}
\end{align}
The normalization constant $ \chi_{\overline{\V{r}}_{n}, \V{a}_{n}, M_{n}} $ combining the two Poisson \acp{pmf} above is given by 
\begin{align} 
	\chi_{\overline{\V{r}}_{n}, \V{a}_{n}, M_{n}} & = \left(\mathrm{e}^{-\mu_{\mathrm{n}}} /M_n! \right) \left( (\mu_{\mathrm{n}}/{\mu_{\mathrm{fa}}}_{n})^{|\mathcal{N}_{\overline{\V{r}}_n}|}  {\mu_{\mathrm{fa}}}_{n}^{-|\mathcal{D}_{\V{a}_n,\underline{\V{r}}_n}|}  \right)  \nn \\
	& \hspace*{30mm} \times \left( \mathrm{e}^{-{\mu_{\mathrm{fa}}}_{n}}  {\mu_{\mathrm{fa}}}_{n}^{M_{n}}\right)
	\label{eq:normConst_FAR}
\end{align}
where the first part (the terms in the first brackets) is fixed after observing the current measurements given the assumption that the mean number of newly detected \acp{pbo} $ \rv{\mu}_{\mathrm{n}} $ is a known constant. The second part can be merged with factors in the sets $ \mathcal{N}_{\overline{\V{r}}_n} $ and $ \mathcal{D}_{\V{a}_n,\underline{\V{r}}_n} $, respectively. The third part equals to $ n({\mu_{\mathrm{fa}}}_{n})^{(K_{n-1} + M_{n})} $ where $ n({\mu_{\mathrm{fa}}}_{n}) \triangleq ( \mathrm{e}^{-{\mu_{\mathrm{fa}}}_{n}} {\mu_{\mathrm{fa}}}_{n}^{M_{n}} )^{1/(K_{n-1} + M_{n})} $ is the \ac{far}-related normalization constant. The two exclusion functions $ \Psi(\V{a}_n,\V{b}_n) $ and $ \Gamma_{\V{a}_{n}}(\overline{r}_{m,n}) = 0 $ ensure that $ p(\overline{\V{r}}_{n}, \V{a}_{n}, \V{b}_{n}, M_{n}|{\mu_{\mathrm{fa}}}_{n},\underline{\V{y}}_{n}) \neq 0 $ if and only if a measurement is generated by only one \ac{pbo} (either a legacy or a new one), and a \ac{pbo} generates no more than one measurement. 

The product of the prior \ac{pdf} of new \acp{pbo} \eqref{eq:priorPDF_newPSMC} and the joint conditional prior \ac{pmf} \eqref{eq:priorPDF_DA} after merging factors can be written up to the normalization constant as
\begin{align}
	& p(\overline{\V{r}}_{n}, \V{a}_{n}, \V{b}_{n}, M_{n}|{\mu_{\mathrm{fa}}}_{n},\underline{\V{y}}_{n}) f(\overline{\V{x}}_{n} | \overline{\V{r}}_{n}, M_{n})\nn \\
	& \propto \left( \psi(\bm{a}_n, \bm{b}_n) \prod_{ k \in \mathcal{D}_{\V{a}_n,\underline{\V{r}}_n} } \dfrac{n({\mu_{\mathrm{fa}}}_{n}) p_{\mathrm{d}}(\underline{u}_{k,n})} {{\mu_{\mathrm{fa}}}_{n}} 
	\right. \nn \\
	& \quad \left. \times 
	\prod_{k' \in \overline{\mathcal{D}}_{\V{a}_n,\underline{\V{r}}_n} } n({\mu_{\mathrm{fa}}}_{n}) \big( \bar{1}(a_{k',n}) - \underline{r}_{k',n} p_{\mathrm{d}}(\underline{u}_{k',n}) \big) \right) \nn \\ 
	& \quad \times \left( \prod_{ m\in \mathcal{N}_{\overline{\V{r}}_n} } \dfrac{n({\mu_{\mathrm{fa}}}_{n})\mu_{\mathrm{n}} f_{\mathrm{n}}(\overline{\V{x}}_{m,n})}{{\mu_{\mathrm{fa}}}_{n}} \Gamma_{\V{a}_{n}}(\overline{r}_{m,n}) \right. \nn \\
	& \quad \left. \times 
	\prod_{m' \in \overline{\mathcal{N}}_{\overline{\V{r}}_n} } n({\mu_{\mathrm{fa}}}_{n}) f_{\mathrm{D}}(\overline{\V{x}}_{m',n}) \right).
	\label{eq:priorPDF_DA_newPSMC_combine} 
\end{align} 
With some simple manipulations using the definitions of exclusion functions $ \Psi(\V{a}_n,\V{b}_n) $ and $ \Gamma_{\V{a}_{n}}(\overline{r}_{m,n}) $ (see Section~\ref{sec:DA}), Eq.~\eqref{eq:priorPDF_DA_newPSMC_combine} can be rewritten as the product of factors related to the legacy \acp{pbo} and to the new \acp{pbo}, respectively, i.e.,
\begin{align}
	& p(\overline{\V{r}}_{n}, \V{a}_{n}, \V{b}_{n}, M_{n}|{\mu_{\mathrm{fa}}}_{n},\underline{\V{y}}_{n}) f(\overline{\V{x}}_{n} | \overline{\V{r}}_{n}, M_{n}) \nn \\
	& \hspace*{2mm} \propto \left(\prod_{k = 1}^{K_{n-1}} g_{1}(\underline{\V{y}}_{k,n}, a_{k,n}, {\mu_{\mathrm{fa}}}_{n}; M_{n}) \prod_{m = 1}^{M_{n}} \psi(a_{k,n},b_{m,n}) \right) \nn \\
	& \hspace*{4mm}\times\left(\prod_{m' = 1}^{M_{n}} h_{1}(\overline{\V{y}}_{m',n}, b_{m',n}, {\mu_{\mathrm{fa}}}_{n}; M_{n})\right)
	\label{eq:priorPDF_DA_newPSMC}
\end{align}
with $ g_{1}(\underline{\V{y}}_{k,n}, a_{k,n}, {\mu_{\mathrm{fa}}}_{n}; M_{n})\rmv\rmv=\rmv\rmv g_{1}(\underline{\V{x}}_{k,n}, \underline{r}_{k,n}, a_{k,n}, {\mu_{\mathrm{fa}}}_{n}; M_{n}) $ given by 
\begin{align}
	& g_{1}(\underline{\V{x}}_{k,n}, \underline{r}_{k,n} = 1, a_{k,n}, {\mu_{\mathrm{fa}}}_{n}; M_{n}) \nn \\
	& \hspace*{15mm} \triangleq
	\begin{cases}
		\dfrac{n({\mu_{\mathrm{fa}}}_{n}) p_{\mathrm{d}}(\underline{u}_{k,n})} {{\mu_{\mathrm{fa}}}_{n}}, 		& a_{k,n} = m \\
		n({\mu_{\mathrm{fa}}}_{n}) \big(1 - p_{\mathrm{d}}(\underline{u}_{k,n})\big),								& a_{k,n} = 0
	\end{cases}
	\label{eq:g1}
\end{align}
and $ g_{1}(\underline{\V{x}}_{k,n}, \underline{r}_{k,n} = 0, a_{k,n}, {\mu_{\mathrm{fa}}}_{n}; M_{n}) \triangleq \bar{1}(a_{k,n}) n({\mu_{\mathrm{fa}}}_{n}) $, and $h_{1}(\overline{\V{y}}_{m,n}, b_{m,n}, {\mu_{\mathrm{fa}}}_{n}; M_{n}) = h_{1}(\overline{\V{x}}_{m,n}, \overline{r}_{m,n},b_{m,n}, {\mu_{\mathrm{fa}}}_{n}; M_{n}) $ is given by 
\begin{align}
	& h_{1}(\overline{\V{x}}_{m,n}, \overline{r}_{m,n} = 1 , b_{m,n}, {\mu_{\mathrm{fa}}}_{n}; M_{n}) \nn \\
	& \hspace*{15mm} \triangleq
	\begin{cases}
		0, 																							& b_{m,n} = k \\
		\dfrac{n({\mu_{\mathrm{fa}}}_{n})\mu_{\mathrm{n}} f_{\mathrm{n}}(\overline{\V{x}}_{m,n})}{{\mu_{\mathrm{fa}}}_{n}},	& b_{m,n} = 0
	\end{cases}
	\label{eq:h1} 
\end{align}
and $ h_{1}(\overline{\V{x}}_{m,n}, \overline{r}_{m,n} = 0 , b_{m,n}, {\mu_{\mathrm{fa}}}_{n}; M_{n}) \triangleq n({\mu_{\mathrm{fa}}}_{n}) $. 

Finally, by inserting (\ref{eq:priorPDF_DA_newPSMC}) into (\ref{eq:jointPriorPDF_global}), the joint prior \ac{pdf} can be rewritten as 
\begin{align}
	&f(\underline{\V{y}}_{1:n}, \overline{\V{y}}_{1:n}, \V{a}_{1:n}, \V{b}_{1:n}, {\V{\mu}_{\mathrm{fa}}}_{1:n}, \V{m}_{1:n}) \nn \\
	& \propto f({\mu_{\mathrm{fa}}}_{1}) \prod_{l = 1}^{M_{1}} h_{1}(\overline{\V{y}}_{l,1}, b_{l,1} ,{\mu_{\mathrm{fa}}}_{1}; M_{1}) \nn \\
	& \hspace*{2mm} \times \prod_{n' = 2}^{n} f({\mu_{\mathrm{fa}}}_{n'} | {\mu_{\mathrm{fa}}}_{n'-1}) \left(\prod_{k' = 1}^{K_{n'-1}} f(\underline{\V{y}}_{k',n'}|\V{y}_{k',n'-1})\right)  \nn \\
	& \hspace*{2mm} \times \left(\prod_{k = 1}^{K_{n'-1}}g_{1}(\underline{\V{y}}_{k,n'}, a_{k,n'}, {\mu_{\mathrm{fa}}}_{n'}; M_{n'}) \prod_{m = 1}^{M_{n'}} \psi(a_{k,n'},b_{m,n'})\right) \nn \\
	& \hspace*{2mm} \times \left(\prod_{m' = 1}^{M_{n'}} h_{1}(\overline{\V{y}}_{m',n'}, b_{m',n'}, {\mu_{\mathrm{fa}}}_{n'}; M_{n'})\right)\ist.
	\label{eq:jointPriorPDF_global_factorized}
\end{align}

\subsection{Joint Likelihood Function}
\label{app:JointLHF}

Assume that the measurements $\RV{z}_{n}$ are independent across $n$, the conditional \ac{pdf} of $\RV{z}_{1:n}$ given $\underline{\RV{y}}_{1:n}$, $\overline{\RV{y}}_{1:n}$, $ \RV{a}_{1:n} $, $ \RV{b}_{1:n} $, and the number of measurements $\RV{m}_{1:n}$ is given by 
\begin{align}
	&f(\V{z}_{1:n}|\underline{\V{y}}_{1:n}, \overline{\V{y}}_{1:n}, \V{a}_{1:n}, \V{b}_{1:n}, \V{m}_{1:n}) \nn \\
	& \quad = \prod_{n'=1}^{n} f(\V{z}_{n'}|\underline{\V{y}}_{n'}, \overline{\V{y}}_{n'}, \V{a}_{n'}, \V{b}_{n'}, M_{n'})\ist.
	\label{eq:LHF1} 
\end{align}
Note that $ f(\V{z}_{1}|\underline{\V{y}}_{1}, \overline{\V{y}}_{1}, \V{a}_{1}, \V{b}_{1}, M_{1}) = f(\V{z}_{1}|\overline{\V{y}}_{1}, \V{a}_{1}, \V{b}_{1}, M_{1}) $ since no legacy \acp{pbo} exist at time $ n = 1 $. The conditional \ac{pdf} $ f(\V{z}_{m,n}|\V{x}_{k,n}) $ characterizing the statistical relation between the measurements $\RV{z}_{m,n}$ and the \ac{pbo} states $\RV{x}_{k,n}$ is a central element in the conditional \ac{pdf} of the measurement vector $\RV{z}_{n}$ given $\underline{\RV{y}}_{n}$, $\overline{\RV{y}}_{n}$, $ \RV{a}_{n} $, $ \RV{b}_{n} $, and the number of the measurements $\rv{M}_{n}$. Assuming that the measurements $\RV{z}_{m,n}$ are conditionally independent across $m$ given $\underline{\RV{y}}_{k,n}$, $\overline{\RV{y}}_{m,n}$, $\rv{a}_{k,n}$, $\rv{b}_{m,n}$, and $\rv{M}_n$ \cite{BarShalom_AlgorithmHandbook}, Eq.~\eqref{eq:LHF1} factorizes as 
\begin{align}
	& f(\V{z}_{1:n}|\underline{\V{y}}_{1:n}, \overline{\V{y}}_{1:n}, \V{a}_{1:n}, \V{b}_{1:n}, \V{m}_{1:n}) \nn \\
	& \hspace*{2mm} = C(\V{z}_{1}) \left(\prod_{m \in \mathcal{N}_{\overline{\V{r}}_1}} \dfrac{f(\V{z}_{m,1}|\overline{\V{x}}_{m,1})}{f_{\mathrm{fa}}(\V{z}_{m,1})}\right) \nn \\ 
	& \hspace*{6mm} \times \prod_{n' = 2}^{n} C(\V{z}_{n'}) \left(\prod_{k \in \mathcal{D}_{\V{a}_{n'},\underline{\V{r}}_{n'}}}  \dfrac{f(\V{z}_{a_{k,n'},n'}|\underline{\V{x}}_{k,n'})} {f_{\mathrm{fa}}(\V{z}_{a_{k,n'},n'})} \right) \nn \\
	& \hspace*{6mm} \times \left(\prod_{m \in \mathcal{N}_{\overline{\V{r}}_{n'}}} \dfrac{f(\V{z}_{m,n'}|\overline{\V{x}}_{m,n'})} {f_{\mathrm{fa}}(\V{z}_{m,n'})}\right), 
	\label{eq:LHF_global}
\end{align}
and the conditional \ac{pdf} at each time $ n \geq 2 $ factorizes as \cite{BarShalom_AlgorithmHandbook} 
\begin{align}
	& f(\V{z}_{n}|\underline{\V{y}}_{n}, \overline{\V{y}}_{n}, \V{a}_{n}, \V{b}_{n}, M_{n}) \nn \\
	& \hspace*{15mm} = C(\V{z}_{n}) \left(\prod_{k \in \mathcal{D}_{\V{a}_n,\underline{\V{r}}_n}}  \dfrac{f(\V{z}_{a_{k,n},n}|\underline{\V{x}}_{k,n})}{f_{\mathrm{fa}}(\V{z}_{a_{k,n},n})}\right)  \nn \\
	& \hspace*{19mm} \times \left(\prod_{m \in \mathcal{N}_{\overline{\V{r}}_n}} \dfrac{f(\V{z}_{m,n}|\overline{\V{x}}_{m,n})}{f_{\mathrm{fa}}(\V{z}_{m,n})}\right)\ist.
	\label{eq:LHF2} 
\end{align} 
Since the normalization factor $ C(\V{z}_{n}) = \prod_{m=1}^{M_n}f_{\mathrm{fa}}(\V{z}_{m,n}) $ depending on $ \V{z}_{n} $ and $ M_n $ is fixed after the current measurement $ \V{z}_{n} $ is observed, the likelihood function in \eqref{eq:LHF2} can be rewritten up to the normalization constant as
\begin{align}
	& f(\V{z}_{n}|\underline{\V{y}}_{n}, \overline{\V{y}}_{n}, \V{a}_{n}, \V{b}_{n}, M_{n}) \nn \\
	& \propto \left(\prod_{k = 1}^{K_{n-1}} g_{2}(\underline{\V{y}}_{k,n}, a_{k,n}; \V{z}_{n})\right) \hspace*{-1mm}
	\vphantom{\prod_{k = 1}^{K_{n-1}}} \left( \prod_{m = 1}^{M_{n}} h_{2}(\overline{\V{y}}_{m,n}, b_{m,n}; \V{z}_{n}) \vphantom{\prod_{k = 1}^{K_{n-1}}}  \right) \rule[-1.8em]{0pt}{0pt}
	\label{eq:LHF3} 
\end{align} 
where the factor related to legacy \ac{pbo} states $ g_{2}(\underline{\V{y}}_{k,n}, a_{k,n};\\ \V{z}_{n}) \hspace*{-0.4mm} = \hspace*{-0.4mm} g_{2}(\underline{\V{x}}_{k,n}, \underline{r}_{k,n}, a_{k,n}; \V{z}_{n}) $ is given by 
\begin{align}
	& g_{2}(\underline{\V{x}}_{k,n}, \underline{r}_{k,n} = 1, a_{k,n}; \V{z}_{n}) \nn \\
	& \hspace*{20mm} \triangleq
	\begin{cases}
		\dfrac{f(\V{z}_{m,n}|\underline{\V{x}}_{k,n})}{f_{\mathrm{fa}}(\V{z}_{m,n})}, 	& a_{k,n} = m \\
		1,																				& a_{k,n} = 0
	\end{cases}
	\label{eq:g2} 
\end{align}
and $ g_{2}(\underline{\V{x}}_{k,n}, \underline{r}_{k,n} = 0, a_{k,n}; \V{z}_{n}) \triangleq 1 $. The factor related to new \ac{pbo} states $h_{2}(\overline{\V{y}}_{m,n}, b_{m,n}; \V{z}_{n})= h_{2}(\overline{\V{x}}_{m,n}, \overline{r}_{m,n}, b_{m,n};\\ \V{z}_{n}) $ is given by 
\begin{align}
	& h_{2}(\overline{\V{x}}_{m,n}, \overline{r}_{m,n} = 1, b_{m,n}; \V{z}_{n}) \nn \\
	& \hspace*{22mm} \triangleq 
	\begin{cases}
		1, 																				& b_{m,n} = k \\
		\dfrac{f(\V{z}_{m,n}|\overline{\V{x}}_{m,n})}{f_{\mathrm{fa}}(\V{z}_{m,n})}, 	& b_{m,n} = 0
	\end{cases}
	\label{eq:h2}
\end{align}
and $ h_{2}(\overline{\V{x}}_{m,n}, \overline{r}_{m,n} = 0, b_{m,n}; \V{z}_{n}) \triangleq 1 $. Inserting \eqref{eq:LHF3} \eqref{eq:g2} and \eqref{eq:h2} into \eqref{eq:LHF1}, the conditional \ac{pdf} can be rewritten as the joint likelihood function 
\begin{align}
	& f(\V{z}_{1:n}|\underline{\V{y}}_{1:n}, \overline{\V{y}}_{1:n}, \V{a}_{1:n}, \V{b}_{1:n}, \V{m}_{1:n})
	\nn \\
	&\hspace*{7mm}\propto \left(\prod_{m = 1}^{M_{1}} h_{2}(\overline{\V{x}}_{m,1}, \overline{r}_{m,1}, b_{m,1}; \V{z}_{1})\right) \nn \\
	& \hspace*{11mm} \times \prod_{n' = 2}^{n} \left(\prod_{k = 1}^{K_{n'-1}} g_{2}(\underline{\V{x}}_{k,n'}, \underline{r}_{k,n'}, a_{k,n'}; \V{z}_{n'})\right) 
	\nn \\ 
	& \hspace*{11mm} \times \left(\prod_{m' = 1}^{M_{n'}} h_{2}(\overline{\V{x}}_{m',n'}, \overline{r}_{m',n'}, b_{m',n'}; \V{z}_{n'})\right)\ist.
	\label{eq:LHF4} 
\end{align}

\subsection{Joint Posterior \ac{pdf}}
\label{app:JointPosterior}

Finally, by substituting the joint prior \ac{pdf} with \eqref{eq:jointPriorPDF_global_factorized} and the joint likelihood function with \eqref{eq:LHF4}, the joint posterior \ac{pdf} \eqref{eq:TheJointPosteriorPDF_1} can be rewritten as 
\begin{align}
	& f(\V{y}_{1:n}, \V{a}_{1:n}, \V{b}_{1:n}, {\V{\mu}_{\mathrm{fa}}}_{1:n}, \V{m}_{1:n} | \V{z}_{1:n})\nn \\ 
	& \hspace*{-1mm} \propto f({\mu_{\mathrm{fa}}}_{1}) \left(\prod_{l = 1}^{M_{1}} h_{1}(\overline{\V{y}}_{l,1}, b_{l,1}, {\mu_{\mathrm{fa}}}_{1}; M_{1}) h_{2}(\overline{\V{y}}_{l,1}, b_{l,1}; \V{z}_{1}) \right) \nn \\
	& \times \prod_{n' = 2}^{n} f({\mu_{\mathrm{fa}}}_{n'}|{\mu_{\mathrm{fa}}}_{n'-1}) \left( \prod_{k' = 1}^{K_{n'-1}} f(\underline{\V{y}}_{k',n'}|\V{y}_{k',n'-1})\right) \nn \\
	& \times \left( \prod_{k = 1}^{K_{n'-1}} g_{1}(\underline{\V{y}}_{k,n'}, a_{k,n'}, {\mu_{\mathrm{fa}}}_{n'}; M_{n'})g_{2}(\underline{\V{y}}_{k,n'}, a_{k,n'}; \V{z}_{n'}) \right. \nn \\
	& \times \left. \prod_{m = 1}^{M_{n'}}\psi(a_{k,n'},b_{m,n'}) \hspace*{-0.5mm} \right) \hspace*{-1.5mm} \left( \prod_{m' = 1}^{M_{n'}} h_{1}(\overline{\V{y}}_{m',n'}, b_{m',n'}, {\mu_{\mathrm{fa}}}_{n'}; M_{n'}) \right. \nn \\
	& \hspace*{1mm} \times \left.  h_{2}(\overline{\V{y}}_{m',n'}, b_{m',n'}; \V{z}_{n'}) \vphantom{\prod_{m' = 1}^{M_{n'}}} \right)\ist.
	\label{eq:TheJointPosteriorPDF_complete} 
\end{align}
The factors related to the legacy \acp{pbo} and to the new \acp{pbo} can be simplified as $ g(\underline{\V{y}}_{k,n}, a_{k,n}, {\mu_{\mathrm{fa}}}_{n}; \V{z}_{n})\triangleq g_{1}(\underline{\V{y}}_{k,n}, a_{k,n}, {\mu_{\mathrm{fa}}}_{n}; M_{n}) g_{2}(\underline{\V{y}}_{k,n}, a_{k,n}; \V{z}_{n})$ and $ h(\overline{\V{y}}_{m,n}, b_{m,n},\\ {\mu_{\mathrm{fa}}}_{n}; \V{z}_{n}) \triangleq h_{1}(\overline{\V{y}}_{m,n}, b_{m,n}, {\mu_{\mathrm{fa}}}_{n}; M_{n}) h_{2}(\overline{\V{y}}_{m,n}, b_{m,n}; \V{z}_{n}) $, respectively (see Fig.~\ref{fig:factorGraph}). 

}{}

\section{Sum-Product Algorithm (SPA) Messages} \label{sec:spa_messages}

\begin{figure*}[t]
	\centering
	\tikzsetnextfilename{factor_graph_full}
	\scalebox{0.96}{
		\includegraphics{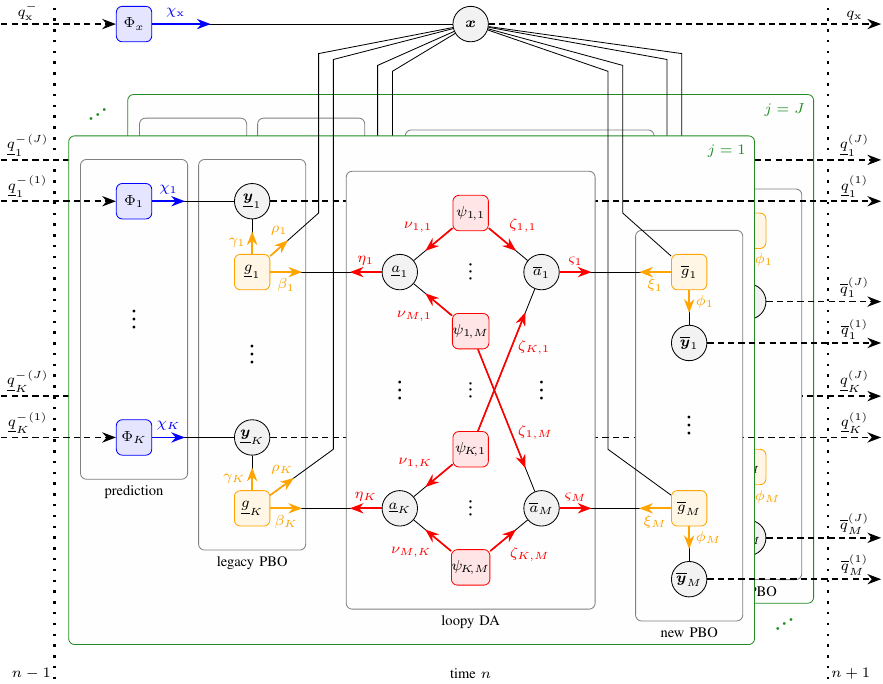}}
	\caption{Factor graph corresponding to the factorization shown in \meqref{eq:factorization1}. Dashed arrows represent messages that are only passed in one direction. The following short notations are used: 
		$ K \triangleq K_{n-1}^{(j)} $, 
		$ M \triangleq M_{n}^{(j)} $; 
		\emph{variable nodes}: 
		$ \underline{a}_{k} \triangleq \underline{a}_{k,n}^{(j)} $, 
		$ \overline{a}_{m} \triangleq \overline{a}_{m,n}^{(j)} $, 
		$ \V{x}\triangleq \V{x}_{n} $,
		$ \underline{\V{y}}_{k} \triangleq \underline{\V{y}}_{k,n}^{(j)} $, 
		$ \overline{\V{y}}_{m} \triangleq \overline{\V{y}}_{m,n}^{(j)} $; 
		\emph{factor nodes}: 
		$ \Phi_{x} \triangleq \Phi_x(\V{x}_{n} | \V{x}_{n} ) $, 
		$ \Phi_{k} \triangleq \Phi_k(\underline{\V{y}}_{k,n}^{(j)} |\V{y}_{k,n-1}^{(j)} ) $, 
		$ \underline{g}_{k} \triangleq \underline{g}( \V{x}_{n}, \underline{\V{\psi}}^{(j)}_{k,n} , \underline{r}^{(j)}_{k,n}, \underline{a}^{(j)}_{k,n}; \V{z}^{(j)}_{n} ) $, 
		$ \overline{g}_{m} \triangleq  \overline{g}( \V{x}_{n}, \overline{\V{\psi}}^{(j)}_{m,n} , \overline{r}^{(j)}_{m,n}, \overline{a}^{(j)}_{m,n}; \V{z}^{(j)}_{n} ) $, 
		$ \psi_{k,m} \triangleq \psi(\underline{a}_{k,n}^{(j)},\overline{a}_{m,n}^{(j)}) $; 
		\emph{prediction}: 
		$ \chi_{k} \triangleq \chi( \underline{\V{\psi}}_{k,n}^{(j)},  \underline{\rv{r}}_{k,n}^{(j)} ) $, 
		$ \chi_\text{x} \triangleq \chi_\text{x}( \V{x}_n)$; 
		\emph{measurement evaluation}: 
		$ \beta_{k} \triangleq \beta(\underline{a}_{k,n}^{(j)}) $, $ \xi_{m} \triangleq \xi(\overline{a}_{m,n}^{(j)}) $; 
		\emph{loopy DA}: 
		$ \nu_{m,k} \triangleq \nu_{m \rightarrow k}(\underline{a}_{k,n}^{(j)}) $, 
		$ \zeta_{k,m} \triangleq \zeta_{k \rightarrow m}(\overline{a}_{m,n}^{(j)}) $, 
		$ \eta_{k} \triangleq \eta(\underline{a}_{k,n}^{(j)}) $, 
		$ \varsigma_{m} \triangleq \varsigma(\overline{a}_{m,n}^{(j)}) $; 
		\emph{measurement update}: 
		$ \gamma_{k} \triangleq \gamma(  \underline{\V{\psi}}_{k,n}^{(j)},  \underline{\rv{r}}_{k,n}^{(j)}  ) $, 
		$ \rho_{k} \triangleq \rho_{k}^{(j)}(  \V{x}_n ) $, 
		$ \phi_{m} \triangleq \phi( \overline{\V{\psi}}^{(j)}_{m,n}, \overline{r}_{m,n}^{(j)}) $, 
		$ \kappa_{m} \triangleq \kappa_{m}^{(j)}(  \V{x}_n ) $; 
		\emph{belief calculation:} 
		$q_\mathrm{x} \triangleq q(\V{x}_{n}), 
		\underline{q}_{k}^{(j)} \triangleq {q}(\underline{\V{y}}_{k,n}^{(j)}),
		\overline{q}_{m}^{(j)} \triangleq	{q}(\overline{\V{y}}_{m,n}^{(j)}),
		q_\mathrm{x}^{-} \triangleq q(\V{x}_{n\minus 1}), 
		\underline{q}_{k}^{- (j)} \triangleq {q}(\underline{\V{y}}_{k,n\minus 1}^{(j)})
		$. 
		\vspace{-4.5mm}}	 
	\label{fig:factorGraphFull}
\end{figure*}

As in \cite{MeyerJTSP2017, MeyerProc2018, LeitingerICC2017, LeitingerTWC2019} we perform loopy message passing (belief propagation) on the factor graph\footnote{Note that the factor graphs shown in \mref{Fig.}{fig:factorGraph} and Fig.~\ref{fig:factorGraphFull} are identical up to the intermediate \ac{spa} messages only shown in Fig.~\ref{fig:factorGraphFull} for visualization of the resulting algorithm.}
shown in Fig.~\ref{fig:factorGraphFull} by means of the sum-product algorithm rules \cite{KschischangTIT2001,Loeliger2004SPM}. 
 Due to the loops in the factor graph, the resulting beliefs $ q(\V{x}_{n}) $, $ {q}(\underline{\V{y}}_{k,n}^{(j)}) = {q}(\underline{\V{\psi}}_{k,n}^{(j)}, \underline{r}_{k,n}^{(j)}) $, 
and
$ {q}(\overline{\V{y}}_{m,n}^{(j)}) = {q}(\overline{\V{\psi}}_{m,n}^{(j)}, \overline{r}_{m,n}^{(j)}) $ are only approximations of the respective posterior marginal \acp{pdf}, and there is no canonical order in which the messages should be computed \cite{KschischangTIT2001}. 
For the proposed algorithm, we choose the order according to the following rules: (i) messages are only sent forward in time; (ii)  iterative message passing is only performed for data association at each time $ n ${, i.e., messages sent from an agent node $\V{x}_n$ do only depend on the predicted agent belief, not on the beliefs sent from the anchors; (iii) along an edge connecting an agent node $\V{x}_n$ and a new \ac{pbo} node $ \overline{\V{y}}_{m,n}^{(j)}$, messages are only sent from the former to the latter.} 

We obtain the following operations (termed ``messages") that are performed at each time $n$ in {the stated order}.

We note that similarly to the ``dummy \acp{pdf}'' introduced in \mref{Sec.}{sec:pbo_states}, we consider messages $\varphi(\cdot)$ of the form $\varphi\big(\V{y}^{(j)}_{k,n}\big) \rmv\rmv=\rmv\rmv \varphi\big( {\V{\psi}}_{k,n}^{(j)} \ist, r^{(j)}_{k,n}\big)$ for non-existing \ac{pbo} states, i.e., for $r^{(j)}_{k,n} \!=\rmv 0$. We define these messages by $\varphi\big( \underline{\V{\psi}}_{k,n}^{(j)} \ist, r^{(j)}_{k,n} \!\!=\!\! 0\big) \triangleq \varphi^{(j)}_{k,n}$. Note that messages are not \acp{pdf} and, thus, are not required to integrate to $1$. 

\subsection{Prediction}
First, a \textit{prediction step} is performed for the agent and all legacy \acp{pbo} $k \rmv\in\rmv \{1,\dots, K_{n-1}^{(j)}\} \triangleq \mathcal{K}_{n-1}^{(j)}$ of all anchors $ j \rmv\in\rmv \mathcal{J} \triangleq \{1,..., J\}$. Based on the \ac{spa} rule, the prediction message for the agent state is given by
\begin{equation}
	\chi_\text{x}( \V{x}_n) =\rmv \int \rmv f(\V{x}_n|\V{x}_{n-1}) \ist q(\V{x}_{n-1}) \ist \mathrm{d}\V{x}_{n-1} \ist,
	\label{eq:stateTransitionMessageAgent}
\end{equation}
and the prediction message for the legacy \acp{pbo} is given by%
\begin{align}
	\chi\big( \underline{\V{\psi}}_{k,n}^{(j)},  \underline{\rv{r}}_{k,n}^{(j)} \big) &=\hspace{-4mm}  \sum_{r^{(j)}_{k,n-1} \in \{0,1\}}\int \rmv\rmv  f\big(  \underline{\V{\psi}}_{k,n}^{(j)},  \underline{\rv{r}}_{k,n}^{(j)}  \big| {\V{\psi}}_{k,n-1}^{(j)},  {\rv{r}}_{k,n-1}^{(j)} \big) \nn \\
	&\hspace{5mm}\times {q}\big( {\V{\psi}}_{k,n-1}^{(j)},  {\rv{r}}_{k,n-1}^{(j)}  \big) \ist \mathrm{d} {\V{\psi}}_{k,n-1}^{(j)}  \ist,
	\label{eq:stateTransitionMessageFeature} 
\end{align}
 where the beliefs of the mobile agent state, $q(\V{x}_{n-1})$,\linebreak and of the PF states, ${q}\big( {\V{\psi}}_{k,n-1}^{(j)},  {\rv{r}}_{k,n-1}^{(j)}   \big)$, were calculated at the preceding time $n \rmv-\! 1$. 
Inserting \meqref{eq:state_transition_pdf1} and \meqref{eq:state_transition_pdf2} for $f\big( \underline{\V{\psi}}_{k,n}^{(j)},  \underline{\rv{r}}_{k,n}^{(j)}  \big| {\V{\psi}}_{k,n\minus 1}^{(j)}, 1\big)$ and $f\big( \underline{\V{\psi}}_{k,n}^{(j)},  \underline{\rv{r}}_{k,n}^{(j)}  \big| {\V{\psi}}_{k,n\minus 1}^{(j)},$ $ 0\big)$, respectively, we obtain for $\underline{\rv{r}}_{k,n}^{(j)}  \!=\! 1$
\begin{equation}
	\chi\big(  \underline{\V{\psi}}_{k,n}^{(j)}  \ist, 1\big) 
	=\ist p_{\mathrm{s}} \rmv\rmv\rmv \int \rmv\rmv\rmv f\big( \underline{\V{\psi}}_{k,n}^{(j)} \big| {\V{\psi}}_{k,n-1}^{(j)} \big) \ist {q}\big( {\V{\psi}}_{k,n-1}^{(j)} \ist, 1 \big) \ist \mathrm{d} {\V{\psi}}_{k,n-1}^{(j)} \ist, 
	\label{eq:stateTransitionMessageFeature_r1}
	\vspace{-1mm}
\end{equation}
and for $\underline{\rv{r}}_{k,n}^{(j)} \!=\rmv 0$ we get $	\chi\big(  \underline{\V{\psi}}_{k,n}^{(j)}  \ist, 0\big)  = \chi_{k,n}^{(j)}\,  f_\mathrm{d}\big( \underline{\V{\psi}}_{k,n}^{(j)} \big)$ with
\begin{align}
	\chi_{k,n}^{(j)} = (1 \!-\rmv p_\mathrm{s}\big) \rmv\int\rmv {q}\big( {\V{\psi}}_{k,n-1}^{(j)} \ist, 1 \big) \ist \mathrm{d}{\V{\psi}}_{k,n-1}^{(j)} +\ist q_{k,n-1}^{(j)}  \ist.
	\label{eq:stateTransitionMessageFeature_r0}\\[-7mm]\nn
\end{align}
We note that $q_{k,n-1}^{(j)}  \triangleq \int \rmv {q}\big( {\V{\psi}}_{k,n-1}^{(j)} \ist, 0 \big)  \ist \mathrm{d} {\V{\psi}}_{k,n-1}^{(j)} $ approximates the probability of non-existence of legacy \ac{pbo} $k$ for anchor $j$ at the previous time step $n\rmv-\rmv1$. 

\subsection{Parallel Update of \ac{pbo} States and Agent State}
	
The following calculations are performed in parallel for all %
anchors $j \rmv\in\rmv \mathcal{J}$. 

	\subsubsection{Measurement evaluation for legacy \acp{pbo}} 
	The messages $\beta\big( \underline{a}_{k,n}^{(j)} \big)$ passed from the factor nodes $\underline{g}\big( \V{x}_{n}, \underline{\V{\psi}}^{(j)}_{k,n} , \underline{r}^{(j)}_{k,n},$ $ \underline{a}^{(j)}_{k,n}; \V{z}^{(j)}_{n} \big)$ to the variable nodes corresponding to the feature-oriented 
	association variables $\underline{a}^{(j)}_{k,n}$ are calculated as
	\begin{align}
		\nn\\[-5mm]
		\beta\big( \underline{a}_{k,n}^{(j)} \big) &\rmv\rmv\rmv=\rmv\rmv\rmv\rmv\rmv\rmv  \int\!\!\! \rmv \rmv \int \rmv\rmv\rmv\rmv 	\chi\big(  \underline{\V{\psi}}_{k,n}^{(j)}  ,\rmv 1\big)  \ist\chi_\text{x}\big( \V{x}_n \big) \ist\underline{g}\big( \V{x}_{n}, \underline{\V{\psi}}^{(j)}_{k,n} ,\rmv 1, \underline{a}^{(j)}_{k,n};\rmv \V{z}^{(j)}_{n} \big) \nn \\
		&\rmv\rmv\rmv \hspace{5mm}\times \mathrm{d}\V{x}_{n} \ist \mathrm{d} \underline{\V{\psi}}^{(j)}_{k,n} \ist+\ist 1_{\{0\}}\big(\underline{a}_{k,n}^{(j)}\big)\ist \chi_{k,n}^{(j)} \ist. 
		\label{eq:bp_measevalutionLF}
		\\[-5mm]\nn
	\end{align}
	for all legacy \acp{pbo} $k \rmv\in\rmv \Set{K}_{n-1}^{(j)}$.
	
	\subsubsection{Measurement evaluation for new \acp{pbo}}
	The messages $\xi\big(\overline{a}^{(j)}_{m,n}\big)$ passed from the factor nodes 	$\overline{g}\big( \V{x}_{n}, \overline{\V{\psi}}^{(j)}_{m,n} , \overline{r}^{(j)}_{m,n}, \overline{a}^{(j)}_{m,n}; \V{z}^{(j)}_{n} \big)$ to the variable nodes corresponding to the measurement-oriented DA variables $\overline{a}^{(j)}_{m,n}$ are calculated as
	\begin{align}
		\xi\big(\overline{a}^{(j)}_{m,n}\big) &=\rmv\rmv\rmv \sum_{\overline{r}^{(j)}_{m,n} \in \{0,1\}} \rmv\rmv\rmv \int\!\!\!\int \rmv 
		\overline{g}\big( \V{x}_{n}, \overline{\V{\psi}}^{(j)}_{m,n} , \overline{r}^{(j)}_{m,n}, \overline{a}^{(j)}_{m,n}; \V{z}^{(j)}_{n} \big) \nn \\
		&\hspace{6mm}\times \chi_\text{x}\big( \V{x}_n \big)  \, \mathrm{d}\V{x}_{n} \ist \mathrm{d} \overline{\V{\psi}}^{(j)}_{m,n} \ist .
		\label{eq:bp_measevalutionNF1}\\[-6.5mm]
		\nn
	\end{align}
 	for all new \acp{pbo}  $m \rmv\in\rmv \{1,\,...\,,M_n^{(j)}\} \triangleq \Set{M}_n^{(j)}$. 
	Using the expression of 
	$\overline{g}\big( \V{x}_{n}, \overline{\V{\psi}}^{(j)}_{m,n} , \overline{r}^{(j)}_{m,n}, \overline{a}^{(j)}_{m,n}; \V{z}^{(j)}_{n} \big) $ stated in \eqref{eq:bp_measevalutionNF1} is easily seen to simplify to $\xi\big(\overline{a}^{(j)}_{m,n}\big) \!=\! 1$
	for $\overline{a}^{(j)}_{m,n} \!\in\! \Set{K}^{(j)}_{n-1}$, and for $\overline{a}^{(j)}_{m,n} \!=\rmv 0$ it becomes
	\begin{align}
		\xi\big(\overline{a}^{(j)}_{m,n}\big) &= 1 + 
		\frac{ {\mu}_{\mathrm{n}} }{ \mu_\mathrm{fa} \ist f_{\mathrm{fa}}\big( \V{z}^{(j)}_{m,n} \big) } 
		\int\!\!\!\int\rmv \chi_\text{x}(\V{x}_n) \ist \ist f_{\mathrm{n}}\big(\overline{\V{\psi}}_{m,n}^{(j)}\big) \nn \\[1.5mm]
		&\hspace{5mm}\times  f\big(\V{z}_{m,n}^{(j)}| \V{x}_n, \overline{\V{\psi}}_{m,n}^{(j)}\big) \, \mathrm{d}\V{x}_{n} \ist \mathrm{d} \overline{\V{\psi}}_{m,n}^{(j)} \ist .
		\label{eq:bp_measevalutionNF}
	\end{align}

	\subsubsection{Iterative data association} \label{sec:iterative_da}
	From $\beta\big( \underline{a}_{k,n}^{(j)} \big)$ and $\xi\big(\overline{a}_{m,n}^{(j)}\big)$ the messages 
	$\eta\big( \underline{a}_{k,n}^{(j)} \big)$ and $\varsigma\big( \overline{a}_{m,n}^{(j)} \big)$ are obtained by means of iterative message passing according to \cite{WilliamsLauTAE2014,MeyerJTSP2017}.
	Iteratively, we first calculate for each $m \rmv\in\rmv \Set{M}_n^{(j)}$  the messages 
	\begin{align}
		\nu_{m\rightarrow k}^{(p)}\big(\underline{a}_{k,n}^{(j)}\big) 
		&=\! \sum^{K_{n-1}^{(j)}}_{\overline{a}_{m,n}^{(j)} = 0} \!\!\xi\big( \overline{a}_{m,n}^{(j)} \big) \ist \psi\big(\underline{a}_{k,n}^{(j)}, \overline{a}_{m,n}^{(j)}\big)  \nn \\[-1mm]
		&\hspace{5mm}\times\rmv\rmv\rmv\rmv\prod_{k' \in \Set{K}^{(j)}_{n-1}\backslash\{k\}} \!\!\! \zeta_{k' \rightarrow m}^{(p-1)}\big(\overline{a}_{m,n}^{(j)}\big) \ist
		\label{eq:featureDARV}
	\end{align}
	passed from variable nodes $ \overline{a}_{m,n}^{(j)}$ over the factor nodes $ \psi(\underline{a}_{k,n}^{(j)},\overline{a}_{m,n}^{(j)})$ to the variable nodes  $\underline{a}^{(j)}_{k,n}$. 
	Then, we calculate for each $k \rmv\in\rmv \Set{K}_{n-1}^{(j)}$  the messages 
	\begin{align}
		\zeta_{k \rightarrow m}^{(p)}\big(\overline{a}_{m,n}^{(j)}\big) 
		&=\! \sum^{M_n^{(j)}}_{\underline{a}_{k,n}^{(j)} = 0} \!\!\beta\big( \underline{a}_{k,n}^{(j)} \big) \ist 
		\psi\big(\underline{a}_{k,n}^{(j)}, \overline{a}_{m,n}^{(j)}\big)  \nn \\[-1mm] 
		&\hspace{5mm}\times\rmv\rmv\rmv\rmv\prod_{m' \in \Set{M}_n^{(j)}\backslash\{m\}} \!\!\! \nu_{m'\rightarrow k}^{(p)}\big(\underline{a}_{k,n}^{(j)}\big) \ist
		\label{eq:measurementDARV} \\[-5mm]
		\nn
	\end{align}
	passed from variable nodes $\underline{a}^{(j)}_{k,n}$ over the factor nodes $ \psi(\underline{a}_{k,n}^{(j)},\overline{a}_{m,n}^{(j)})$ to the variable nodes  $ \overline{a}_{m,n}^{(j)}$.  
	The recursion defined by \eqref{eq:featureDARV} and \eqref{eq:measurementDARV} is repeated for each iteration index $p \rmv \in \rmv \{1,\dots, P\} $ and initialized (for $p \!=\!  0$) by	
	\begin{equation} \label{eq:featureDARVInit}
		\nu_{m\rightarrow k}^{(0)}\big(\underline{a}_{k,n}^{(j)}\big)\rmv =\rmv \sum^{K_{n-1}^{(j)}}_{\overline{a}_{m,n}^{(j)} = 0} \xi\big( \overline{a}_{m,n}^{(j)} \big) \, 
		\psi\big(\underline{a}_{k,n}^{(j)}, \overline{a}_{m,n}^{(j)}\big)  \ist . 
	\end{equation}
	After the last iteration $p \rmv=\rmv P\rmv$, the messages $\eta\big( \underline{a}_{k,n}^{(j)} \big)$ and $\varsigma\big( \overline{a}_{m,n}^{(j)} \big)$ are calculated as
	\begin{align}
		\eta\big( \underline{a}_{k,n}^{(j)} \big) &=\! \prod_{m \in \Set{M}_n^{(j)}} \!\!\! \nu_{m\rightarrow k}^{(P)}\big(\underline{a}_{k,n} ^{(j)}\big) \\[1mm] 
		\varsigma\big( \overline{a}_{m,n}^{(j)} \big) &=\! \prod_{k \in \Set{K}_{n-1}^{(j)}} \!\!\rmv \zeta_{k \rightarrow m}^{(P)}\big(\overline{a}_{m,n}^{(j)}\big) \ist.
		\label{eq:messagesDA}\\[-7mm]\nn
	\end{align}

	Note that the iterative data association scheme presented in this section, is simplified for implementation in accordance with \cite{WilliamsLauTAE2014, MeyerProc2018}. This step significantly reduces the computational cost from $\mathcal{O}({M}_n^{(j)\s 2}\, {K}_{n-1}^{(j) \s 2} )$ to $\mathcal{O}({M}_n^{(j)}\, {K}_{n-1}^{(j)} )$ per iteration, as shown in \cite{WilliamsTAES2014}. 
	
	\subsubsection{Measurement update for the agent} 
	From $\eta\big( \underline{a}_{k,n}^{(j)} \big)$, $\chi\big( \underline{\V{\psi}}_{k,n}^{(j)}, 1\big)$, and $ \chi_{k,n}^{(j)}\ist$, 
	the message $\rho_{k}^{(j)}\big(  \V{x}_n \big) $ related to the agent is obtained as
	\begin{align}
		\hspace{-2mm} \rho_{k}^{(j)}\big(  \V{x}_n \big)  &=\! \sum^{M_n^{(j)}}_{\underline{a}_{k,n}^{(j)} = 0} \!\! \eta\big( \underline{a}_{k,n}^{(j)} \big) 
		\rmv\int \! \underline{g}\big( \V{x}_{n}, \underline{\V{\psi}}^{(j)}_{k,n} ,\rmv 1, \underline{a}^{(j)}_{k,n};\rmv \V{z}^{(j)}_{n} \big) \nn \\
		&\hspace{2mm}\times \chi\big( \underline{\V{\psi}}_{k,n}^{(j)} , 1\big) \mathrm{d} \underline{\V{\psi}}_{k,n}^{(j)}
		\ist+\ist \eta\big( \underline{a}_{k,n}^{(j)} \!\rmv=\! 0\big) \ist\chi_{k,n}^{(j)} \ist
		\label{eq:measurementUpdateAgent}\\[-4mm]\nn
	\end{align}
		for all legacy \acp{pbo} $k \rmv\in\rmv \Set{K}_{n-1}^{(j)}$.
		
	\subsubsection{Measurement update for legacy \acp{pbo}} 
	Similarly, the messages $\gamma\big(  \underline{\V{\psi}}_{k,n}^{(j)},  \underline{\rv{r}}_{k,n}^{(j)}  \big)$ related to the legacy \acp{pbo} are given by
	\begin{align}
		\label{eq:measurementUpdateLegacy1}
	\gamma\big(  \underline{\V{\psi}}_{k,n}^{(j)},  1 \big) &=\! \sum^{M_n^{(j)}}_{\underline{a}_{k,n}^{(j)} = 0} \!\! \eta\big( \underline{a}_{k,n}^{(j)} \big) \rmv\int \! 
	\underline{g}\big( \V{x}_{n}, \underline{\V{\psi}}^{(j)}_{k,n} ,\rmv 1, \underline{a}^{(j)}_{k,n};\rmv \V{z}^{(j)}_{n} \big)  \nn \\ 
		&\hspace{5mm}\times \ist \chi_\text{x}(\V{x}_n) \mathrm{d}\V{x}_n \\[1.5mm]
		\gamma_{k,n}^{(j)} &\triangleq\ist \gamma\big(  \underline{\V{\psi}}_{k,n}^{(j)},  0 \big) = \eta\big( \underline{a}_{k,n}^{(j)} \!\rmv=\! 0\big) \ist
		\label{eq:measurementUpdateLegacy2}
	\end{align}
	for all legacy \acp{pbo} $k \rmv\in\rmv \Set{K}_{n-1}^{(j)}$.
		
	\subsubsection{Measurement update for new \acp{pbo}} 
	Finally, the messages $\phi\big( \overline{\V{\psi}}^{(j)}_{m,n}, \overline{r}_{m,n}^{(j)} \big)$ related to the new \acp{pbo}
	are calculated as
	\begin{align}
		\label{eq:measurementUpdateNew1}
		\phi\big( \overline{\V{\psi}}^{(j)}_{m,n}, 1 \big) &=\ist \varsigma\big( \overline{a}_{m,n}^{(j)} \!\rmv=\! 0\big) \rmv\int\!  
		\overline{g}\big( \V{x}_{n}, \overline{\V{\psi}}^{(j)}_{m,n} , 1 , 0 ; \V{z}^{(j)}_{n} \big)   \, \nn \\
		&\hspace{5mm}\times
		\ist \chi_\text{x}(\V{x}_n) \mathrm{d}\V{x}_n \\[.5mm] 
		\phi_{m,n}^{(j)} &\triangleq\ist \phi\big( \overline{\V{\psi}}^{(j)}_{m,n},0 \big)  =\rmv \sum^{K_{n-1}^{(j)}}_{\overline{a}_{m,n}^{(j)} = 0} \!\! \varsigma\big( \overline{a}_{m,n}^{(j)}\big) \ist.
		\label{eq:measurementUpdateNew2}
	\end{align}
 	for all new \acp{pbo}  $m \rmv\in\rmv  \Set{M}_n^{(j)}$. 

\subsection{Belief Calculation}

Finally, the beliefs approximating the desired marginal posterior \acp{pdf} can be obtained. The belief of the agent state is given %
by 
\begin{equation} \label{eq:beliefAgent}
	q(\V{x}_n) \ist=\ist \frac{1}{C_{\text{x}\s n}}\chi_\text{x}( \V{x}_n) \prod_{j=1}^J \prod_{k \in \Set{K}_{n-1}^{(j)}} \!\!\rmv \rho^{(j)}_k( \V{x}_n) \ist
\end{equation}
with $C_{\text{x}\s n} = \int \prod_{j=1}^J \prod_{k \in \Set{K}_{n-1}^{(j)}} \!\!\rmv \gamma^{(j)}_k( \V{x}_n) \mathrm{d} \V{x}_n  $ being a normalization constant. 
The belief $q(\V{x}_n)$ provides an approximation of the marginal posterior pdf $f(\V{x}_n |\V{z}_{1:n})$, and it is used instead of $f(\V{x}_n |\V{z}_{1:n})$ in \meqref{eq:mmse}.
The beliefs ${q}\big(\underline{\V{y}}_{k,n}^{(j)}\big) = {q}\big(\underline{\V{\psi}}_{k,n}^{(j)}, \underline{r}_{k,n}^{(j)} \big)$ for the augmented states of the legacy \acp{pbo} $\underline{\V{y}}_{k,n}^{(j)}$,  $k \rmv\in\rmv \Set{K}_{n-1}^{(j)}$, are calculated as 
\begin{align}
	\label{eq:beliefLegacy1}
	{q}\big( \underline{\V{\psi}}_{k,n}^{(j)}, 1 \big) &\ist=\ist \frac{1}{\underline{C}_{k,n}^{(j)}} \chi\big( \underline{\V{\psi}}_{k,n}^{(j)}, 1\big) \ist \gamma\big( \underline{\V{\psi}}_{k,n}^{(j)},1\big)\\[1mm]
	{q}_{k,n}^{(j)} &\triangleq\, {q}\big( \underline{\V{\psi}}_{k,n}^{(j)}  ,0\big) \ist=\ist \frac{1}{\underline{C}_{k,n}^{(j)}} \chi_{k,n}^{(j)} \ist \gamma_{k,n}^{(j)}\ist,
	\label{eq:beliefLegacy2}
\end{align}
with $\underline{C}_{k,n}^{(j)} = \int \Big( \chi\big( \underline{\V{\psi}}_{k,n}^{(j)}, 1\big) \ist \gamma\big( \underline{\V{\psi}}_{k,n}^{(j)},1\big)  + \chi_{k,n}^{(j)} \ist \gamma_{k,n}^{(j)} \Big)  \mathrm{d}  \underline{\V{\psi}}_{k,n}^{(j)}  $ being the normalization constant. 
The beliefs ${q}\big(\overline{\V{y}}_{m,n}^{(j)}\big) = {q}\big(\overline{\V{\psi}}_{m,n}^{(j)}, \overline{r}_{m,n}^{(j)}\big) $ for the augmented states of the new \acp{pbo} $\overline{\V{y}}_{m,n}^{(j)}$,  $m \rmv\in\rmv  \Set{M}_n^{(j)}$, are calculated as 
\begin{align}
	\label{eq:beliefNew1}
	{q}\big(\overline{\V{\psi}}_{m,n}^{(j)} ,1\big) &\ist=\ist \frac{1}{\overline{C}_{m,n}^{(j)}} \phi\big( \overline{\V{\psi}}_{m,n}^{(j)} ,1\big) \\
	{q}_{m,n}^{(j)} &\triangleq\, \breve{q}\big( \overline{\V{\psi}}_{m,n}^{(j)} ,0\big) \ist=\ist  \frac{1}{\overline{C}_{m,n}^{(j)}}  \phi_{m,n}^{(j)} \ist, 
	\label{eq:beliefNew2}
\end{align}
with $\overline{C}_{m,n}^{(j)} = \int \Big( \phi\big( \overline{\V{\psi}}_{m,n}^{(j)} ,1\big)  + \phi_{m,n}^{(j)} \Big)  \mathrm{d}  \underline{\V{\psi}}_{k,n}^{(j)}  $ being the normalization constant. 
In particular, ${q}\big( \underline{\V{\psi}}_{k,n}^{(j)}, 1 \big) $ and ${q}\big(\overline{\V{y}}_{m,n}^{(j)} ,1\big)$ approximate the marginal posterior \acp{pdf} 
$f\big({\V{\psi}}_{k,n}^{(j)}, {r}_{k,n}^{(j)}  \!=\! 1\big|\V{z}_{1:n}\big)$, where $k' \!\in \Set{K}_{n-1}^{(j)} \cup \Set{M}_{n}^{(j)}$ used in \meqref{eq:existProb} and \meqref{eq:PSMC_existProb}.

\section{Particle-based Implementation} \label{sec:particle_based_implementation}

\newcommand{\wy}{{{}{w}_{{\mathsf{\bm{\psi}}}}}}
\newcommand{\wx}{{{}{w}_{{{\mathsf{\bm{x}}}}}}}
\newcommand{\wyobar}{{{}\overline{w}_{{\mathsf{\bm{\psi}}}}}}
\newcommand{\wyubar}{{{}\underline{w}_{{\mathsf{\bm{\psi}}}}}}
\newcommand{\wybarhat}{{\hat{\bm{w}}_{{\mathsf{\bm{\psi}}}}}}
\newcommand{\wyhat}{{\hat{w}_{{\mathsf{\bm{\psi}}}}}}
\newcommand{\wxhat}{{\hat{w}_{{{\mathsf{\bm{x}}}}}}}
\newcommand{\wyprime}{{{}{w}^{\s\prime}_{\underline{\mathsf{\bm{\psi}}}}}}
\newcommand{\wxprime}{{{}{w}^{\s\prime}_{{{\mathsf{\bm{x}}}}}}}
\newcommand{\wytilde}{{{}\tilde{w}_{\underline{\mathsf{\bm{\psi}}}}}}
\newcommand{\wxtilde}{{{}\tilde{w}_{{{\mathsf{\bm{x}}}}}}}
\newcommand{\wprime}{{{}{w}^{\s\prime}}}
\newcommand{\wprimeunder}{{{}\underline{w}^{\s\prime}}}
\newcommand{\wprimeprimeover}{{{}\overline{w}^{\s\prime\prime}}}
\newcommand{\wprimeprimeunder}{{{}\underline{w}^{\s\prime\prime}}}
\newcommand{\wprimeprime}{{{}{w}^{\s\prime\prime}}}
\newcommand{\wyprimeprime}{{{}{w}^{\s\prime\prime}_{{\mathsf{\bm{\psi}}}}}}
\newcommand{\wxprimeprime}{{{}{w}^{\s\prime\prime}_{{{\mathsf{\bm{x}}}}}}}
\newcommand{\myn}{n}

Since the integrals involved in the calculations of the messages presented in Sec.~\ref{sec:spa_messages} cannot be obtained analytically, we use a computationally efficient, sequential particle-based message passing implementation following \cite{MeyerJSIPN2016,MeyerJTSP2017,MeyerProc2018} that provides approximate computation. 
In this implementation, each belief $q(\V{x}_n)$, and ${q}\big({\V{y}}_{k,n}^{(j)}\big) = {q}\big({\V{\psi}}_{k,n}^{(j)}, {r}_{k,n}^{(j)} \big)$ for all $k \rmv\in\rmv \Set{K}_{n}^{(j)} \!,\, j \rmv\in\rmv \mathcal{J} $
is represented by a set of particles and corresponding weights %
$\{( {\V{x}}_{n}^{[i]}, \wx{}_{n}^{[i]}) \}_{i=1}^{I} $ and  $\{ ({\V{\psi}}_{k,n}^{(j)\s [i]}, {w_{{\mathsf{\V{\psi}}}}}_{k,n}^{(j)\s [i]}  )\}_{i=1}^{I} $ for all  $k \rmv\in\rmv \Set{K}_{n}^{(j)} \rmv, j \rmv\in\rmv \mathcal{J} $. 
In particular, the beliefs of the augmented \ac{pbo} states ${q}\big({\V{\psi}}_{k,n}^{(j)}, {r}_{k,n}^{(j)} = 1 \big)$ 
are represented by $\{ ({\V{\psi}}_{k,n}^{(j)\s [i]}, \wy{}_{k,n}^{(j)\s [i]}  )\}_{i=1}^{I} $ , and
${q}\big({\V{\psi}}_{k,n}^{(j)}, {r}_{k,n}^{(j)} = 0 \big)$ is 
given implicitly by
the normalization property of ${q}\big({\V{\psi}}_{k,n}^{(j)}, {r}_{k,n}^{(j)} \big)$, i.e., ${q}\big({\V{\psi}}_{k,n}^{(j)}, {r}_{k,n}^{(j)} = 0 \big) = 1 - \int {q}\big({\V{\psi}}_{k,n}^{(j)}, {r}_{k,n}^{(j)} = 1 \big) \ist \mathrm{d} {\V{\psi}}_{k,n}^{(j)}$. 
Contrary to conventional particle filtering \cite{ArulampalamTSP2002,Doucet2005}, the particle weights ${w_{{\mathsf{\V{\psi}}}}}_{k,n}^{(j)\s [i]}$, $i \in \{1,\dots,I\}$ do not sum to one, but define the existence probability estimate \cite{MeyerJTSP2017} 
\begin{equation}
	{p_{\text{e}}}_{n,k}^{(j)} \ist\triangleq\ist \sum^{I}_{i=1} {w_{{\mathsf{\V{\psi}}}}}_{k,n}^{(j)\s [i]} \approx\rmv \rmv\rmv\rmv \int \rmv\rmv\rmv q({\V{\psi}}_{k,n}^{(j)}\ist, r_{k,n}^{(j)} \!=\! 1) \ist\mathrm{d}{\V{\psi}}_{k,n}^{(j)} \ist.
	\label{eq:approxExistProb}
\end{equation}
Note that since the belief $q({\V{\psi}}_{k,n}^{(j)}\ist, r_{k,n}^{(j)} ) $ 
approximates the joint posterior distribution $f({\V{\psi}}_{k,n}^{(j)}\ist, r_{k,n}^{(j)} \big|\V{z}_{1:n})$, it follows that the sum of weights $	{p_{\text{e}}}_{n,k}^{(j)} $ is approximately equal to $p( r_{k,n}^{(j)}\!\!~=\!\!~1\big|\V{z}_{1:n}) $.
\subsection{Prediction} \label{sec:particle_prediction}
The beliefs $q(\V{x}_{n-1})$, and ${q}\big({\V{\psi}}_{k,n-1}^{(j)}, {r}_{k,n-1}^{(j)}  \!=\! 1 \big)$ for all  $k \rmv\in\rmv \Set{K}_{n-1}^{(j)} \rmv, j \rmv\in\rmv \mathcal{J} $ of the previous time step $ n-1 $, are represented by $ I $ particles and corresponding weights, i.e., 
$\{( {\V{x}}_{n-1}^{[i]}, \wx{}_{n-1}^{[i]}) \}_{i=1}^{I} $ and  $\{ ({\V{\psi}}_{k,n-1}^{(j)\s [i]}, {w_{{\mathsf{\V{\psi}}}}}_{k,n-1}^{(j)\s [i]}  )\}_{i=1}^{I} $ for all  $k \rmv\in\rmv \Set{K}_{n-1}^{(j)} \rmv, j \rmv\in\rmv \mathcal{J} $. 
In line with the sampling importance resampling particle filter\cite{ArulampalamTSP2002}, weighted particles 	$\{ ( {\V{x}}_{n}^{\prime\s  [i]}, \wxprime{}_{n}^{[i]} ) \}_{i=1}^{I} $ and  $\{ ( \underline{\V{\psi}}_{k,n}^{\prime\s (j)\s [i]}, \wyprime{}_{k,n}^{(j)\s [i]} ) \}_{i=1}^{I}$ for all  $k \rmv\in\rmv \Set{K}_{n-1}^{(j)} \rmv, j \rmv\in\rmv \mathcal{J} $, representing the messages $\chi_\text{x}\big( {\bm{x}}_n \big)$ and $\chi\big(\underline{\V{\psi}}_{k,n}^{(j)}, \underline{r}_{k,n}^{(j)}  \!=\! 1 \big)$ in \eqref{eq:stateTransitionMessageAgent} and \eqref{eq:stateTransitionMessageFeature_r1} 
are determined in parallel as follows: 
For each particle ${\V{x}}_{n-1}^{[i]}$ with  $ i \in \{1,\dots, I\} $, one particle ${\V{x}}_{n}^{\prime\s [i]}$ with corresponding weights $  \wxprime{}_{n}^{[i]} =  \wx{}_{n-1}^{[i]} $ is drawn from the proposal distribution  $  f({\bm{x}}_n |{\bm{x}}_{n\minus 1}^{[i]})$, where $\sum^{I}_{i=1} \wxprime{}_{n}^{[i]}  = 1$.
Also, for each particle ${\V{\psi}}_{k,n-1}^{(j)\s [i]}$ with  $ i \in \{1,\dots, I\} $, one particle  $\underline{\V{\psi}}_{k,n}^{\prime\s (j)\s [i]}$ with corresponding weights $  \wyprime{}_{k,n}^{(j)\s [i]} = p_{\mathrm{s}}\, \wy{}_{k,n-1}^{(j)\s [i]} $ is drawn from the proposal distribution $f(\underline{\bm{y}}_n^{(j)}| {\bm{y}}_{n\minus 1}^{(j)\s [i]}) $ for all  $k \rmv\in\rmv \Set{K}_{n-1}^{(j)} \rmv, j \rmv\in\rmv \mathcal{J} $. 
Here ${p_{\text{e}}}_{n,k}^{\prime\s (j)} \ist\triangleq\ist \sum^{I}_{i=1}\wyprime{}_{k,n}^{(j)\s [i]}$ according to \eqref{eq:approxExistProb}.

\subsection{Measurement Evaluation}

The following calculations are performed in parallel for all anchors $j \rmv\in\rmv \mathcal{J}$. 

\subsubsection{Measurement evaluation for legacy \acp{pbo}}
	For all  $k \rmv\in\rmv \Set{K}_{n-1}^{(j)} $ we determine 
	from the weighted particles $\{ ( {\V{x}}_{n}^{\prime\s  [i]}, \wxprime{}_{n}^{[i]} ) \}_{i=1}^{I} $ and  $\{ ( \underline{\V{\psi}}_{k,n}^{\prime\s (j)\s [i]}, \wyprime{}_{k,n}^{(j)\s [i]} ) \}_{i=1}^{I}$ that represent the messages $\chi_\text{x}\big( {\bm{x}}_n \big)$ and $\chi\big(\underline{\V{\psi}}_{k,n}^{(j)}, \underline{r}_{k,n}^{(j)}  \!=\! 1 \big)$ a \textit{``stacked state"}  \cite{MeyerPhd2015,MeyerJSIPN2016,LeitingerTWC2019} given by $\{ ( {\V{x}}_{n}^{\prime\s  [i]}, \underline{\V{\psi}}_{k,n}^{\prime\s (j)\s [i]}, \wprime{}_{k,n}^{(j)\s [i]} ) \}_{i=1}^{I} $ that represents the joint distribution $\chi_\text{x}\big( {\bm{x}}_n \big)\, \chi\big(\underline{\V{\psi}}_{k,n}^{(j)}, \underline{r}_{k,n}^{(j)}  \!=\! 1 \big)$. The according weights are determined as\footnote{
	Note that \eqref{eq:renormalizationunder} reduces to $\wprime{}_{k,n}^{(j)\s [i]}  =  I \, \wyprime{}_{k,n}^{(j)\s [i]} \, \wxprime{}_{n}^{[i]}$ for $\{( {\V{x}}_{n-1}^{[i]}, $ $ \wx{}_{n-1}^{[i]}= 1/I ) \}_{i=1}^{I} $ and  $\{ ({\V{\psi}}_{k,n-1}^{(j)\s [i]}, {w_{{\mathsf{\V{\psi}}}}}_{k,n-1}^{(j)\s [i]} = {p_{\text{e}}}_{n-1,k}^{(j)}/I )\}_{i=1}^{I} $ )} 
	\begin{equation} \label{eq:renormalizationunder}
		\wprimeunder{}_{k,n}^{(j)\s [i]} \rmv =  %
		\rmv {p_{\text{e}}}_{n,k}^{\prime\s (j)} \,
		\frac{
			 \wyprime{}_{k,n}^{(j)\s [i]} \,\wxprime{}_{n}^{[i]}
		}{
		 \sum_{i^{\prime}=1}^{I} \rmv \wyprime{}_{k,n}^{(j)\s [i^{\prime}]} \wxprime{}_{n}^{[i^{\prime}]} 
		}
		  \ist .
	\end{equation}
	While the product in the numerator $\wyprime{}_{k,n}^{(j)\s [i]} \,\wxprime{}_{n}^{[i]}$ of \eqref{eq:renormalizationunder} naturally arises due to multiplication of the independent beliefs, the denominator ensures ${p_{\text{e}}}_{n,k}^{\prime\s (j)} \ist\triangleq\ist \sum^{I}_{i=1}\wprimeunder{}_{k,n}^{(j)\s [i]}$. Then, an approximation $\tilde{\beta}\big(\underline{a}_{k,n}^{(j)} \big)$ of the message ${\beta}\big( \underline{a}_{k,n}^{(j)} \big)$ 
	in \eqref{eq:bp_measevalutionLF} can be calculated as 
	\begin{align}
			\tilde{\beta}\big( \underline{a}_{k,n}^{(j)} \big) &\ist = \sum^{I}_{i=1} \underline{g}\big( \V{x}^{\prime\s [i]}_{n}, \underline{\V{\psi}}^{\prime\s [i]\s (j)}_{k,n} ,\rmv  \underline{r}_{k,n}^{(j)}  \!=\! 1, \underline{a}^{(j)}_{k,n};\rmv \V{z}^{(j)}_{n} \big) \ist \wprimeunder{}_{k,n}^{(j)\s [i]}  \nn\\[-2mm]
		&\hspace{6mm}+\ist 1_{\{0\}}\big( \underline{a}_{k,n}^{(j)}  \big) \Bigg(\rmv 1- \sum^{I}_{i=1}\wyprime{}_{k,n}^{(j)\s [i]}  \rmv \Bigg) \ist. \label{eq:approxBeta}
	\end{align}
	Here, 
	$\sum^{I}_{i=1} \underline{g}\big( \V{x}^{\prime\s [i]}_{n}, \underline{\V{\psi}}^{\prime\s [i]\s (j)}_{k,n} ,\rmv 1, \underline{a}^{(j)}_{k,n};\rmv \V{z}^{(j)}_{n} \big) \ist \wprimeunder{}_{k,n}^{(j)\s [i]}  $ 
	provides a Monte Carlo approximation of 
	$\int\!\! \rmv \rmv \int \rmv\rmv\rmv\rmv 	\chi\big(  \underline{\V{\psi}}_{k,n}^{(j)}\rmv \rmv\rmv ,\rmv 1\big)  \ist\chi_\text{x}\big(  \V{x}_n \big)  \ist\underline{g}\big( \V{x}_{n}, $ $ \underline{\V{\psi}}^{(j)}_{k,n} ,\rmv 1, \underline{a}^{(j)}_{k,n};\rmv \V{z}^{(j)}_{n} \big)  \mathrm{d}\V{x}_{n} \ist \mathrm{d} \underline{\V{\psi}}^{(j)}_{k,n}$ in \eqref{eq:bp_measevalutionLF}, 
	and the expression  
	$ 1_{\{0\}}\big( \underline{a}_{k,n}^{(j)}  \big) \big(\rmv 1- \sum^{I}_{i=1} \wyprime{}_{k,n}^{(j)\s [i]}  \rmv \big) $ 
	provides an approximation of $ 1_{\{0\}} \big(\underline{a}_{k,n}^{(j)}\big)\ist \chi_{k,n}^{(j)} $.
	We note that %
	$1- \sum^{I}_{i=1}\wyprime{}_{k,n}^{(j)\s [i]} $ can be interpreted as a ``predicted nonexistence probability,'' which approximates $\chi_{k,n}^{(j)}$ \cite{MeyerJSIPN2016}. 
	
\subsubsection{Measurement evaluation for new \acp{pbo}} \label{sec:particle_meas_eval_new_pbo}
	For each particle $\{ ( {\V{x}}_{n}^{\prime\s  [i]}, \wxprime{}_{n}^{[i]} ) \}_{i=1}^{I} $ representing the message $\chi_\text{x}\big( {\bm{x}}_n \big)$ we draw for all  $m \rmv\in\rmv \Set{M}_n^{(j)}$ one particle from the proposal distribution  $f_{\mathrm{n}}\big(\overline{\V{\psi}}_{m,n}^{(j)}\big)$ yielding the stacked state $\{ ( {\V{x}}_{n}^{\prime\s  [i]}, \overline{\V{\psi}}_{m,n}^{\prime\s (j)\s [i]} , \wxprime{}_{n}^{[i]}  ) \}_{i=1}^{I} $ that represents the joint distribution $\chi_\text{x}(\V{x}_n) \ist \ist f_{\mathrm{n}}\big(\overline{\V{\psi}}_{m,n}^{(j)}\big)$. Since, $\sum^{I}_{i=1} \wxprime{}_{n}^{[i]}  = 1$ the particle-based approximation is properly normalized. 
	Then, for all  $m \rmv\in\rmv \Set{M}_n^{(j)}$ an approximation $\tilde{\xi}\big(\overline{a}^{(j)}_{m,n} \big)$ of the messages $\xi\big(\overline{a}^{(j)}_{m,n} \big)$ as given in \eqref{eq:bp_measevalutionNF} can be calculated and for $\overline{a}^{(j)}_{m,n} = 0$ as 
	\begin{align}
		\tilde{\xi}\big(\overline{a}^{(j)}_{m,n}\big) &= 1 + 
		\frac{ {\mu}_{\mathrm{n}} }{ \mu_\mathrm{fa} \ist f_{\mathrm{fa}}\big( \V{z}^{(j)}_{m,n} \big) } 
		 \nn\\
		 &\hspace{8mm}\times 
		\sum^{I}_{i=1} \rmv f\big(\V{z}_{m,n}^{(j)}| \V{x}_{n}^{\prime\s  [i]} ,  \overline{\V{\psi}}_{m,n}^{\prime\s (j)\s [i]} \big) \, \wxprime{}_{n}^{[i]}  \ist .
		\label{eq:bp_measevalutionNF}
	\end{align}

\subsection{Iterative data association}

	The approximate messages  $\tilde{\beta}\big( \underline{a}_{k,n}^{(j)} \big)$ and $\tilde{\xi}\big(\overline{a}_{m,n}^{(j)}\big)$ are used for iterative message passing, i.e., they substitute the messages 
	${\beta}\big( \underline{a}_{k,n}^{(j)} \big)$ and ${\xi}\big(\overline{a}_{m,n}^{(j)}\big)$ in \eqref{eq:featureDARV},  \eqref{eq:measurementDARV} and \eqref{eq:featureDARVInit}. 
	After convergence of the data association loop, we obtain approximate messages $\tilde{\eta}\big( \underline{a}_{k,n}^{(j)} \big)$ and $\tilde{\varsigma}\big( \overline{a}_{m,n}^{(j)} \big)$ from \eqref{eq:messagesDA}.

\subsection{Measurement Update}	
\subsubsection{Measurement update and belief for legacy \acp{pbo}} 
	We start by rewriting the belief $	{q}\big(  \underline{\V{\psi}}_{k,n}^{(j)},  \underline{\rv{r}}_{k,n}^{(j)}  = 1 \big)$ in \eqref{eq:beliefLegacy1} by inserting \eqref{eq:measurementUpdateLegacy1} to get 
	\begin{align}
		&{q}\big(  \underline{\V{\psi}}_{k,n}^{(j)},  \underline{\rv{r}}_{k,n}^{(j)} = 1 \big) \nn \\\nn
			&\hspace{5mm} =  \ist \rmv \frac{1}{\underline{C}_{k,n}^{(j)}} \rmv\rmv \int\rmv \rmv\rmv \sum^{M_n^{(j)}}_{\underline{a}_{k,n}^{(j)} = 0} \!\! \eta\big( \underline{a}_{k,n}^{(j)} \big)  \,
		\underline{g}\big( \V{x}_{n}, \underline{\V{\psi}}^{(j)}_{k,n} ,\rmv 1, \underline{a}^{(j)}_{k,n};\rmv \V{z}^{(j)}_{n} \big)  \nn \\ 
		&\hspace{10mm}\times \ist \chi_\text{x}(\V{x}_n) \chi\big(\underline{\V{\psi}}_{k,n}^{(j)}, \underline{r}_{k,n}^{(j)}  \!=\! 1 \big) \, \mathrm{d}\V{x}_n \ist .
		\label{eq:measurementUpdateLegacy1Particle}
	\end{align}
	Using the stacked state $\{ ( {\V{x}}_{n}^{\prime\s  [i]}, \underline{\V{\psi}}_{k,n}^{\prime\s (j)\s [i]}, \wprime{}_{k,n}^{(j)\s [i]} ) \}_{i=1}^{I} $ from \eqref{eq:renormalizationunder} that represents the joint distribution $\chi_\text{x}\big( {\bm{x}}_n \big)\, \chi\big(\underline{\V{\psi}}_{k,n}^{(j)}, \underline{r}_{k,n}^{(j)}  \!=\! 1 \big)$ we determine a particle-based representation $\{ ( {\V{x}}_{n}^{\prime\prime\s  [i]}, \underline{\V{\psi}}_{k,n}^{\prime\prime\s (j)\s [i]}, \wprimeprimeunder{}_{k,n}^{(j)\s [i]} ) \}_{i=1}^{I} $ by means of importance sampling \cite{Doucet2005}. This particle-based representation  approximates the joint distribution given by the terms inside the integral in \eqref{eq:measurementUpdateLegacy1Particle}. Thus, we calculate for each particle $ i \in \{1,\dots, I\} $ with ${{\V{x}}_{n}^{\prime\prime\s  [i]} \triangleq \V{x}}_{n}^{\prime\s  [i]} $ and $\underline{\V{\psi}}_{k,n}^{\prime\prime\s (j)\s [i]} \triangleq \underline{\V{\psi}}_{k,n}^{\prime\s (j)\s [i]} $ the importance weight, given as 
	\begin{align}
	\wprimeprimeunder{}_{k,n}^{(j)\s [i]} &= \rmv \wprime{}_{k,n}^{(j)\s [i]}\, \rmv\rmv \rmv\rmv\sum^{M_n^{(j)}}_{\underline{a}_{k,n}^{(j)} = 0} \!\! \tilde{\eta}\big( \underline{a}_{k,n}^{(j)} \big) \nn\\ %
	&\hspace{8mm}\times 
		\underline{g}\big( {\V{x}}_{n}^{\prime\prime\s  [i]}, \underline{\V{\psi}}_{k,n}^{\prime\prime\s (j)\s [i]},\rmv 1, \underline{a}^{(j)}_{k,n};\rmv \V{z}^{(j)}_{n} \big) \ist .
	\end{align}
	By noting that the integral in of \eqref{eq:measurementUpdateLegacy1Particle} is a marginalization, we simply drop $ \{ (  {\V{x}}_{n}^{\prime\prime\s  [i]}  ) \}_{i=1}^{I} $ of the obtained particle-based representation to get $\{ ( \underline{\V{\psi}}_{k,n}^{\prime\prime\s (j)\s [i]}, \wprimeprimeunder{}_{k,n}^{(j)\s [i]} ) \}_{i=1}^{I} $. 
	The normalization constant  $\underline{C}_{k,n}^{(j)}$ in \eqref{eq:measurementUpdateLegacy1Particle} is approximated as $ \tilde{\underline{C}}_{k,n}^{(j)} = 	\sum^{I}_{i=1} \rmv \wprimeprimeunder{}_{k,n}^{(j)\s [i]} + (1 - \sum^{I}_{i=1} {\wyprime{}}_{k,n}^{(j)\s [i]} )\,  \tilde{\eta}\big( \underline{a}_{k,n}^{(j)} \!\rmv=\! 0\big) $ with $ (1 - \sum^{I}_{i=1} {\wyprime{}}_{k,n}^{(j)\s [i]} ) \tilde{\eta}\big( \underline{a}_{k,n}^{(j)} \!\rmv=\! 0\big)$ approximating $\chi_{k,n}^{(j)}$ and $\tilde{\eta}\big( \underline{a}_{k,n}^{(j)} \!\rmv=\! 0\big)$ approximating ${\gamma}_{k,n}^{(j)}$. Finally, we determine the normalized representation $\{ ( \underline{\V{\psi}}_{k,n}^{(j)\s [i]}, \wyubar{}_{k,n}^{(j)\s [i]} ) \}_{i=1}^{I} $, with $\underline{\V{\psi}}_{k,n}^{(j)\s [i]} \triangleq \underline{\V{\psi}}_{k,n}^{\prime\prime\s (j)\s [i]} $ by calculating  $\wyubar{}_{k,n}^{(j)\s [i]} = \wprimeprimeunder{}_{k,n}^{(j)\s [i]} / \tilde{\underline{C}}_{k,n}^{(j)} $, which is the desired approximation of  $q\big(  \underline{\V{\psi}}_{k,n}^{(j)},  \underline{\rv{r}}_{k,n}^{(j)} = 1 \big)$. 
	
\subsubsection{Measurement update and belief for new \acp{pbo}}: 
	We start by rewriting the belief ${q}\big( \overline{\V{\psi}}^{(j)}_{m,n}, \overline{r}_{m,n}^{(j)}  = 1 \big)$ in \eqref{eq:beliefNew1} by inserting \eqref{eq:measurementUpdateNew1} to get 
	\begin{align}
		&q\big( \overline{\V{\psi}}^{(j)}_{m,n}, \overline{r}_{m,n}^{(j)} = 1 \big) \nn \\\nn
		&\hspace{10mm} =  \frac{1}{\overline{C}_{m,n}^{(j)}} \int \rmv
		\frac{ {\mu}_{\mathrm{n}} \ist \varsigma\big( \overline{a}_{m,n}^{(j)} \!\rmv=\! 0\big) }{ \mu_\mathrm{fa} \ist f_{\mathrm{fa}}\big( \V{z}^{(j)}_{m,n} \big) } \,
		  f\big(\V{z}_{m,n}^{(j)}| \V{x}_n, \overline{\V{\psi}}_{m,n}^{(j)}\big) \nn \\[1.5mm]
		&\hspace{28mm}\times  \chi_\text{x}(\V{x}_n) \ist \ist f_{\mathrm{n}}\big(\overline{\V{\psi}}_{m,n}^{(j)}\big) \, \mathrm{d}\V{x}_{n} \ist .
		\label{eq:bp_measevalutionNFParticle}
	\end{align}
	Using the stacked state  $\{ ( {\V{x}}_{n}^{\prime\s  [i]}, \overline{\V{\psi}}_{m,n}^{\prime\s (j)\s [i]} , \wxprime{}_{n}^{[i]}  ) \}_{i=1}^{I} $ from the measurement evaluation of new \ac{pbo}\footnote{As the proposal density $ f_{\mathrm{n}}\big(\overline{\V{\psi}}_{m,n}^{(j)}\big) $ usually is uninformative and, thus, requires a large number of particles for representation, we perform a systematic resampling step \cite{ArulampalamTSP2002} on $\{ ( {\V{x}}_{n}^{\prime\s  [i]}, \overline{\V{\psi}}_{m,n}^{\prime\s (j)\s [i]} , \wxprime{}_{n}^{[i]}  ) \}_{i=1}^{I} $ before evaluating the weights in \eqref{eq:weightsNewPBO}.} 
	in Sec.~\ref{sec:particle_meas_eval_new_pbo} 
	that represents the joint distribution $\chi_\text{x}(\V{x}_n) \ist \ist f_{\mathrm{n}}\big(\overline{\V{\psi}}_{m,n}^{(j)}\big)$ 
	we again determine a particle-based representation $\{ ( {\V{x}}_{n}^{\prime\prime\s  [i]}, \overline{\V{\psi}}_{m,n}^{\prime\prime\s (j)\s [i]} , \wprimeprimeover{}_{m,n}^{(j)\s [i]}  ) \}_{i=1}^{I} $ by means of importance sampling \cite{Doucet2005} that approximates the joint distribution given by the terms inside the integral in \eqref{eq:bp_measevalutionNFParticle}. Thus, we calculate for each particle $ i \in \{1,\dots, I\} $ with ${{\V{x}}_{n}^{\prime\prime\s  [i]} \triangleq \V{x}}_{n}^{\prime\s  [i]} $ and $\overline{\V{\psi}}_{m,n}^{\prime\prime\s (j)\s [i]} \triangleq \overline{\V{\psi}}_{m,n}^{\prime\s (j)\s [i]} $  the importance weight, given as 
	\begin{equation} \label{eq:weightsNewPBO}
	 \wprimeprimeover{}_{m,n}^{(j)\s [i]}  = \wxprime{}_{n}^{[i]}  \ist	\frac{ {\mu}_{\mathrm{n}} \ist \tilde{\varsigma}\big( \overline{a}_{m,n}^{(j)} \!\rmv=\! 0\big) }{ \mu_\mathrm{fa} \ist f_{\mathrm{fa}}\big( \V{z}^{(j)}_{m,n} \big) } \,
	 f\big(\V{z}_{m,n}^{(j)}|   {\V{x}}_{n}^{\prime\s  [i]}, \overline{\V{\psi}}_{m,n}^{\prime\s (j)\s [i]}  \big)   \ist. 
	\end{equation}
	Again, the integral in \eqref{eq:bp_measevalutionNFParticle} is a marginalization. Thus, we drop $ \{ (  {\V{x}}_{n}^{\prime\prime\s  [i]}  ) \}_{i=1}^{I} $ to get $\{ ( \overline{\V{\psi}}_{m,n}^{\prime\prime\s (j)\s [i]} ,  \wprimeprimeover{}_{m,n}^{(j)\s [i]} ) \}_{i=1}^{I} $
	The normalization constant  $\overline{C}_{m,n}^{(j)}$ in  \eqref{eq:bp_measevalutionNFParticle} is approximated as 
	$ \tilde{\overline{C}}_{m,n}^{(j)} = 	\sum^{I}_{i=1} \rmv  \wprimeprimeover{}_{m,n}^{(j)\s [i]} + \sum^{K_{n-1}^{(j)}}_{\overline{a}_{m,n}^{(j)} = 0} \!\! \tilde{\varsigma}\big( \overline{a}_{m,n}^{(j)}\big)$ with $\sum^{K_{n-1}^{(j)}}_{\overline{a}_{m,n}^{(j)} = 0} \!\! \tilde{\varsigma}\big( \overline{a}_{m,n}^{(j)}\big)$ approximating ${\phi}_{m,n}^{(j)}$. 
	Finally, we determine the normalized representation $\{ ( \overline{\V{\psi}}_{m,n}^{(j)\s [i]}, \wyobar{}_{k,n}^{(j)\s [i]} ) \}_{i=1}^{I} $, with $\overline{\V{\psi}}_{m,n}^{(j)\s [i]} \triangleq \overline{\V{\psi}}_{m,n}^{\prime\prime\s (j)\s [i]} $ by calculating  $\wyobar{}_{k,n}^{(j)\s [i]} = \wprimeprimeover{}_{m,n}^{(j)\s [i]} / \tilde{\overline{C}}_{m,n}^{(j)} $,
	which represents the desired approximation of $q\big( \overline{\V{\psi}}^{(j)}_{m,n}, \overline{r}_{m,n}^{(j)}~=~1 \big)$. 

\begin{figure*}[!t]		
	\begin{align}
		\hspace{-2mm}   \rho_{k}^{(j)}\big(  \V{x}_n \big) %
		&= \rmv\int \rmv \sum^{M_n^{(j)}}_{m = 1} \!\! 
		\frac{ \eta\big( \underline{a}_{k,n}^{(j)} \rmv\rmv\rmv =  \rmv\rmv\rmv m \big)  \pd  \ist  f( \zu | u_{k,n}^{(j)})  }{\mu_\mathrm{fa} \ist f_{\mathrm{fa}}(\V{z}_{m,n}^{(j)}) } f_\text{N}(\zd; d({b}^{(j)}_{k,n},\bm{p}_n), \sigma_{\text{d}}^{2} (u^{(j)}_{k,n}) )   %
		\nn\\ & \hspace{80mm}  +	(1 \!-\rmv \pd )  \chi\big( \underline{\V{\psi}}_{k,n}^{(j)} , 1\big) \mathrm{d} \underline{\V{\psi}}_{k,n}^{(j)}
		\ist+\ist \eta\big( \underline{a}_{k,n}^{(j)} \!\rmv=\! 0\big) \ist\chi_{k,n}^{(j)} \ist 
		\label{eq:measurementUpdateAgentInterpretation}\\[0mm]
		\rho_{0}^{(j)} \big(  \V{x}_n \big)	& \approx \sum^{M_n^{(j)}}_{m = 1} \! 
		\frac{ \eta\big( \underline{a}_{k,n}^{(j)} \rmv\rmv\rmv =  \rmv\rmv\rmv m \big)  p_{\text{d}}(\hat{u}_{k,n}^{(j)})  \ist  f( \zu | \hat{u}_{k,n}^{(j)})  }{\mu_\mathrm{fa} \ist f_{\mathrm{fa}}(\V{z}_{m,n}^{(j)}) } \, f_\text{N}(\zd; d(0,\bm{p}_n), \sigma_{\text{d}}^{2} (\hat{u}_{k,n}^{(j)}) )  %
		+	(1 \!-\rmv p_{\text{d}}(\hat{u}_{k,n}^{(j)}) )
		\ist+\ist \eta\big( \underline{a}_{k,n}^{(j)} \!\rmv=\! 0\big) \ist\chi_{k,n}^{(j)} \ist 
		\label{eq:measurementUpdateAgentInterpretationApprox}%
	\end{align}
	\hrulefill
	\vspace*{-3mm}
\end{figure*}	
\subsubsection{Measurement update for the agent}
	Again we start by rewriting the belief  ${q}\big( {\V{x}}_{n} \big)$ by inserting the messages \eqref{eq:measurementUpdateAgent} for all legacy \acp{pbo} $k \rmv\in\rmv \Set{K}_{n-1}^{(j)}$ into \eqref{eq:beliefLegacy1} to get
	\begin{align} 
	q(\V{x}_n) \ist &= \ist \frac{1}{C_{\text{x}\s n}} \rmv\int \rmv\rmv\rmv\cdots\rmv\rmv\rmv \int \!  \chi_\text{x}( \V{x}_n) \prod_{j=1}^J \prod_{k \in \Set{K}_{n-1}^{(j)}} \!\!\rmv 
	 \chi\big( \underline{\V{\psi}}_{k,n}^{(j)} , 1\big)  \nn \\
	&\hspace{6mm} \Bigg( \sum^{M_n^{(j)}}_{\underline{a}_{k,n}^{(j)} = 0} \!\! \eta\big( \underline{a}_{k,n}^{(j)} \big) 
	\underline{g}\big( \V{x}_{n}, \underline{\V{\psi}}^{(j)}_{k,n} ,\rmv 1, \underline{a}^{(j)}_{k,n};\rmv \V{z}^{(j)}_{n} \big) \nn \\
	&\hspace{10mm}  
	\ist+\ist \eta\big( \underline{a}_{k,n}^{(j)} \!\rmv=\! 0\big) \ist\chi_{k,n}^{(j)} \Bigg) \mathrm{d} \underline{\V{\psi}}_{1,n}^{(1)} \rmv\cdots  \, \mathrm{d} \underline{\V{\psi}}_{K_n^{(j)}\rmv\rmv,n}^{(J)} \ist . \label{eq:agentBeliefXParticle}
	\end{align}
	Here, we facilitate the stacked state 
	$\{ ( {\V{x}}_{n}^{\prime\s  [i]}, \,
	\underline{\V{\psi}}_{1,n}^{\prime\s (1)\s [i]} 
	\, \dots \, \underline{\V{\psi}}_{K_n^{(J)}\rmv\rmv\rmv,n}^{\prime\s (J)\s [i]}, \,
	\wprime{}_{n}^{[i]} ) \}_{i=1}^{I} $. It represents the joint distribution of the agent state and all \ac{pbo} states $k \rmv\in\rmv \Set{K}_{n-1}^{(j)}$ of all anchors $j \rmv\in\rmv \mathcal{J}$, given as $\chi_\text{x} \big( {\bm{x}}_n \big) \prod_{j=1}^J \prod_{k \in \Set{K}_{n-1}^{(j)}} \chi\big(\underline{\V{\psi}}_{k,n}^{(j)}, \underline{r}_{k,n}^{(j)}  \!=\! 1 \big)$. 
	The according weights are determined as 
	\begin{equation} \label{eq:renormalizationx}
		\wprime{}_{n}^{[i]} \rmv = \rmv  
		\frac{
			\rmv\rmv\rmv\rmv\rmv  \wxprime{}_{n}^{[i]} \, \prod_{j=1}^J \prod_{k \in \Set{K}_{n-1}^{(j)}} \wyprime{}_{k,n}^{(j)\s [i]} 
		}{
			\sum_{i^{\prime}=1}^{I} \wxprime{}_{n}^{[i^{\prime}]} \rmv  \prod_{j=1}^J \prod_{k \in \Set{K}_{n-1}^{(j)}}  \rmv\rmv \wyprime{}_{k,n}^{(j)\s [i^{\prime}]}  
		}	\ist .
	\end{equation}
	We determine a particle-based representation 
		$\{ ( {\V{x}}_{n}^{\prime\prime\s  [i]}, \,
	\underline{\V{\psi}}_{1,n}^{\prime\prime\s (1)\s [i]} 
	\, \dots \, \underline{\V{\psi}}_{K_n^{(J)}\rmv\rmv\rmv,n}^{\prime\prime\s (J)\s [i]}, \,
	\wprimeprime{}_{n}^{[i]} ) \}_{i=1}^{I} $ 
	that approximates the joint distribution given by the terms inside the integral in \eqref{eq:agentBeliefXParticle} by means of importance sampling. We calculate for each particle $ i \in \{1,\dots, I\} $ with ${{\V{x}}_{n}^{\prime\prime\s  [i]} \triangleq \V{x}}_{n}^{\prime\s  [i]} $ and $\underline{\V{\psi}}_{k,n}^{\prime\prime\s (j)\s [i]} \triangleq \underline{\V{\psi}}_{k,n}^{\prime\s (j)\s [i]} $ for all $k \rmv\in\rmv \Set{K}_{n-1}^{(j)} ,  j \rmv\in\rmv \mathcal{J}$ an importance weight, given as 
	\begin{equation} 
		\wprimeprime{}_{n}^{[i]}  = 	\wprime{}_{n}^{[i]} \, \prod_{j=1}^J \, \prod_{k \in \Set{K}_{n-1}^{(j)}} \rmv\rmv\rmv\rmv  \wxprimeprime{}_{k,n}^{(j)\s [i]} 	
	\end{equation}
	with  $\wxprimeprime{}_{k,n}^{(j)\s [i]} $ containing the contribution to the overall weight \ac{wrt} each individual \ac{pbo} given as
	\begin{align} 
	 	  \wxprimeprime{}_{k,n}^{(j)\s [i]} &=  \sum^{M_n^{(j)}}_{\underline{a}_{k,n}^{(j)} = 0} \!\! \tilde{\eta}\big( \underline{a}_{k,n}^{(j)} \big) \, 
		 \underline{g}\big( {\V{x}}_{n}^{\prime\prime\s  [i]}, \underline{\V{\psi}}_{k,n}^{\prime\prime\s (j)\s [i]} ,\rmv 1, \underline{a}^{(j)}_{k,n};\rmv \V{z}^{(j)}_{n} \big) \nn \\
		&\hspace{6mm} \ist+\ist \tilde{\eta}\big( \underline{a}_{k,n}^{(j)} \!\rmv=\! 0\big) \ist \big(\rmv 1- \sum^{I}_{i=1} \wyprime{}_{k,n}^{(j)\s [i]}  \rmv \big) \ist .
		\\[-4mm]\nn
	\end{align}
		Again, the integral in \eqref{eq:measurementUpdateNew1} is a marginalization. 
		Thus, we simply drop 
		$ \{ ( 
		\underline{\V{\psi}}_{1,n}^{\prime\s (1)\s [i]} 
		\, \dots \, \underline{\V{\psi}}_{K_n^{(J)}\rmv\rmv\rmv,n}^{\prime\s (J)\s [i]} ) \}_{i=1}^{I} $ 
		to get $\{ (  {\V{x}}_{n}^{\prime\prime\s  [i]},   \wprimeprime{}_{n}^{(j)\s [i]} ) \}_{i=1}^{I} $, which represents the marginalized state. 
		The normalization constant  $C_{\text{x}\s n}$ in  \eqref{eq:agentBeliefXParticle} is approximated as 
		$ \tilde{C}_{\text{x}\s n} = 	\sum^{I}_{i=1} \rmv \wprimeprime{}_{n}^{[i]} $.  
		Finally, we determine the normalized representation $\{( {\V{x}}_{n-1}^{[i]}, \wx{}_{n-1}^{[i]}) \}_{i=1}^{I} $ , with ${\V{x}}_{n-1}^{[i]} \triangleq  {\V{x}}_{n}^{\prime\prime\s  [i]} $ by calculating  $\wx{}_{n-1}^{[i]} = \wprimeprime{}_{n}^{(j)\s [i]} /  \tilde{C}_{\text{x}\s n} $,
		which represents the desired approximation of $q\big( \V{x}\big)$.

\subsection{State Estimation, Detection and Resampling} \label{sec:particle_state_estimation}
The weighted particles $\{( {\V{x}}_{n}^{[i]}, \wx{}_{n}^{[i]}) \}_{i=1}^{I} $ and  $\{ ({\V{\psi}}_{k,n}^{(j)\s [i]}, {w_{{\mathsf{\V{\psi}}}}}_{k,n}^{(j)\s [i]}  )\}_{i=1}^{I} $ that represent the marginal distributions can now be used to approximate the quantities of interest from \mref{Sec.}{sec:factor_graph}. The \ac{mmse} estimates of the agent state in \meqref{eq:mmse} is calculated according to 
\begin{equation}
\hat{{\bm{x}}}^\text{MMSE}_{n} \approx \sum_{i=1}^{I} {\V{x}}_{n}^{[i]} \,  \wx{}_{n}^{[i]}  \ist .
\end{equation}
The existence probability of a \ac{pbo} $p( r_{k,n}^{(j)}~=~1\big|\V{z}_{1:n}) $ is approximated using the sum of weights $	{p_{\text{e}}}_{n,k}^{(j)} $ according to \eqref{eq:approxExistProb}. For detected \acp{pbo} the \ac{mmse} estimates of the \ac{pbo} state in \meqref{eq:mmsepbo} are approximated as
\begin{equation}
	\hat{\bm{\psi}}^{(j)\s\text{MMSE}}_{k,n} \approx \frac{1}{{p_{\text{e}}}_{n,k}^{(j)} } \sum_{i=1}^{I} {\V{\psi}}_{k,n}^{(j)\s [i]} {w_{{\mathsf{\V{\psi}}}}}_{k,n}^{(j)\s [i]}   \ist .
\end{equation}

To avoid particle degeneracy \cite{ArulampalamTSP2002}, a resampling step\footnote{We suggest to use ``systematic" resampling for efficiency \cite{ArulampalamTSP2002}.} is performed as a preparation for the next time step $ n+1 $ to obtain equally weighted particles $\{( \tilde{\V{x}}_{n}^{[i]}, \wxtilde{}_{n}^{[i]} = 1/I ) \}_{i=1}^{I} $ and  $\{ ( \tilde{\V{\psi}}_{k,n}^{(j)\s [i]}, \wytilde_{k,n}^{(j)\s [i]} = 	{p_{\text{e}}}_{n,k}^{(j)}/I )\}_{i=1}^{I} $ for all $k \rmv\in\rmv \Set{K}_{n-1}^{(j)} ,  j \rmv\in\rmv \mathcal{J}$, 
which are used instead of $\{( {\V{x}}_{n}^{[i]}, \wx{}_{n}^{[i]}) \}_{i=1}^{I} $ and  $\{ ({\V{\psi}}_{k,n}^{(j)\s [i]}, \wy_{k,n}^{(j)\s [i]}  )\}_{i=1}^{I} $ as an input for the prediction step (see Sec.~\ref{sec:particle_prediction}). This procedure is in accordance with sampling importance resampling (SIR) particle filter \cite{ArulampalamTSP2002}.

\section{Analysis of the Agent Message}\label{sec:message_interpretation}

Here, we provide an analysis of the message given in \eqref{eq:measurementUpdateAgent}, i.e., the agent measurement update $\rho_{k}^{(j)}\big(  \V{x}_n \big) $. By inserting the proposed system model according to the main text \mref{Sec.}{sec:system_model} we obtain \eqref{eq:measurementUpdateAgentInterpretation}.

\begin{figure}[t]
	\centering
	\tikzsetnextfilename{likelihood_position}
	\includegraphics{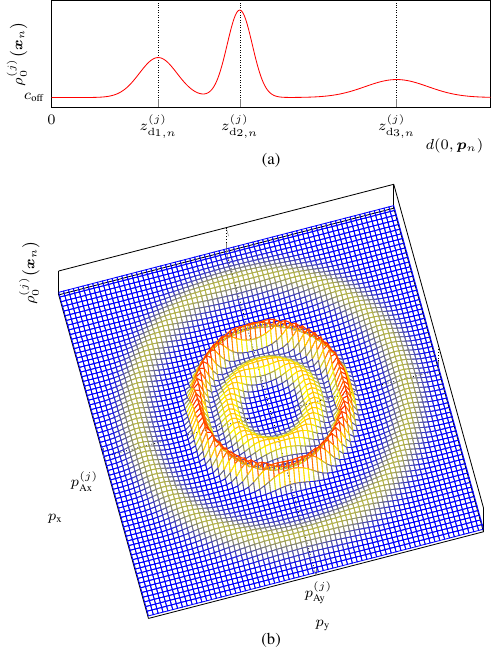}
	\caption{Graphical representation of \eqref{eq:measurementUpdateAgentInterpretationApprox} according to Sec.~\ref{sec:message_interpretation}. $\sigma_{\text{d}}^{2} (u^{(j)}_{k,n})$}
	\label{fig:likelihood_position}
\end{figure}

In particular, we consider the message $\rho_{0}^{(j)} \big(  \V{x}_n \big)$ of the explicit \ac{los} component at $k=0$. Remember from \mref{Sec.}{sec:pbo_states} that ${b}^{(j)}_{0,n}\equiv 0$ and $v_{\text{b}\s 1,n}^{(j)} \equiv 0$ , i.e., $\chi\big( \underline{\V{\psi}}_{k,n}^{(j)} ,1 ) = \chi\big( \underline{\V{u}}_{k,n}^{(j)} ,1 )$. 
For simplicity of the analysis, we neglect the uncertainty of the predicted \ac{pbo} belief, i.e., $\chi\big( \underline{\V{u}}_{k,n}^{(j)} ,1 ) = \delta (\hat{u}_{k,n}^{(j)} - {u}_{k,n}^{(j)})$ with $\delta(\cdot)$ being the Dirac delta distribution, and obtain \eqref{eq:measurementUpdateAgentInterpretationApprox}. %
Fig.~\ref{fig:likelihood_position} shows a graphical representation of \eqref{eq:measurementUpdateAgentInterpretationApprox}, where we assume to obtain for anchor $j$ at time $n$ the measurements {${z_\mathrm{d}^{(j)}}_{\rmv\rmv\rmv\rmv\rmv\rmv\rmv  1,n} $},{${z_\mathrm{d}^{(j)}}_{\rmv\rmv\rmv\rmv\rmv\rmv\rmv  2,n}$},{${z_\mathrm{d}^{(j)}}_{\rmv\rmv\rmv\rmv\rmv\rmv\rmv  3,n}$}, i.e., $M_n^{(j)}=3$. 
The left-hand side sum of equation \eqref{eq:measurementUpdateAgentInterpretationApprox} denotes a Gaussian mixture model with respective weights given as $\frac{ \eta ( \underline{a}_{k,n}^{(j)} \, =\, m )  \pd  \ist  f( \zu | u_{k,n}^{(j)})  }{\mu_\mathrm{fa} \ist f_{\mathrm{fa}}(\V{z}_{m,n}^{(j)}) }$ that are constant \ac{wrt} the agent state $\V{x}_n$. Note that these weights that are altered by the inferred amplitude  state via the amplitude likelihood and the detection probability and the state of the other \ac{pbo} via loopy data association lead to ``soft" treatment of the (distance) measurements in line with \cite{ContiProcIEEE2019}. The right-hand side expression $(1 \!-\rmv p_{\text{d}}(\hat{u}_{k,n}^{(j)}) )
\ist+\ist \eta\big( \underline{a}_{k,n}^{(j)} \!\rmv=\! 0\big) \ist\chi_{k,n}^{(j)} $ is a constant offset \ac{wrt} the agent state $\V{x}_n$ (referred to as $c_\text{off}$ in Fig.~\ref{fig:likelihood_position}) and depends on both, the detection probability and the existence probability of \acp{pbo}. It leads to the belief \eqref{eq:measurementUpdateAgentInterpretationApprox} being a \textit{heavy-tailed} function. This property is known from literature to yield ``robust" models that are resilient to model mismatch\cite[Sec. 2.3.7]{Bishop2006}. Also, heavy-tailed functions offer a narrow value range, which is advantageous for numerical implementation. 

\ifthenelse{0=1}
{

\section{Initial States} \label{sec:init_states}
The initial distributions $f(\V{x}_{0})$ and $f(\underline{\V{y}}_{0,}) = \prod_{j=1}^{J}  
f(\underline{\V{y}}^{(j)}_{1,0})$ are determined heuristically, assuming an initial measurement vector $\V{z}_0$ containing $\V{M}_0$ measurements to be available. 
It is assumed to factorize as
${f}({\bm{x}}_{0}) = 
{f}({\bm{p}}_{0})
{f}({\bm{v}}_{0})$.

For all anchors $j \rmv\in\rmv \mathcal{J}$, we assume that the joint \ac{pbo} state only contains the \ac{los} component, i.e., $\mathcal{K}_n^{(j)} = \{ 1 \}$, while \acp{pbo} corresponding to \acp{mpc} are initialized as new \acp{pbo} during filter operation (at times $n\geq 1$). As discussed in the main text \mref{Sec.}{sec:pbo_states} the bias of the \ac{los} component is zero, i.e., $b_{1,0}^{(j)} \equiv 0,\iist v_{\text{b}\s 1,0}^{(j)} \equiv 0$. We initialize the normalized amplitude \acp{pdf} as 
${f}(u_{1,0}^{(j)}) \sim f( \zuZero | u_{\s \text{init}}^{(j)})\, f(u_{\s \text{init}}^{(j)}) $ where $f(u_{\s \text{init}}^{(j)})$ is drawn from a uniform distribution given as $f(u_{\s \text{init}}^{(j)}) \triangleq f_\mathrm{U}(x;0,u_\text{max})$ and $u_\text{max}$ is the maximum amplitude as given in the main text \mref{Sec.}{sec:common_setup}. 
The existence variables are initialized uniformly distributed as $p({r}^{(j)}_{1,0} ) = p_\mathrm{UD}({r}^{(j)}_{1,0}; \{0.5,0.5\})$.

The agent position state is initialized as $f(\bm{p}_0) \sim \prod_{j=1}^{J}  \prod_{m=1}^{M^{(j)}_0} f( \zdZero | {b}^{(j)}_{1,0}~\!=~\!0,  \bm{p}_{\s \text{init}},\zuZeroMax ) \, f(\bm{p}_{\s \text{init}}) $, where $\zuZeroMax$ is the maximum normalized amplitude measurement in $\V{z}_0^{(j)}$. The proposal distribution $f(\bm{p}_{\s \text{init}})$ is drawn uniformly on two-dimensional discs around each anchor $j$, which are bounded by the maximum possible distance $d_\text{max}$ and a sample is drawn from each of the $J$ discs with equal probability. 
We assume the velocity vector ${\bm{v}}_{0}$ to be zero mean, Gaussian, with covariance matrix $\sigma_\text{v}^2\,  \bm{I}_2 $ and $\sigma_\text{v} = 6\, \text{m/s}$, as we do not know in which direction we are moving.
After drawing from the proposal distributions $ f(\bm{p}_{\s \text{init}})$ and $ f(u_{\s \text{init}}^{(j)}) $, we perform a resampling step similarly as discussed in Sec.~\ref{sec:particle_state_estimation} that avoids particle degeneracy to obtain particle-based representations of ${f}({\bm{p}}_{0})$ and ${f}(u_{1,0}^{(j)})$. Note that for numerical implementation this can be realized by drawing samples directly from the measurement space. 

}{}

\section{Implementation Details of the \acs{gptrack} Method} \label{sec:reference_methods_details}

{For the \ac{aednn} generating the feature measurements, we use feed-forward networks with three convolutional layers for both, encoder and decoder. The encoder is set up as $27\times{}17-\text{ELU}, 27\times{}13-\text{ReLU}, 16\times{}5-\text{ELU}$, which denotes the number of convolutional kernels times filter size and the respective activations, and applies max pooling of size 2 after all activation functions. It uses the magnitudes of the baseband signal vector $|\bm{r}_n^{(j)}|$ as an input and has a latent space of $4$ variables. The decoder network mirrors the encoder network and the \ac{mse} of measured and predicted signal magnitudes is used as loss function. For the ``anomaly detection" \ac{aednn} the authors use feed-forward networks with three dense layers for both, encoder and decoder. The encoder uses the stacked real and imaginary parts of the baseband signal vector as an input. It consists of 100, 80, and 60 neurons, respectively, all with ReLU activation functions, and has two latent variables. The decoder again mirrors the encoder. As suggested, we used a beta variational \ac{aednn}\cite{Kingma2019} with regularization hyper parameter set to $\beta=10^{-3}$ and \ac{mse} as data reconstruction loss.} We used the suggested ``time index signal strength indicator" for predicting the anomaly score and compared to the optimum detection threshold being set to the intersection point of the histograms of the agent trajectory data (which is not available in reality). 
For implementation we used Python along with TensorFlow/Keras and optimized using Adam with learning rate of $2\cdot 10^{-3}$. To implement the \ac{gpr}-based measurement model we utilized MATLAB's \ac{gpr} toolbox, where we employed the suggested ``Matern52" kernel function \cite{Kram2022GPR}.

\ifthenelse{0=1}
{

\section{Additional Information for Experiment 6 (Real Radio Measurements)} \label{sec:measurement_photos_nxp}

This section provides additional information to the measurement scenario for real radio measurements in Experiment 6 \mref{Sec.}{sec:performance_real}. 
Fig.~\ref{fig:car_ant} shows the measurement hall providing a wide, open space that includes a demonstration car (Lancia Thema 2011), furniture, and metallic surfaces, thereby representing a typical multipath-prone industrial environment. 
The two full \ac{olos} situations were created using an obstacle consisting of a metal plate covered with attenuators as shown in Fig.~\ref{fig:absorber}. 

\begin{figure}[t]	
	\centering
	\setlength{\belowcaptionskip}{1pt}	
	\captionsetup[subfloat]{farskip=8pt,captionskip=4pt} 
	\subfloat[\label{fig:car_ant}]{\includegraphics[height=0.20\textwidth]{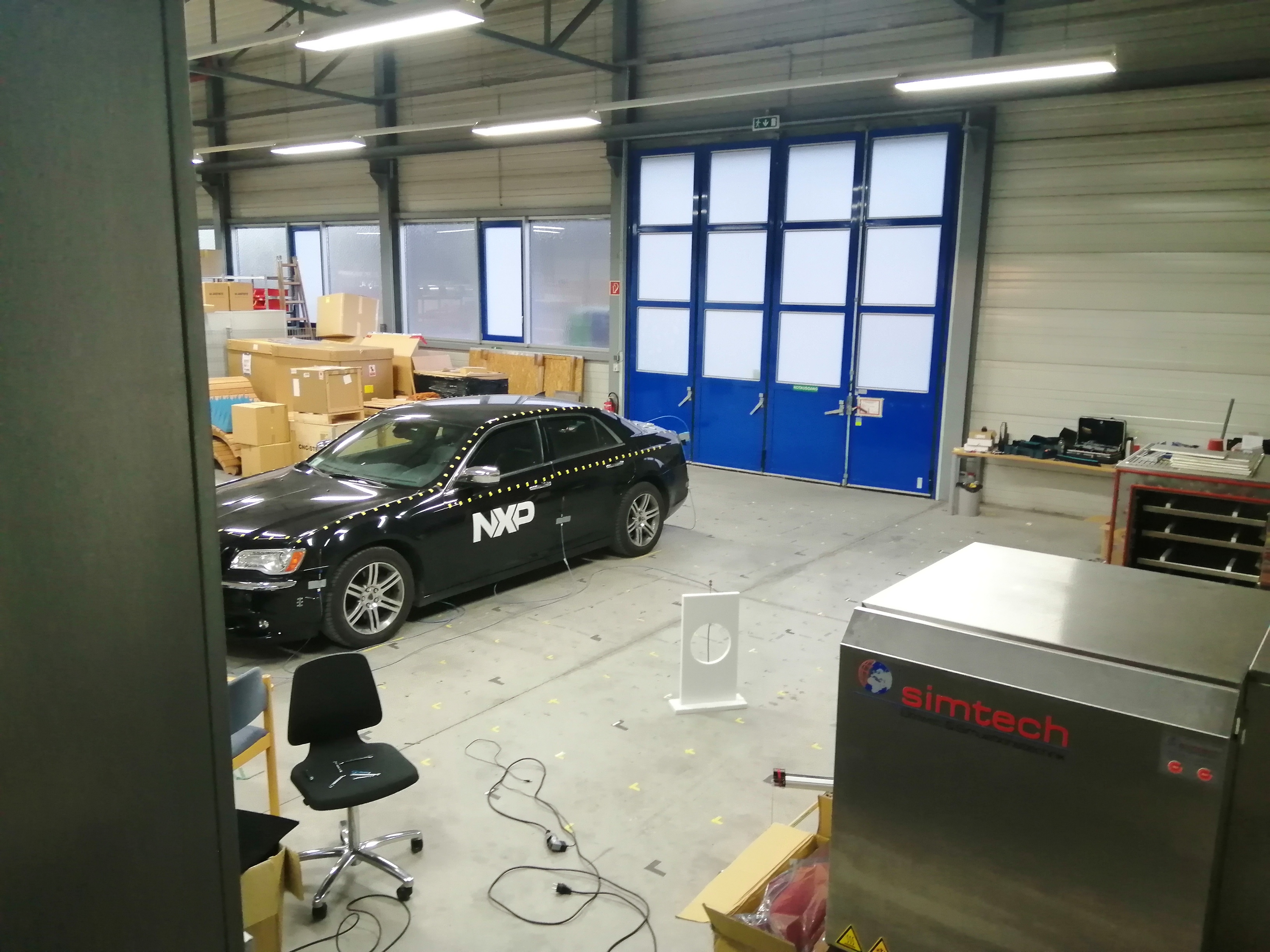}}\vspace{2mm}
	\subfloat[\label{fig:absorber}]{\includegraphics[height=0.20\textwidth]{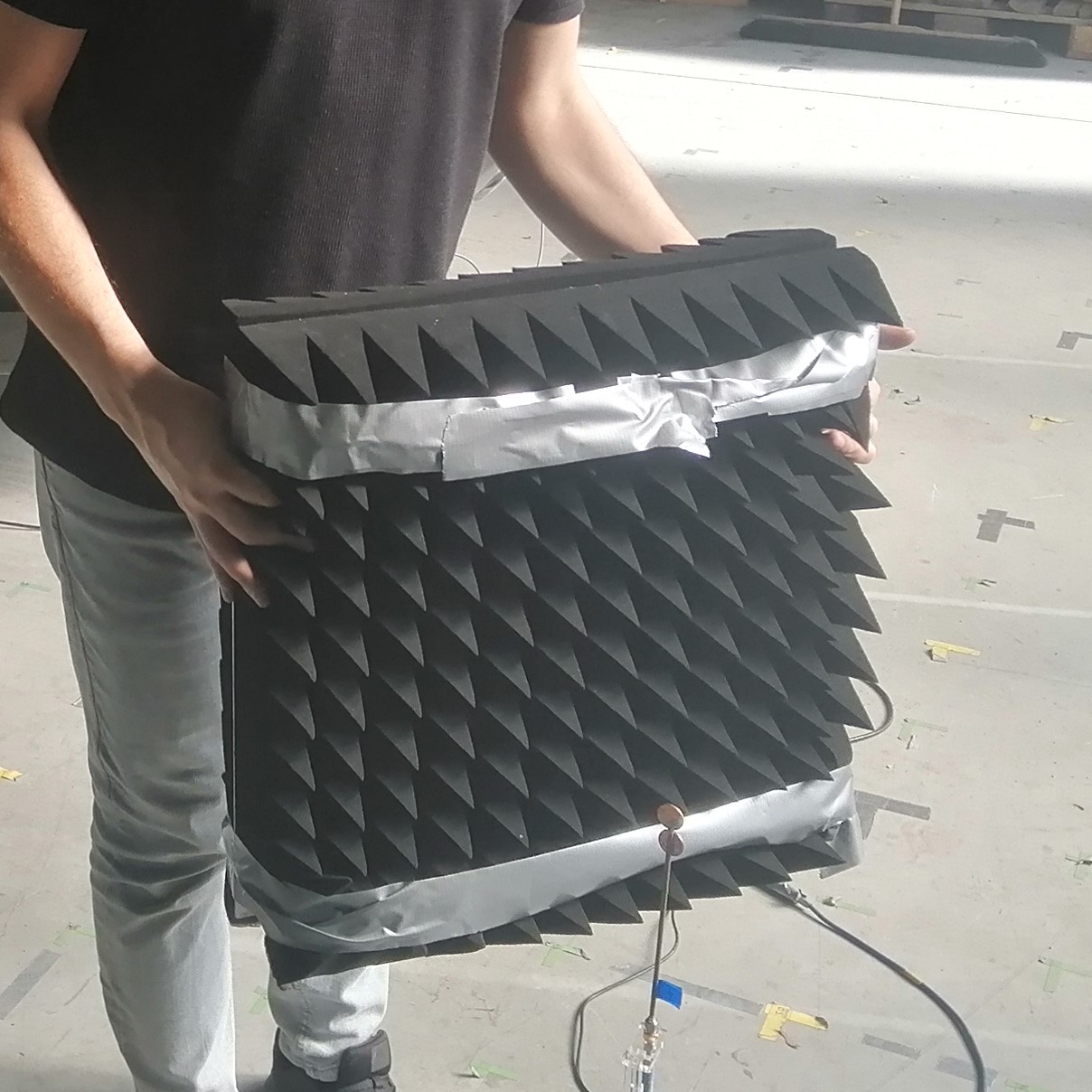}}
	\vspace{-3mm}
	\caption{Measurement setup of Experiment 6. We show pictures of (a) the overall scenario and (b) the OLOS setup used.
	}\label{fig:measurement_photos_nxp}
	\vspace{-4mm}
\end{figure}

\section{Runtime} \label{sec:execution_time}
Table~\ref{tbl:execution_times} shows the average runtime of the proposed algorithm and compares it to the runtime of all reference methods (see \mref{Table}{tbl:reference_methods}). All runtimes are estimated using Matlab implementations executed on an AMD Ryzen Threadripper 1900X 8-Core Processor with up to $4\,\text{GHz}$ for all scenarios investigated. We also show the average number of measurements (over all anchors and time steps) ${M}_\text{mean}$ and the number of anchors $J$, which together with then number of particles used (see \mref{Sec.}{sec:reference_methods}) determine the algorithm complexity per time step. 
Note that we used the ``subsampled-reduced" dataset (see \mref{Fig.}{fig:track_geometric}) to determine the runtime of the \acs{delaybias} method. 
The runtime of the proposed method is the same order of magnitude as that of \ac{mpslam}, slightly outperforming it due to the higher number of objects (\acp{pbo} vs. potential \acp{va}) initialized by \ac{mpslam}. 
The runtime of \acs{delaybias} is the lowest, as it only requires one \ac{gpr}-based bias calculation per measurement. Therefore, the time-limiting component is the particle filter. acs{pdaai} and \acs{cluster} offer a significantly lower runtime than the proposed method due to the lower complexity of the inference model as well as the lower number of particles required for inference. Finally, \acs{gptrack} shows the highest runtime. It requires to evaluate the \ac{gpr}-based mapping once per particle, anchor and feature. Its computational complexity is given as $\mathcal{O}(N_\text{T}^{\s 2} J I F)$ \cite{Rasmussen2006GP}, where $N_\text{T}$ is the number of training samples and $F$ is the size of the feature space of the \ac{aednn}.
Note that we did not include the runtime of the reprocessing algorithms (\ac{ceda} variants, \acp{aednn}) in this comparison.

\begin{table}[h]
	\renewcommand{\baselinestretch}{1}\small\normalsize
	\setlength{\tabcolsep}{3pt} %
	\renewcommand{\arraystretch}{1} %
	\centering
	\footnotesize
	\caption{Algorithm runtimes and characteristic values of all investigated scenarios.}\label{tbl:execution_times}
	\begin{tabular}{ | r c c c c c c |} 
		\toprule
		\textbf{method}  &  \textbf{Ex. 1} &   \textbf{Ex. 2} &   \textbf{Ex. 3} &   \textbf{Ex. 4}  &   \textbf{Ex. 5}  &   \textbf{Ex. 6}  \\ %
		\midrule
		\textbf{proposed}  &$ 460\,\mathrm{ms}$  & $ 244\,\mathrm{ms}$ & $231\,\mathrm{ms}$ & $224\,\mathrm{ms}$ & $170\,\mathrm{ms}$ & $312\,\mathrm{ms}$ \\
		\textbf{\acs{pdaai}} & $ 32\,\mathrm{ms}$  & $ 28\,\mathrm{ms}$ & $-$ & $-$ & $-$ & $27\,\mathrm{ms}$  \\
		\textbf{\acs{mpslam}}  & $750\,\mathrm{ms}$ 	& $283\,\mathrm{ms}$ &   $-$ & $-$ & $-$ & $340\,\mathrm{ms}$  \\
		\textbf{\acs{cluster}} & $ 54\,\mathrm{ms}$  & $ 42\,\mathrm{ms}$ & $ 41\,\mathrm{ms}$ & $-$ & $-$ & $46\,\mathrm{ms}$  \\
		\textbf{\acs{delaybias}} & $ 6\,\mathrm{ms}$  & $ 5\,\mathrm{ms}$ & $-$  & $-$ & $-$ & $-$ \\
		\textbf{\acs{gptrack}} & $ 1340\,\mathrm{ms}$  & $ 1330\,\mathrm{ms}$ &  $-$ & $-$ & $-$ & $-$\\
		\midrule 
		${M}_\text{mean}\times J$ & $8.6 \times 3 $ & $4.2 \times 3$ & $4.1 \times 3$ & $4 \times 3$  & $3.7 \times 4$  & $4.3 \times 4$  \\
		\bottomrule
	\end{tabular}
	\vspace{-2.5mm}
\end{table}

\section{Alternative Channel Estimation and Detection Algorithm} \label{sec:ceda_sbl}

In this section we derive a channel estimation an detection algorithm based on the stochastic maximum likelihood approach\cite{} and Sparse Bayesian Learning (SBL)\cite{}. The presented method is inspired by \cite{HansenSBL,ErikStefan} in the sense that we use super-resolution estimates of the dispersion parameters (i.e., the delay/distance estimate), but uses the classical marginal maximum likelihood approach, as shown in \cite{FaulTipping}.

\alex{show crossing scenario where agent walks through the mean of two walls, the anchor is in perpendicular direction \ac{wrt} the walls}

We start by redefining the discrete-time specular signal vector \eqref{eq:signal_model_sampled} for notational convenience as
\begin{align}\label{eq:stack_recvgen}
	\RV{r}_n^{(j)} = \vm{S}(\tilde{\vm{\tau}}_n^{(j)})\tilde{\vm{\alpha}}_n ^{(j)}+ \RV{w}_n^{(j)},%
\end{align}
where $\tilde{\vm{\alpha}}_n^{(j)} = [\tilde{\alpha}_{n,0}^{(j)} \; ... \; \tilde{\alpha}_{n, \tilde{K}_n^{(j)}}^{(j)} ]^{\text{T}}$ are the complex amplitudes and $\tilde{\vm{{\tau}}}_n^{(j)} = [\tilde{\tau}_{n,0}^{(j)} \; ...  \; \tilde{\tau}_{n,\tilde{K}_n^{(j)}}^{(j)}   ]^{\text{T}}$ are the delays of all $\tilde{K}_n^{(j)} + 1$ signal components, including the \ac{los} component and $\tilde{K}_n^{(j)}$ \acp{mpc} and $\vm{S}(\tilde{\vm{\tau}}_{n}^{(j)}) = [\vm{s}({\tilde{\tau}}_{n,0}^{(j)}) \, ...  \, \vm{s}({\tilde{\tau}}_{\tilde{K}_n^{(j)} }^{(j)} )]$ is the signal matrix. Since the proposed \ac{ceda} operates independently on each radio signal snapshot, we omit the indices for time $n$ and anchor $j$ in the following for brevity of notation.

Using \eqref{eq:stack_recvgen} the model \ac{lhf} of a single signal snapshot can be written as  
\begin{align}
	f(\vm{r};\vm{\tau},\vm{\alpha},\sigma^2) &= \frac{e^{-(\vm{r} - \vm{S}(\vm{\tau})\vm{\alpha})^\text{H} (\vm{r} - \vm{S}(\vm{\tau})\vm{\alpha})\sigma^{-2}}}{(\pi\sigma^2)^{N_{\text{s}}}}. \label{eq:likelihood}
\end{align}
Based on \eqref{eq:likelihood} we formulate a deterministic maximum likelihood (ML) estimator for delays of multiple components, with the complex amplitudes and the noise variance as nuisance parameters.
\begin{algorithm}[t!]\label{alg:a1}\footnotesize%
	\textbf{Initialization:}
	\begin{itemize}\setlength\itemsep{1mm}
		\item $m = 0$ and $\hat{\vm{\tau}}_0 = [\,]$\;
	\end{itemize}
	\textbf{Iterations:} \\
		\Do{$ u_{\text{ML}}(\vm{r}_{\mathrm{res}}) < \gamma$}
		{
			$m~\leftarrow~m + 1$\;
			\lIf{m=1}{set $\vm{r}_{\mathrm{res}} \leftarrow \vm{r}$}
			\lElse{
				compute $\vm{r}_{\mathrm{res}} = \vm{r} -\vm{S}(\hat{\vm{\tau}}_{m-1})\s\hat{\vm{\alpha}}_{m-1}$}%
		add component using $\hat{\tau}_{m} = \argmax\limits_{\tau_m} \frac{|\vm{r}_{\mathrm{res}}^{\text{H}}\vm{s}(\tau_m)|^2}{\vm{s}(\tau_m)^{\text{H}}\vm{s}(\tau_m)}$\;
		$\hat{\vm{\tau}}_{m}~\leftarrow$ prepend $\hat{\tau}_{m}$ to $\hat{\vm{\tau}}_{m-1}$\;\vspace{0.7mm}
		compute $\hat{\sigma}^2 = \frac{1}{{N_{\text{s}}}-1} \norm{\vm{r}_{\mathrm{res}}}{2}$\;\vspace{0.5mm}
		compute $\hat{\V{\alpha}}_m$ using $\hat{\bm{\tau}}_{m}$ in \eqref{eq:alpha_hat}\;\vspace{0.5mm}
	}%
\caption{Snapshot-based \ac{ceda}}
\end{algorithm}

Taking the natural logarithm of \eqref{eq:likelihood} enables formulating the maximization problem as \cite{Kay1993}
\begin{equation}
\{\hat{\vm{\tau}},\hat{\vm{\alpha}},\hat{\sigma}^2\} = \argmax_{\vm{\tau},\vm{\alpha},\sigma^2}\Big( \rmv \minus {N_{\text{s}}} \ln{\sigma^2} - \frac{\|\vm{r} - \vm{S}(\vm{\tau})\vm{\alpha}\|^2}{\sigma^2}\Big)\rmv\rmv\label{eq:log_likelihood}
\end{equation}
where a ``hat" denotes ML parameter estimates.
Taking the gradient \ac{wrt} $\vm{\alpha}$ we obtain a closed form solution %
given as \cite{ZiskindTASSP1988} \vspace{-0.5mm}
\begin{align}
\hat{\vm{\alpha}}%
= (\vm{S}(\vm{\tau})^{\text{H}}\vm{S}(\vm{\tau}))^{-1} \vm{S}(\vm{\tau})^{\text{H}}\vm{r}\label{eq:alpha_hat}
\end{align}
only depending on the delays.
Inserting \eqref{eq:alpha_hat} into \eqref{eq:log_likelihood} removes the amplitude dependency from the {(log-)likelihood}.
The maximization problem becomes
\begin{align}
\{\hat{\vm{\tau}},\hat{\sigma}^2\} &= \argmax_{\vm{\tau},\sigma^2}\big( -{N_{\text{s}}} \ln{\sigma^2} - \|\vm{r}\|^2 {\sigma}^{-2} \nonumber \\
&\quad+ \vm{r}^{\text{H}}\vm{S}(\vm{\tau})(\vm{S}(\vm{\tau})^{\text{H}}\vm{S}(\vm{\tau}))^{-1}\vm{S}(\vm{\tau}) ^{\text{H}}\vm{r} {\sigma}^{-2} \big). \label{eq:concentrated_log_likelihood}
\end{align}
To solve for  $\V{\tau}$ we simplify \eqref{eq:concentrated_log_likelihood}, by assuming the individual signal components to be uncorrelated \cite{Fleury1999}, i.e., $\bm{S}(\vm{\tau})^\text{H}\, \bm{S}(\vm{\tau}) = \mathrm{diag}\{[\norm{\V{s}(\tau_0)}{2}\, ... \, \norm{\V{s}(\tau_{K})}{2}]\}$. 
Thus, we can decompose the optimization problem \ac{wrt} $\V{\tau}$ into individual terms.
Following an expectation maximization scheme similar to \cite{Fleury1999}, we can solve the equation iteratively, in a bottom-up manner. The expectation term for iteration $m$ reads
\begin{equation}
\vm{r}_{\mathrm{res}} = \vm{r} -\vm{S}(\hat{\vm{\tau}}_{m-1})\,\hat{\vm{\alpha}}_{m-1}%
\end{equation}
and the maximization terms are
\begin{equation} \label{eq:ml_tau}
\hat{\tau}_m = \argmax_{\tau_m}  \frac{ |\vm{r}_{\mathrm{res}}^{\text{H}} \bm{s}(\tau_m)|^2}{\norm{\bm{s}(\tau_m)}{2} }
\end{equation}
and
\begin{equation}
\hat{\sigma}^{2} = \frac{1}{{N_{\text{s}}}-1} \norm{\vm{r}_{\mathrm{res}}}{2} .
\end{equation}

We solve \eqref{eq:ml_tau} by successively performing grid-based optimization with the grid set to $T_\text{s} / 3$ and applying a continuous unconstrained optimizer \cite{Lagarias1998}.
Following \cite{NadlerTSP2011}, we search for components until the \ac{glrt} for a single signal component in noise, as given in \eqref{eq:ml_normalized_amplitude}, falls below the detection threshold $\gamma$, which is a constant to be chosen. Note that the maximum in \eqref{eq:ml_normalized_amplitude} is approximated using the current estimates $\hat{\tau}_m$ and $\hat{\sigma}$. See \cite{LeitingerAsilomar2020} on how to determine $\gamma$ out of a fixed value for the false alarm probability per signal snapshot.

An overview of the resulting algorithm is shown in Algorithm~\ref{alg:a1}, which represents a search-and-substract approach in the sense of \cite{RichterPhD2005}. Note that the presented scheduling is suboptimal \ac{wrt} the joint update of $\vm{\alpha}$ in \eqref{eq:alpha_hat} but offers the advantage of the  execution time being in the range of tens of milliseconds even with a large number of detected signal components.%

}{}

 \acrodef{mimo}[MIMO]{multiple input multiple output}
 \acrodef{awgn}[AWGN]{additive white Gaussian noise}
 \acrodef{bw}[BW]{bandwidth}
 \acrodef{blt}[BLT]{bluetooth}
 \acrodef{cdf}[CDF]{cumulative distribution function}
 \acrodef{crlb}[CRLB]{Cram\'er-Rao lower bound}
 \acrodef{dmc}[DMC]{dense multipath component}
 \acrodef{dut}[DUT]{device under test}
 \acrodef{eirp}[EIRP]{equivalent isotropic radiated power}
 \acrodefplural{esl}[ESLs]{electronic shelf labels} 
 \acrodef{los}[LOS]{line-of-sight}
 \acrodef{mf}[MF]{matched filter}
 \acrodef{ml}[ML]{maximum likelihood}
 \acrodef{mpc}[MPC]{multipath component}
 \acrodef{nlos}[NLOS]{non-line-of-sight}
 \acrodef{pcb}[PCB]{printed circuit board}
 \acrodef{pdf}[PDF]{probability density function}
 \acrodef{reb}[REB]{ranging error bound}
 \acrodef{rss}[RSS]{received signal strength}
 \acrodef{smc}[SMC]{specular multipath component}
 \acrodef{snr}[SNR]{signal-to-noise-ratio}
 \acrodef{sinr}[SINR]{signal-to-interference-plus-noise-ratio}
 \acrodef{tdoa}[TDOA]{time difference of arrival}
 \acrodef{tka}[TKA]{trusted keyless access}
 \acrodef{toa}[TOA]{time-of-arrival}
 \acrodef{aoa}[AOA]{angle-of-arrival}
 \acrodef{uwb}[UWB]{ultra wide band}
 \acrodef{mie}[MIE]{method of interval estimation}
 \acrodef{mc}[MC]{Monte Carlo}
 \acrodef{mse}[MSE]{mean squared error}
 \acrodef{ci}[CI]{confidence interval}
 \acrodef{cl}[CL]{confidence level}
 \acrodef{pdp}[PDP]{power delay profile}
 \acrodef{dps}[DPS]{delay power spectrum}
 \acrodef{dm}[DM]{dense multipath}
 \acrodef{nlike}[NLIKE]{normalized likelihood}
 \acrodef{zzb}[ZZB]{Ziv-Zakai bound}
 \acrodef{ut}[UT]{unscented transform}
 \acrodef{glrt}[GLRT]{generalized likelihood ratio test}
 \acrodef{mse}[MSE]{mean squared error}
 \acrodef{rmse}[RMSE]{root mean squared error}
 \acrodef{nnlike}[NNLIKE]{normalized noise-free likelihood}
 \acrodef{stdv}[STDV]{standard deviation}
 \acrodef{rv}[RV]{random variable}
 \acrodef{bp}[BP]{belief propagation}
 \acrodef{pda}[PDA]{probabilistic data association}
 \acrodef{mp}[MP]{multipath}
 \acrodef{pmf}[PMF]{probability mass function}
 \acrodef{pdaf}[PDAF]{probabilistic data association filter}
 \acrodef{pdaai}[AIPDA]{amplitude-information \ac{pda}}
 \acrodef{olos}[OLOS]{obstructed line-of-sight}
 \acrodef{spa}[SPA]{sum-product algorithm}
 \acrodef{mmse}[MMSE]{minimum mean-square error}
 \acrodef{lhf}[LHF]{likelihood function}
 \acrodef{fa}[FA]{false alarm}
 \acrodef{ceda}[CEDA]{channel estimation and detection algorithm} 
 \acrodef{pcrlb}[P-CRLB]{posterior Cram\'er-Rao lower bound}
 \acrodef{mpslam}[MP-SLAM]{multipath-based SLAM}
 \acrodef{va}[VA]{virtual anchor}
 \acrodef{dnr}[DNR]{dense-to-noise ratio}
 \acrodef{pbo}[PBO]{potential bias object}
 \acrodef{npbo}[NPBO]{new \ac{pbo}}
 \acrodef{lpbo}[LPBO]{legacy \ac{pbo}}
 \acrodef{aednn}[AE-DNN]{autoencoder deep neural network}   
 \acrodef{gpr}[GPR]{Gaussian process regression}  
 \acrodef{cluster}[CLUSTER]{{\color{red}error}}  
 \acrodef{delaybias}[ML-BIAS]{{\color{red}error}}  
 \acrodef{gptrack}[GP-TRACK]{{\color{red}error}}  
 \acrodef{chslam}[CH-SLAM]{{\color{red}error}}  
 \acrodef{wrt}[w.r.t.]{with respect to} 

\bibliographystyle{IEEEtran}
\bibliography{IEEEabrv,References,TempRefs}